\documentclass[apj]{emulateapj}
\received{August 2014}
\revised{September 2014}
\accepted{20 October 2014}
\shorttitle{SNRs in the LMC with Herschel}
\shortauthors{Laki\'cevi\'c et al.}
\usepackage{longtable}
\usepackage{epsfig}
\usepackage{rotating}
\usepackage{verbatim}
\usepackage{appendix}
\begin{document}
\title{The influence of supernova remnants on the interstellar medium in the Large Magellanic Cloud seen at 20--600 $\mu$m wavelengths}
       \author{Ma\v{s}a Laki\'cevi\'c\altaffilmark{1}, Jacco Th. van Loon\altaffilmark{1}, Margaret Meixner\altaffilmark{2,3}, Karl Gordon\altaffilmark{2,4}, 
        Caroline Bot\altaffilmark{5}, Julia Roman-Duval\altaffilmark{2}, Brian Babler\altaffilmark{6}, Alberto Bolatto\altaffilmark{7}, Chad Engelbracht\altaffilmark{8}, 
        Miroslav Filipovi\'c\altaffilmark{9}, Sacha Hony\altaffilmark{10}, Remy Indebetouw\altaffilmark{11,12}, Karl Misselt\altaffilmark{8}, Edward Montiel\altaffilmark{8,13},
        K. Okumura\altaffilmark{10}, Pasquale Panuzzo\altaffilmark{10,14}, Ferdinando Patat\altaffilmark{15}, Marc Sauvage\altaffilmark{10}, Jonathan Seale\altaffilmark{2,16},
        George Sonneborn\altaffilmark{17}, Tea Temim\altaffilmark{17,18}, Dejan Uro\v{s}evi\'c\altaffilmark{19,20} \& Giovanna Zanardo\altaffilmark{21}} 
        
       \altaffiltext{1}{Lennard-Jones Laboratories, Keele University, ST5 5BG, UK; m.lakicevic@keele.ac.uk}
       \altaffiltext{2}{Space Telescope Science Institute, 3700  San Martin Dr., Baltimore, MD 21218, USA}      
       \altaffiltext{3}{Department of Physics and Astronomy, Johns Hopkins University, 366 Bloomberg Center, 3400 N. Charles Street, Baltimore, MD 21218, USA}  
       \altaffiltext{4}{Sterrenkundig Observatorium, Universiteit Gent, Gent, Belgium} 
       \altaffiltext{5}{Observatoire astronomique de Strasbourg, Universit\'{e} de Strasbourg, CNRS, UMR 7550, 11 rue de l'universit\'{e}, F-67000 Strasbourg, France}    
       \altaffiltext{6}{Department of Astronomy, 475 North Charter St., University of Wisconsin, Madison, WI 53706, USA}          
       \altaffiltext{7}{Laboratory of Millimeter Astronomy, University of Maryland, College Park, MD 29742, USA} 
       \altaffiltext{8}{Steward Observatory, University of Arizona, 933 North Cherry Ave., Tucson, AZ 85721, USA}
       \altaffiltext{9}{University of Western Sydney, Locked Bag 1797, Penrith South DC, NSW 1797, Australia}
       \altaffiltext{10}{CEA, Laboratoire AIM, Irfu/SAp, Orme des Merisiers, F-91191 Gif-sur-Yvette, France}
       \altaffiltext{11}{Department of Astronomy, University of Virginia, P.O.\ Box 400325, Charlottesville, VA 22903, USA}
       \altaffiltext{12}{National Radio Astronomy Observatory, 520 Edgemont Road, Charlottesville, VA 22903, USA}
       \altaffiltext{13}{Louisiana State University, Department of Physics \& Astronomy, 233-A Nicholson Hall, Tower Dr., Baton Rouge, LA 70803, USA}
       \altaffiltext{14}{CNRS, Observatoire de Paris - Lab. GEPI, Bat. 11, 5, place Jules Janssen, 92195 Meudon CEDEX, France}
       \altaffiltext{15}{European Organization for Astronomical Research in the Southern Hemisphere (ESO), Karl-Schwarzschild-Stra{\ss}e 2, 85748 Garching bei M\"unchen, Germany} 
       \altaffiltext{16}{The Johns Hopkins University, Department of Physics and Astronomy, 366 Bloomberg Center, 3400 N. Charles Street, Baltimore, MD 21218, USA}
       \altaffiltext{17}{NASA Goddard Space Flight Center, Code 665, Greenbelt, MD 20771, USA}
       \altaffiltext{18}{CRESST, University of Maryland, College Park, MD 20742, USA}
       \altaffiltext{19}{Department of Astronomy, Faculty of Mathematics, University of Belgrade, Studentski trg 16, 11000 Belgrade, Serbia}
       \altaffiltext{20}{Isaac Newton Institute of Chile, Yugoslavia Branch}  
       \altaffiltext{21}{International Centre for Radio Astronomy Research (ICRAR), M468, University of Western Australia, Crawley, WA 6009, Australia}

\begin{abstract}
We present the analysis of supernova remnants (SNRs) in the Large Magellanic
Cloud (LMC) and their influence on the environment at far-infrared (FIR) and
submillimeter wavelengths. We use new observations obtained with the {\it Herschel}
Space Observatory and archival data obtained with the {\it Spitzer} Space
Telescope, to make the first FIR atlas of these objects. The SNRs are
not clearly discernible at FIR wavelengths, however their influence becomes
apparent in maps of dust mass and dust temperature, which we constructed by
fitting a modified black-body to the observed spectral energy
distribution in each sightline. Most of the dust that is seen is pre-existing
interstellar dust in which SNRs leave imprints. The temperature maps clearly
reveal SNRs heating surrounding dust, while the mass maps indicate the removal of 
3.7$^{+7.5}_{-2.5}$ M$_{\odot}$ of dust per SNR. This agrees with the calculations 
by others that significant amounts of dust are sputtered by SNRs. Under the assumption 
that dust is sputtered and not merely pushed away, we estimate a dust destruction rate in
the LMC of $0.037^{+0.075}_{-0.025}$ M$_\odot$ yr$^{-1}$ due to SNRs, yielding an average
lifetime for interstellar dust of $2^{+4.0}_{-1.3}\times10^7$ yr. We conclude that
sputtering of dust by SNRs may be an important ingredient in models of galactic
evolution, that supernovae may destroy more dust than they produce, and that they therefore may not be 
net producers of long lived dust in galaxies.

\end{abstract}

\keywords{ISM: clouds
-- dust
-- ISM: supernova remnants 
-- Magellanic Clouds
-- infrared: ISM
-- submillimeter: ISM}

\vspace{.2in}
\section{Introduction}
Supernov{\ae} (SNe) could be significant dust producers in galaxies, since
around 0.1--1 M$_\odot$ of dust can be produced in their ejecta as observations of 
some SNRs \citep{Barlow10,matsuura,Gomez12a} and theoretical models suggest \citep{Bianchi07,Nozawa10}. However, the amounts of dust seen 
at high redshift are difficult to reconcile with dust forming in SNe alone 
\citep{Silvia10, Rowlands14}. There have been ample detections of dust created 
in SN ejecta shortly after the explosion \citep{Elmhamdi03, Fox09, matsuura}, and 
in young SNRs such as Cas\,A \citep{Nozawa10}, Crab \citep{Gomez12a,Temim12,Temim13}, E\,0102 
\citep{Stanimirovic05, Sandstrom09}, N\,132D and G\,11.2$-$0.3 \citep{Rho09}, but 
the inferred masses are generally well below theoretical predictions. While \citet{Gomez12b} noticed 
the lack of dust production in Type Ia supernova remnants (SNRs), and dust produced from Ic and Ib SNe has not been 
observed, there are indications for dust to be formed from IIn and IIP SNe \citep{Gall10}.
However, for many SNe it is not certain whether the dust was pre-existing or formed in ejecta and some SNe were not 
detected \citep{Szalai13}. SNe and SNRs also sputter dust in 
the surrounding interstellar medium (ISM) and pre-burst circumstellar medium 
(CSM). While it is well established that dust grows in evolved stars (e.g., AGB stars, see \citealt{Boyer12}), 
it is not yet clear whether the net result of SNe and SNRs is a supply or removal 
of interstellar dust, and hence alternative solutions for dust growth are being considered, e.g., in the 
ISM \citep{Zhukovska14}.

The Large Magellanic Cloud (LMC) is a convenient place to study populations of SNRs because there is little
foreground and internal contamination from interstellar clouds, the distances
to all SNRs in the LMC are essentially identical and well known, and the LMC
is close enough ($\approx 50$ kpc, \citealt{Walker12}) to resolve the far-infrared (FIR) and submillimeter
(submm) emission of remnants with diameters $>9$ pc ($>90$\% of known
objects). Hence, SNRs in the LMC have been the subject of many detailed studies at
all wavelengths. Here, we use the submm data obtained as part of
the HERITAGE (HERschel Inventory of The Agents of Galaxy Evolution) survey, \citep{Meixner13},
covering 100--500 $\mu$m, together with archival {\it Spitzer} Space
Telescope data at 24 and 70 $\mu$m from the Surveying the Agents of Galaxy Evolution
(SAGE) LMC survey \citep{Meixner06}, to quantify the influence of SNRs on the
ISM of the LMC.

Many SNRs are detected at infrared (IR) wavelengths in the Galaxy and in the Magellanic Clouds \citep{Reach06, Seok08, Seok13, Williams10}. 
The radiation at $\lambda\gtrsim$ 24 $\mu$m is mostly dust from swept-up 
ISM collisionally heated by the hot plasma generated by SNR shocks, while emission at shorter wavelengths originates from
ionic/molecular lines, polycyclic aromatic hydrocarbons emission (PAH), or synchrotron 
emission \citep{Seok13}. The IR emission from SNRs may also include fine-structure line emission from hot plasma and/or shocks 
\citep{vanloon10} and the contributions from small grains which are stochastically heated and which may otherwise be rather cold. 

However, the only detections of Magellanic SNRs at submm wavelengths ($\lambda\gtrsim$ 100 $\mu$m) are SN\,1987A, due to dust formed in the ejecta
\citep{lakicevic, lakicevicm1, lakicevicm2, matsuura, indebetouw} and LHA\,120-N\,49,
explained by 10 M$_\odot$ of dust in an interstellar cloud heated up by the
forward shock \citep{Otsuka10}. The submm emission from SNRs could include a
non-thermal (synchrotron) component from a strongly magnetized plasma, for
instance if a pulsar wind nebula (PWN) is present \citep{Temim12}.


The most common ways for sputtering of grains in SNRs are thermal
sputtering, when energetic particles knock atoms off the grain surface
\citep{Casoli98}, more often in fast shocks, $v >150$ km s$^{-1}$ and 
grain--grain collisions, dominant in slower shocks, $\leq 50$--80 km s$^{-1}$ \citep{Jones94}, often 
called shattering. Sputtering is most effective on small grains \citep{Andersen11}, 
resulting in a deficit of small grains in SNRs compared to the ISM. Shattering is destroying primarily big 
grains. Big grains (BG) become small (SG) which 
produces increased SG-to-BG ratio \citep{Andersen11}.

\citet{Sankrit10} showed that $\sim35$\% of dust is sputtered in the Cygnus\,Loop,
a Galactic SNR, by modeling the flux ratio at 70 and 24 $\mu$m in the
post-shock region. Just behind the shock this ratio was 14, compared to 22
further out of the remnant, which could be understood in terms of the
destruction and heating of the dust by the shocks of this middle-aged SNR
\citep[$\sim10,000$ yr --][]{Blair05}. \citet{Arendt10} showed that the
interaction of the shock in young SNR Puppis\,A (3,700 yr) with a molecular
cloud has led to the destruction of $\approx 25$\% of the population of very
small grains (PAH). \citet{Micelotta10} 
explored processing of PAHs in interstellar shocks by ion and electron collisions, 
finding that interstellar PAHs do not survive in shocks of velocities greater than 100 
km s$^{-1}$. Various other studies like \citet{Borkowski06a}, \citet{WilliamsB06,Williams10}, also
found that significant amounts of dust were sputtered in the shocks of
young SNRs. 

For a homogeneous ISM and under the assumption that silicon and carbon
grains are equally mixed, \citet{Dwek07} obtained that the mass of the ISM, which is completely
cleared of dust by one single SNR can be as high as 1200 M$_{\odot}$. On the other hand, \citet{Bianchi07}
predict 0.1 M$_{\odot}$ of dust produced in SN to survive the passage of the reverse shock.

This paper is organised as follows: in Section~\ref{datamethods} we introduce
the sample of SNRs in the LMC, the data we analyse and the methods we use; 
in Section~\ref{results} we present the results, in
particular regarding the surface brightness, flux ratios, and dust mass 
and temperature maps; in Section~\ref{discussion} we discuss the implications, in terms of dust
removal and the ISM properties within which the SNRs evolve. We summarise
our conclusions in Section~\ref{conclusions}.

\section{Data and methods}\label{datamethods}

\subsection{Sample of objects} \label{sample}

\begin{table}
\caption{Complete sample of SNRs in the LMC.\label{tbl5}}
\footnotesize
\begin{tabular}{lccr}
\hline\hline
Name & RA [h m s]\tablenotemark{a} & Dec [$^{\circ}$ $^{\prime}$ $^{\prime\prime}$]\tablenotemark{a} &
$D$ [$^{\prime\prime}$]\tablenotemark{a} \\
\hline
J\,0448.4$-$6660                & 04 48 22 & $-$66 59 52 & 220 \\
J\,0449.3$-$6920                & 04 49 20 & $-$69 20 20 & 133 \\
B\,0449$-$693\tablenotemark{b}  & 04 49 40 & $-$69 21 49 & 120 \\
B\,0450$-$6927                  & 04 50 15 & $-$69 22 12 & 210 \\
B\,0450$-$709                   & 04 50 27 & $-$70 50 15 & 357 \\
LHA\,120-N\,4 (N\,4)            & 04 53 14 & $-$66 55 13 & 252 \\
0453$-$68.5                     & 04 53 38 & $-$68 29 27 & 120 \\
B\,0454$-$7000                  & 04 53 52 & $-$70 00 13 & 420 \\
LHA\,120-N\,9 (N\,9)            & 04 54 33 & $-$67 13 13 & 177 \\
LHA\,120-N\,11L (N\,11L)        & 04 54 49 & $-$66 25 32 &  87 \\
LHA\,120-N\,86 (N\,86)          & 04 55 37 & $-$68 38 47 & 348 \\
LHA\,120-N\,186D (N\,186D)      & 04 59 55 & $-$70 07 52 & 150 \\
DEM\,L71                        & 05 05 42 & $-$67 52 39 &  72 \\
LHA\,120-N\,23 (N\,23)          & 05 05 55 & $-$68 01 47 & 111 \\
J\,0506.1$-$6541                & 05 06 05 & $-$65 41 08 & 408 \\
B\,0507$-$7029                  & 05 06 50 & $-$70 25 53 & 330 \\
RXJ\,0507$-$68\tablenotemark{b} & 05 07 30 & $-$68 47 00 & 450 \\
J0508$-$6830\tablenotemark{d}   & 05 08 49 & $-$68 30 41 & 123 \\
LHA\,120-N\,103B (N\,103B)      & 05 08 59 & $-$68 43 35 &  28 \\
0509$-$67.5                     & 05 09 31 & $-$67 31 17 &  29 \\
J0511$-$6759\tablenotemark{d}   & 05 11 11 & $-$67 59 07 & 108 \\
DEM\,L109                       & 05 13 14 & $-$69 12 20 & 215 \\
J0514$-$6840\tablenotemark{d}   & 05 14 15 & $-$68 40 14 & 218 \\
J0517$-$6759\tablenotemark{d}   & 05 17 10 & $-$67 59 03 & 270 \\
LHA\,120-N\,120 (N\,120)        & 05 18 41 & $-$69 39 12 & 134 \\
0519$-$69.0                     & 05 19 35 & $-$69 02 09 &  31 \\
0520$-$69.4                     & 05 19 44 & $-$69 26 08 & 174 \\
J\,0521.6$-$6543                & 05 21 39 & $-$65 43 07 &  162$^{e}$ \\
LHA\,120-N\,44 (N\,44)          & 05 23 07 & $-$67 53 12 & 228 \\
LHA\,120-N\,132D (N\,132D)      & 05 25 04 & $-$69 38 24 & 114 \\
LHA\,120-N\,49B (N\,49B)        & 05 25 25 & $-$65 59 19 & 168 \\
LHA\,120-N\,49 (N\,49)          & 05 26 00 & $-$66 04 57 &  84 \\
B\,0528$-$692                   & 05 27 39 & $-$69 12 04 & 147 \\
DEM\,L\,204                     & 05 27 54 & $-$65 49 38 & 303 \\
HP99498\tablenotemark{c}        & 05 28 20 & $-$67 13 40 &  97 \\
B\,0528$-$7038                  & 05 28 03 & $-$70 37 40 &  60 \\
DEM\,L203                       & 05 29 05 & $-$68 32 30 & 667 \\
DEM\,L214                       & 05 29 51 & $-$67 01 05 & 100 \\
DEM\,L214\tablenotemark{c}      & 05 29 52 & $-$66 53 31 & 120 \\
DEM\,L218                       & 05 30 40 & $-$70 07 30 & 213 \\
LHA\,120-N\,206 (N\,206)        & 05 31 56 & $-$71 00 19 & 192 \\
0532$-$67.5                     & 05 32 30 & $-$67 31 33 & 252 \\
B\,0534$-$69.9                  & 05 34 02 & $-$69 55 03 & 114 \\
DEM\,L238                       & 05 34 18 & $-$70 33 26 & 180 \\
SN\,1987A                       & 05 35 28 & $-$69 16 11 &   2 \\
LHA\,120-N\,63A (N\,63A)        & 05 35 44 & $-$66 02 14 &  66 \\
Honeycomb                       & 05 35 46 & $-$69 18 02 & 102 \\
DEM\,L241                       & 05 36 03 & $-$67 34 35 & 135 \\
DEM\,L249                       & 05 36 07 & $-$70 38 37 & 180 \\
B\,0536$-$6914                  & 05 36 09 & $-$69 11 53 & 480 \\
DEM\,L256                       & 05 37 27 & $-$66 27 50 & 204 \\
B\,0538$-$6922                  & 05 37 37 & $-$69 20 23 & 169 \\
0538$-$693\tablenotemark{b}     & 05 38 14 & $-$69 21 36 & 169 \\
LHA\,120-N\,157B (N\,157B)      & 05 37 46 & $-$69 10 28 & 102 \\
LHA\,120-N\,159 (N\,159)        & 05 39 59 & $-$69 44 02 &  78 \\
LHA\,120-N\,158A (N\,158A)      & 05 40 11 & $-$69 19 55 &  60 \\
DEM\,L299                       & 05 43 08 & $-$68 58 18 & 318 \\
DEM\,L316B                      & 05 46 59 & $-$69 42 50 &  84 \\
DEM\,L316A                      & 05 47 22 & $-$69 41 26 &  56 \\
0548$-$70.4                     & 05 47 49 & $-$70 24 54 & 102 \\
J\,0550.5$-$6823                & 05 50 30 & $-$68 22 40 & 312 \\ 
\hline
\end{tabular}
\tablenotetext{a}{From \citet{Badenes10}, if not written different.}
\tablenotetext{b}{This SNR is from the \citet{Blair06} catalogue.}
\tablenotetext{c}{This SNR is from Filipovi\'c's unpublished catalogue.}
\tablenotetext{d}{From \citet{Maggi14}.}
\tablenotetext{e}{From \citet{Desai10}.}
\end{table}

\begin{table*}
\begin{center}
\caption{The LMC SNRs with known age and type from the literature.\label{tbl2}}
\begin{tabular}{lccrrrllr}
\hline\hline
Name & RA [h m s]\tablenotemark{a} & Dec [$^{\circ}$ $^{\prime}$ $^{\prime\prime}$]\tablenotemark{a} &
$D$ [pc]\tablenotemark{a} & age [yr] & ref. age &
$F_{1.4{\rm GHz}}$ [Jy]\tablenotemark{a} & type & ref. type \\
\hline
0450$-$70.9    & 04 50 27 &$-$70 50 15 & 86   &$ \leq 45000$ & 26          & 0.56 & CC?  & 27  \\
B\,0453$-$68.5 & 04 53 38 &$-$68 29 27 & 29   & 13000        & 1           & 0.11 & CC   & 1 \\
N\,9           & 04 54 33 &$-$67 13 13 & 43   & 30000        & 24          & 0.06 & Ia?  & 24 \\
N\,11L         & 04 54 49 &$-$66 25 32 & 21   & 11000        & 2           & 0.11 & CC   & 9 \\
N\,86          & 04 55 37 &$-$68 38 47 & 84   & 86000        & 2           & 0.26 & CC?  & 2 \\
DEM\,L71       & 05 05 42 &$-$67 52 39 & 17   & 4400         & 3           & 0.01 & Ia   & 3 \\
N\,23          & 05 05 55 &$-$68 01 47 & 27   & 6300         & 3,4         & 0.35 & CC   & 5 \\
N\,103B        & 05 08 59 &$-$68 43 35 & 7    & 1000         & 6           & 0.51 & Ia   & 6 \\
B\,0509$-$67.5 & 05 09 31 &$-$67 31 17 & 8    & 400          & 7           & 0.08 & Ia   & 7 \\
N\,120         & 05 18 41 &$-$69 39 12 & 32   & 7300         & 8           & 0.35 & CC   & 9 \\
B\,0519$-$69.0 & 05 19 35 &$-$69 02 09 & 8    & 600          & 7           & 0.1  & Ia   & 7 \\
N\,44          & 05 23 07 &$-$67 53 12 & 55   & 18000        & 10          & 0.14 & CC?  & 28\\
N\,132D        & 05 25 04 &$-$69 38 24 & 28   & 2750         & 3           & 3.71 & CC   & 11\\
N\,49B         & 05 25 25 &$-$65 59 19 & 41   & 10900        & 3           & 0.32 & CC   & 12\\
N\,49          & 05 26 00 &$-$66 04 57 & 20   & 6600         & 13          & 1.19 & CC   & 14\\
N\,206         & 05 31 56 &$-$71 00 19 & 46   & 25000        & 15          & 0.33 & CC   & 15\\
0534$-$69.9    & 05 34 02 &$-$69 55 03 & 27   &10000         & 16          & 0.08 & Ia   & 16\\
DEM\,L238      & 05 34 18 &$-$70 33 26 & 43   & 13500        & 17          & 0.06 & Ia   & 17\\
SN\,1987A      & 05 35 28 &$-$69 16 11 & 0.5    & 27         & 18          & 0.05 & CC   & 18\\
N\,63A         & 05 35 44 &$-$66 02 14 & 16   & 3500         & 19          & 1.43 & CC   & 19\\
DEM\,L249      &05 36 07  &$-$70 38 37  & 43   & 12500       & 17          & 0.05 & Ia   & 17\\
N\,157B        &05 37 46  &$-$69 10 28  & 25   &5000         & 20          & 2.64 & CC   & 20 \\
N\,159         &05 39 59  &$-$69 44 02  & 19   &18000        & 21          & 1.9  & CC   & 21 \\
B\,0540$-$69.3 &05 40 11  &$-$69 19 55  & 15   & 800         & 22          &0.88  & CC   & 5 \\
DEM\,L316A     &05 47 22  &$-$69 41 26  & 14   & 33000       & 23          & 0.33 & Ia   & 23 \\
B\,0548$-$70.4 &05 47 49  &$-$70 24 54  & 25   & 7100        & 3           & 0.05 & CC   & 3 \\
DEM\,L241      &05 36 03  &$-$67 34 35 & 33   & 12000        & 25          & 0.29 & CC   & 25 \\
\hline
\end{tabular}
\tablenotetext{a}{From \citet{Badenes10}.}
\tablerefs{
(1) \citet{Haberl12};
(2) \citet{Williams99};
(3) \citet{Williams10};
(4) \citet{Someya10};
(5) \citet{Hayato06};
(6) \citet{Lewis03};
(7) \citet{Rest05};
(8) \citet{Rosado93};
(9) \citet{Chu88};
(10) \citet{Chu93};
(11) \citet{Vogt11};
(12) \citet{Park03a};
(13) \citet{Park03b};
(14) \citet{Otsuka10};
(15) \citet{Williams05};
(16) \citet{Hendrick03};
(17) \citet{Borkowski06b};
(18) \citet{Zanardo10};
(19) \citet{Hughes98};
(20) \citet{Wang01};
(21) \citet{Seward10};
(22) \citet{Badenes09};
(23) \citet{Williams2005};
(24) \citet{Seward06};
(25) \citet{Seward12};
(26) \citet{Williams04};
(27) The SNR appears centrally filled in X-rays and radio, but no point source is detected in either radio or X-ray observations \citep{Williams04}.
(28) The remnant is near H\,{\sc ii} complexes and OB associations \citep{Chu93}.
}
\end{center}
\end{table*}

We examined 61 SNRs in the LMC, i.e.\ all objects that are relatively
certain to be SNRs and that have accurate positions and dimensions selected
using all existing survey catalogues: \citet{Williams99}, \citet{Blair06}, \citet{Seok08}, 
\citet{Payne08}, \citet{Badenes10}, \citet{Desai10}, \citet{Maggi14} and an unpublished catalogue of Filipovi\'c et al. 
The complete sample of targets, their positions and sizes are given in Table~\ref{tbl5}, with more details for SNRs 
for which ages and types have been determined in Table~\ref{tbl2}.

The explosion type is often not known or is uncertain and the assumed core-collapse (CC) SNRs in our sample 
may still harbour some Ia, while the group of assumed Ias are the SNRs that are considered to be Ias in the literature.

\subsection{FIR and submm data}\label{makx}

We use FIR and submm data from the {\it Herschel} Space Observatory
open time key programme, HERschel Inventory of The Agents of Galaxy Evolution
\citep[HERITAGE --][]{Meixner13}, of the LMC, comprising images
obtained with SPIRE (Spectral and Photometric Imaging Receiver) at 250, 350,
and 500 $\mu$m and with PACS (Photodetector Array Camera and Spectrometer) at
100 and 160 $\mu$m. We complement this with IR images from the Surveying the Agents of a Galaxy's
Evolution (SAGE) project, of the LMC \citep{Meixner06}, obtained with the MIPS 
instrument onboard the {\it Spitzer} Space Telescope at 24 and 70 $\mu$m.
 
\begin{figure*}
\centerline{\vbox{
\hbox{
\psfig{figure=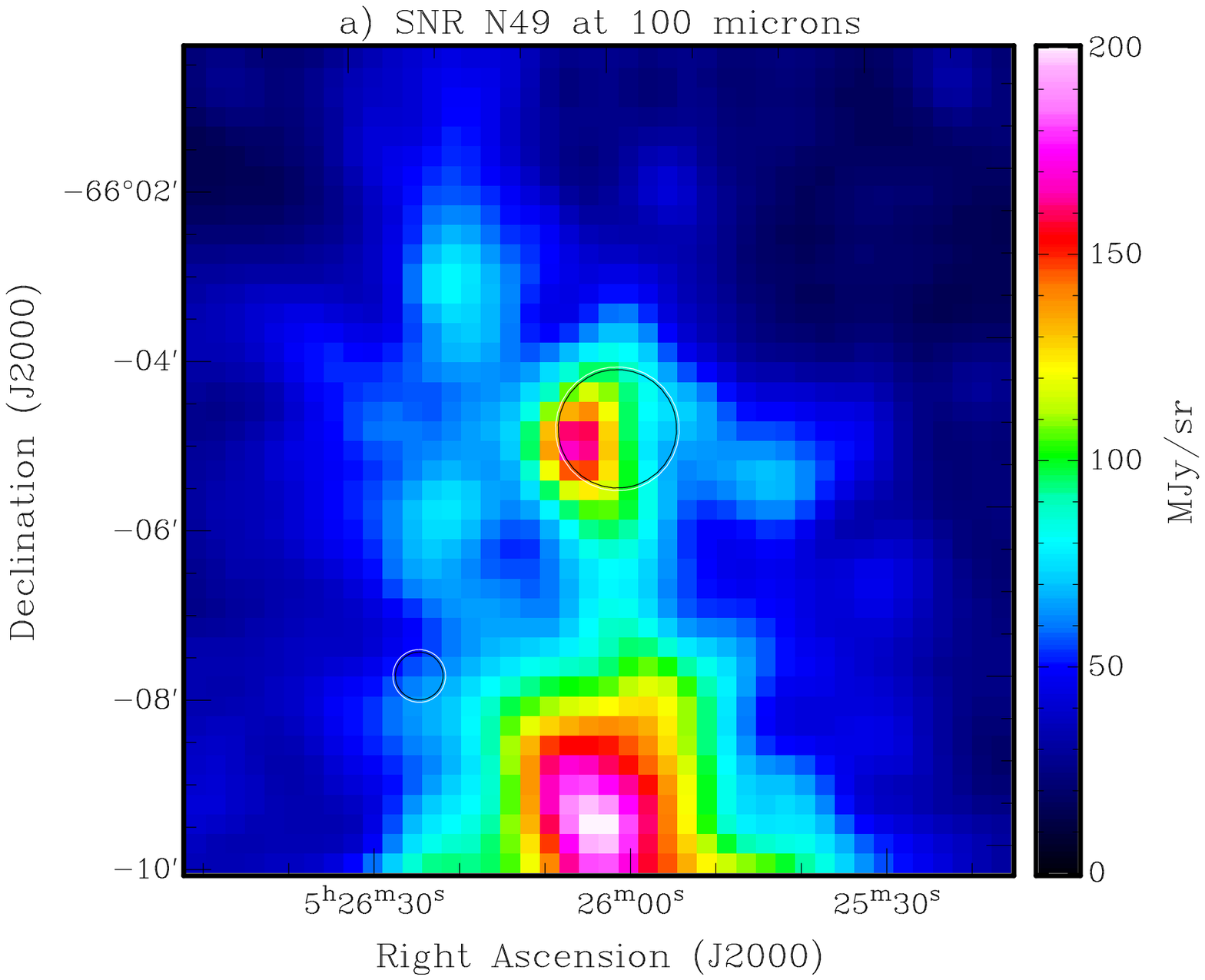,width=60mm}   
\psfig{figure=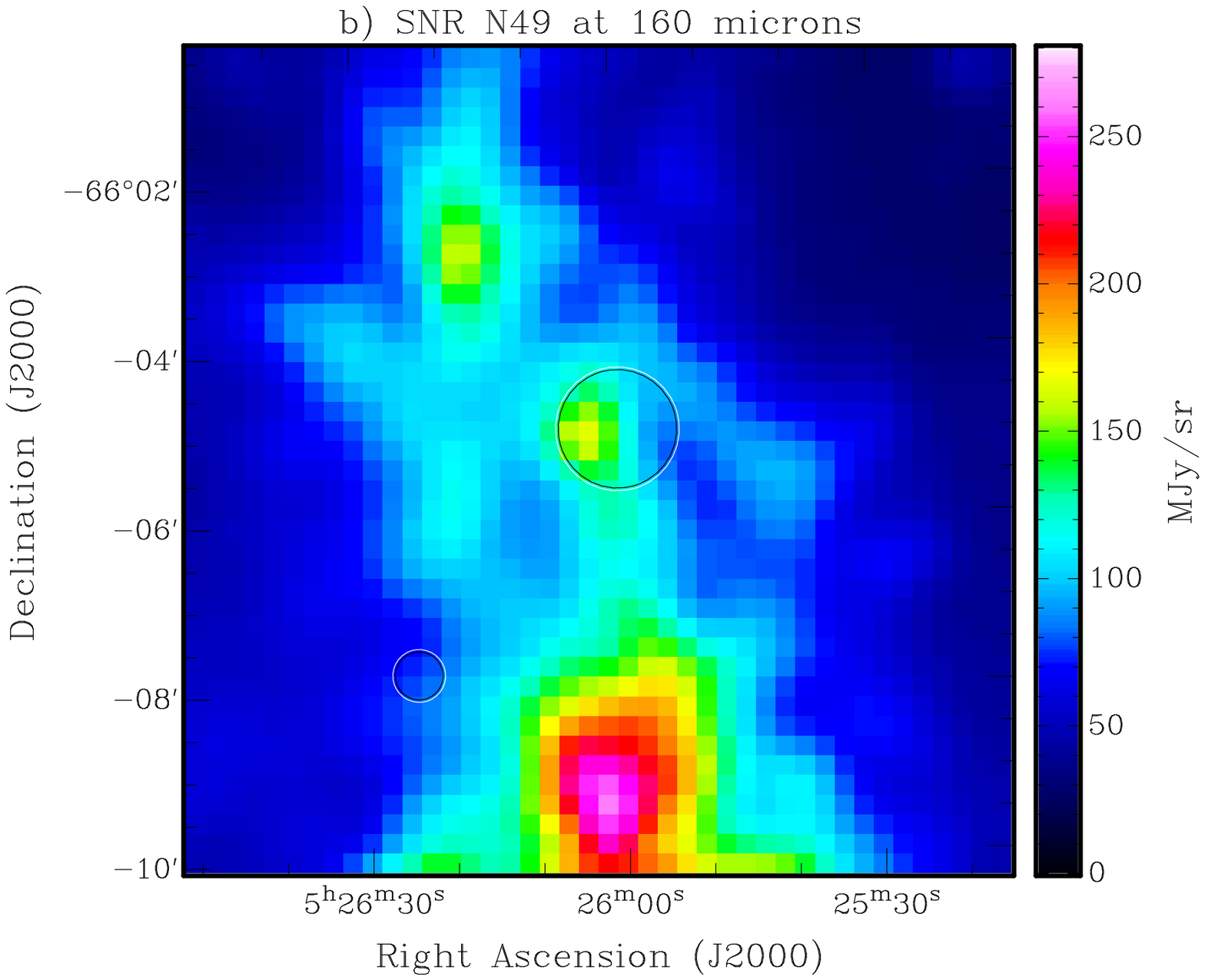,width=60mm}
\psfig{figure=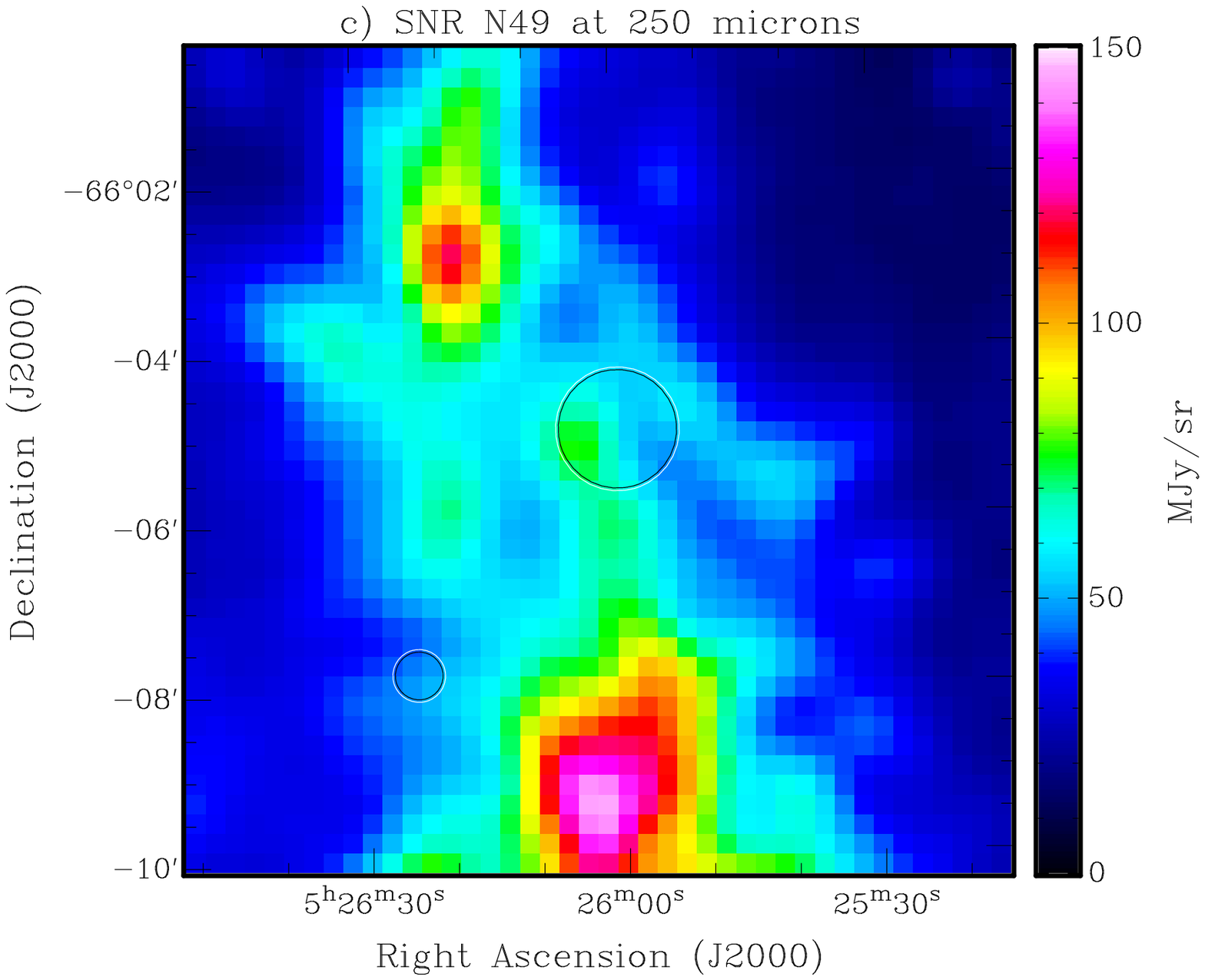,width=60mm}}
\hbox{
\psfig{figure=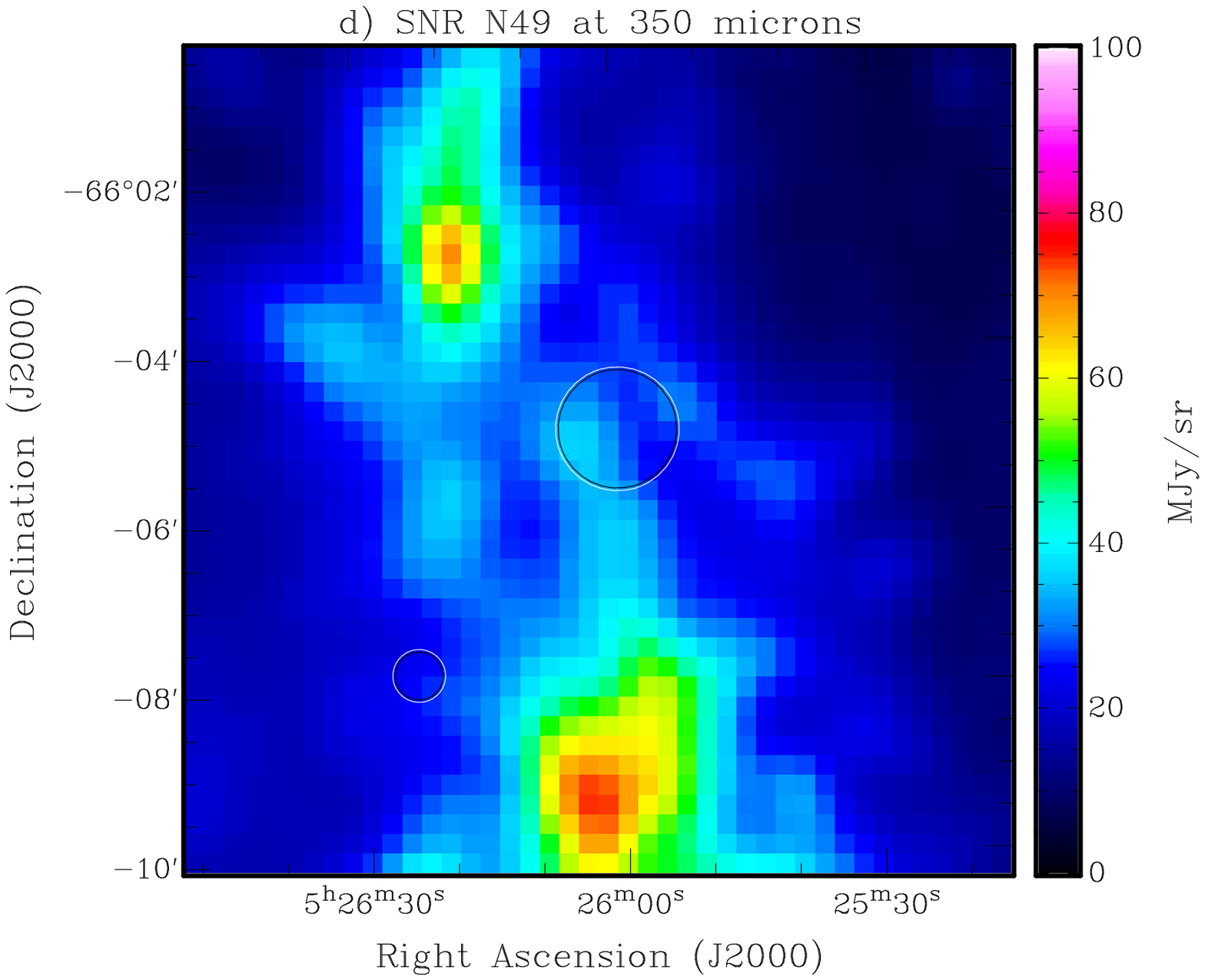,width=60mm}
\psfig{figure=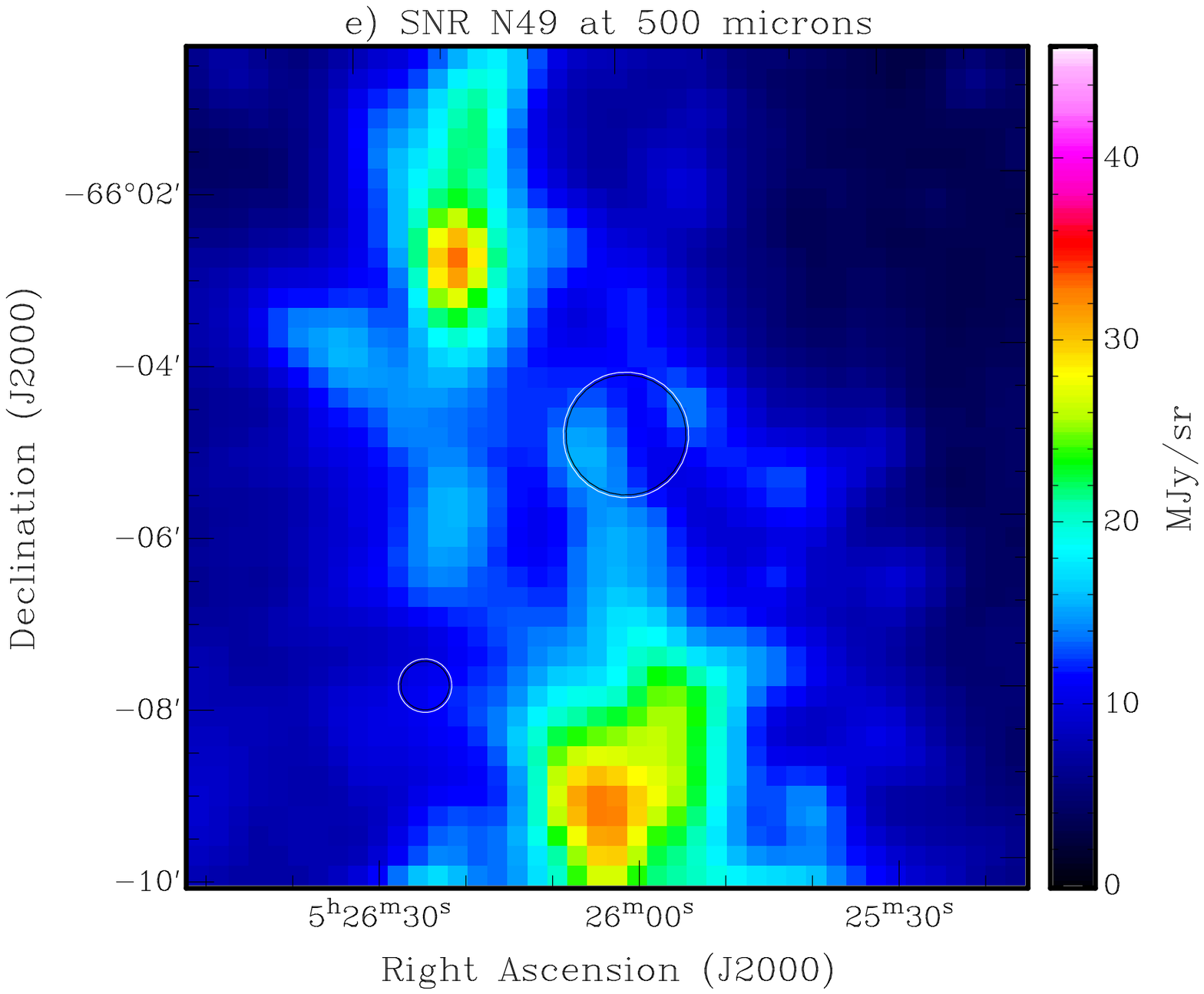,width=60mm}}  
}}
\caption[]{Images of SNR N\,49: {\it a:} at 100, {\it b:} at 160, {\it c:} at 250, {\it d:} at 350 and {\it e:} at 500 $\mu$m, shown on the resolution of $36\rlap{.}^{\prime\prime}3$. Beam size is presented by the circle in the lower left corner and the
remnant dimension and position by the circle in the middle. \label{N49_first}}
\end{figure*}

For making mass and temperature maps we use {\it Herschel} images at 100, 160, 250, 350 and 500 $\mu$m from \citet{Gordon14}. These images have the same resolution 
($36\rlap{.}^{\prime\prime}3$), which is the limiting resolution of the 500 $\mu$m band, are projected to the same pixel size (14$^{\prime\prime}$)
and are subtracted for the residual foreground Milky Way cirrus emission and the sea of 
unresolved background galaxies. In Figure~\ref{N49_first} we show an example of these data for one object -- SNR N\,49. 
For the ratio maps and the rest of this paper we also use similar maps of MIPS data at 24 and 70 $\mu$m.

The errors in the MIPS data are calculated by adding in quadrature the flux
calibration uncertainties of 2\% and 5\% (for 24 and 70 $\mu$m data) and background 
noise measured away from the galaxy. Likewise, errors in the PACS and SPIRE images were 
found by combining in quadrature, respectively 10\% and 8\%
uncertainties of the absolute flux calibration \citep{Meixner13} with the background
noise described by \citet{Gordon14}. In their work it is conservatively assumed that the uncertainties between bands were not
correlated and the impact of this assumption on the resulting dust masses is discussed.

\section{Results}\label{results}

\subsection{Evolution of the FIR, submm and radio surface brightness and
diameter with SNR age}\label{EvolutionFIRRadio}

We first examine the FIR and radio surface brightness with regard to SNR diameters, to check if there is any evolution in the time. In this
analysis we do not apply colour corrections. The radio surface brightness is obtained from
\begin{equation}
\Sigma_\nu\left[{\rm W m^{-2} Hz^{-1} sr^{-1}}\right]\ =\ 1.505\times10^{-19}\
\frac{F_\nu [{\rm Jy}]}{\Theta[^{\prime}]^2},
\end{equation}
where $F_\nu$ is the flux density and $\Theta$ the angular diameter of a source \citep{Vukotic09}. Surface brightness
of SNRs is decreasing with diameter at radio wavelengths because the objects spread, cool down and mix with ISM. 

\begin{table*}
\begin{center}
\caption{Relations between surface brightness ($\Sigma$) and SNR diameters. For these linear relations we tabulate the correlation coefficient, constant $A$ and slope 
$B$, where log $y=A + B$ log $x$. \label{tbl1}}
\begin{tabular}{llllllll}
\hline\hline
                            & 
 \multicolumn{3}{c}{$\log D$ [pc]}                     & &  \multicolumn{3}{c}{$\log\Sigma_{1.4{\rm GHz}}$} \\     
\cline{2-4} \cline{6-8}\\
$\log\Sigma$                       &               c. corr. &             $A$ &                    $B$ &  &               c. corr. &             $A$ &                    $B$ \\
\hline
$\log\Sigma_{24}$  &          $-0.3 $     & $-18.53 \pm 0.78$ & $ -0.85 \pm 0.55$     & &           0.7       & $-6.61\pm2.71$   &  $0.66    \pm0.14$             \\
$\log\Sigma_{160}$ &          $-0.14$     & $-17.88\pm0.58$   & $ -0.29 \pm 0.41$     & &           0.59      & $-10.49\pm2.19$  &  $0.39     \pm0.11$               \\
$\log\Sigma_{350}$ &          $-0.09$     & $-18.67\pm0.51$   & $ -0.16 \pm 0.36$     & &           0.55      & $-12.46\pm1.98 $ &  $0.3      \pm0.1$              \\
$\log\Sigma_{500}$ &          $-0.08$     & $-23.08\pm0.49$   & $ -0.14 \pm 0.34$     & &           0.56      & $-17.09\pm1.89$  &  $0.31    \pm0.09$               \\
$\log\Sigma_{1.4{\rm GHz}}$  &          $-0.7$      & $-16.94\pm0.63$   & $ -2.1 \pm0.4$        & &                     &                  &                          \\                         
\hline
\end{tabular}
\end{center}
\end{table*}

\begin{figure*}
\centerline{\vbox{
\hbox{
\psfig{figure=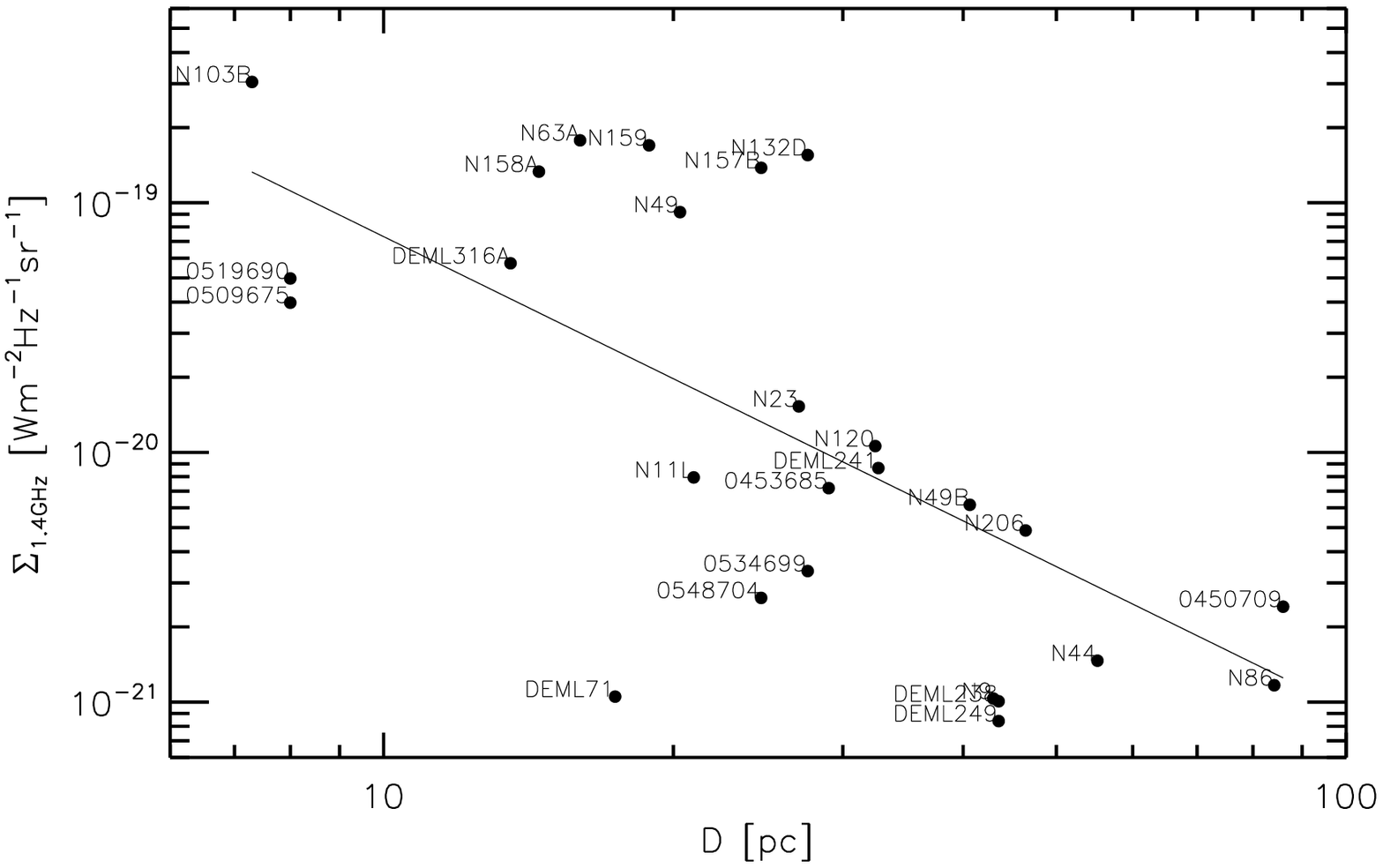,width=90mm}  
\psfig{figure=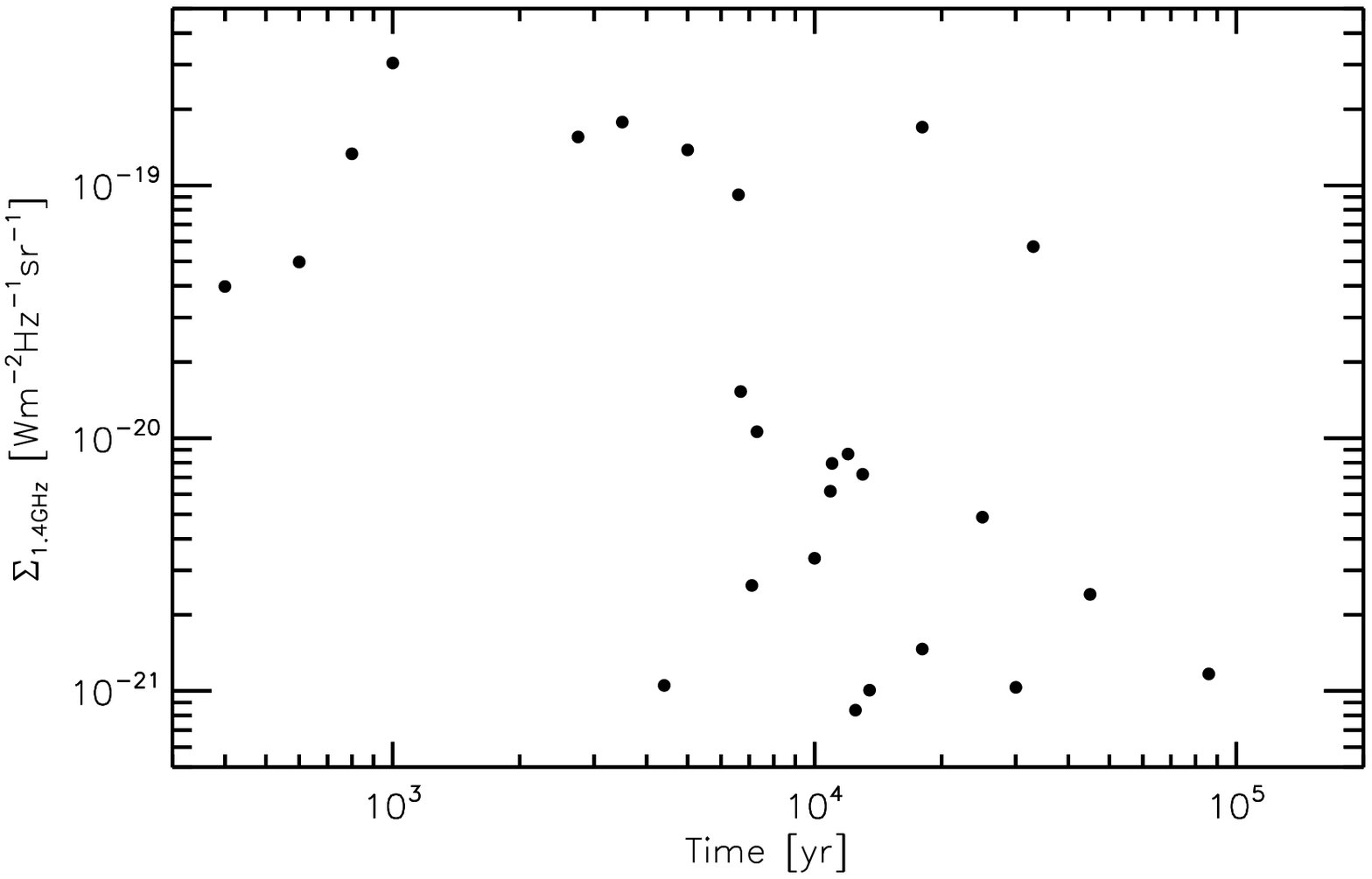,width=90mm}} 
\hbox{
\psfig{figure=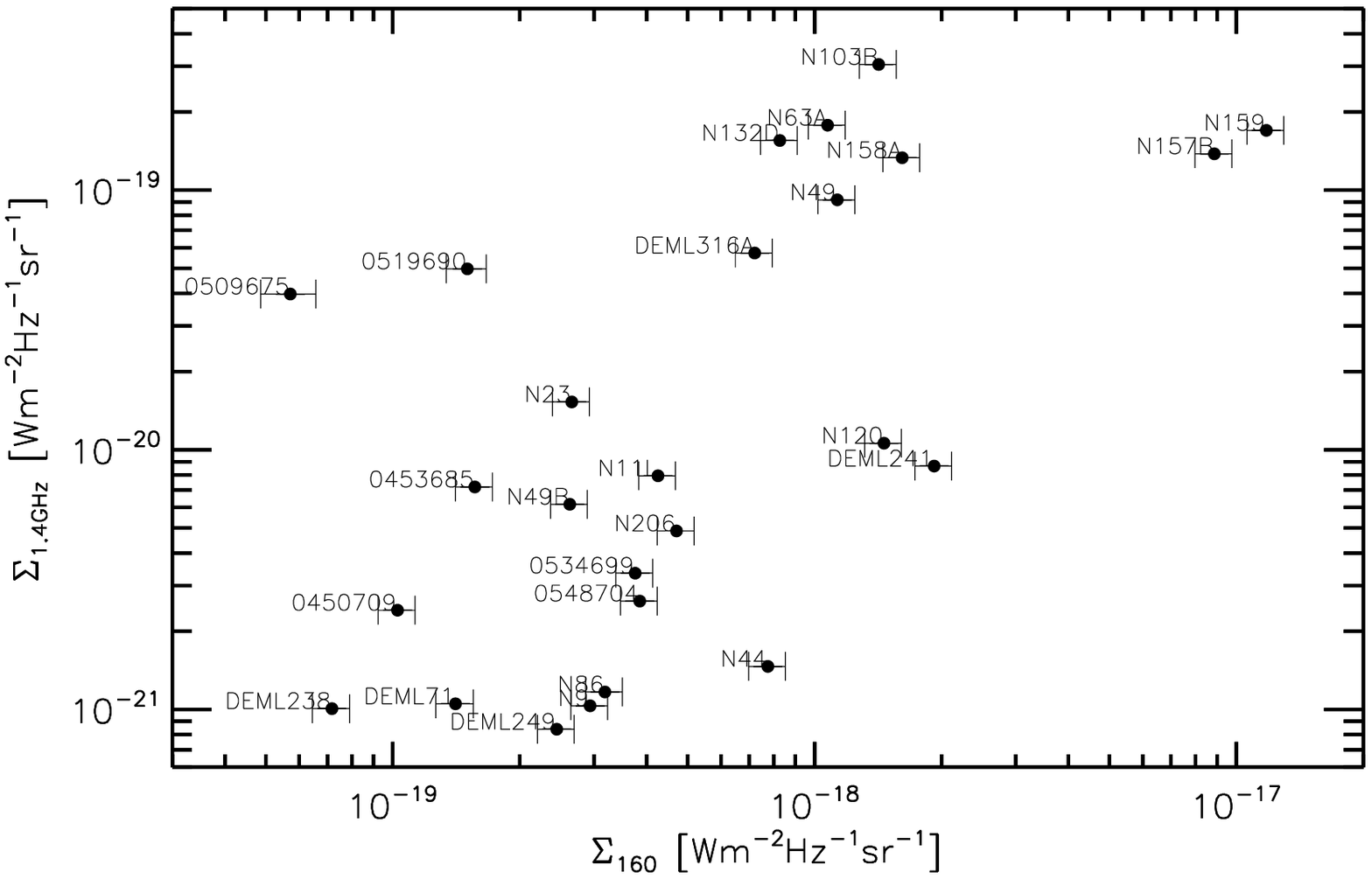,width=90mm}  
\psfig{figure=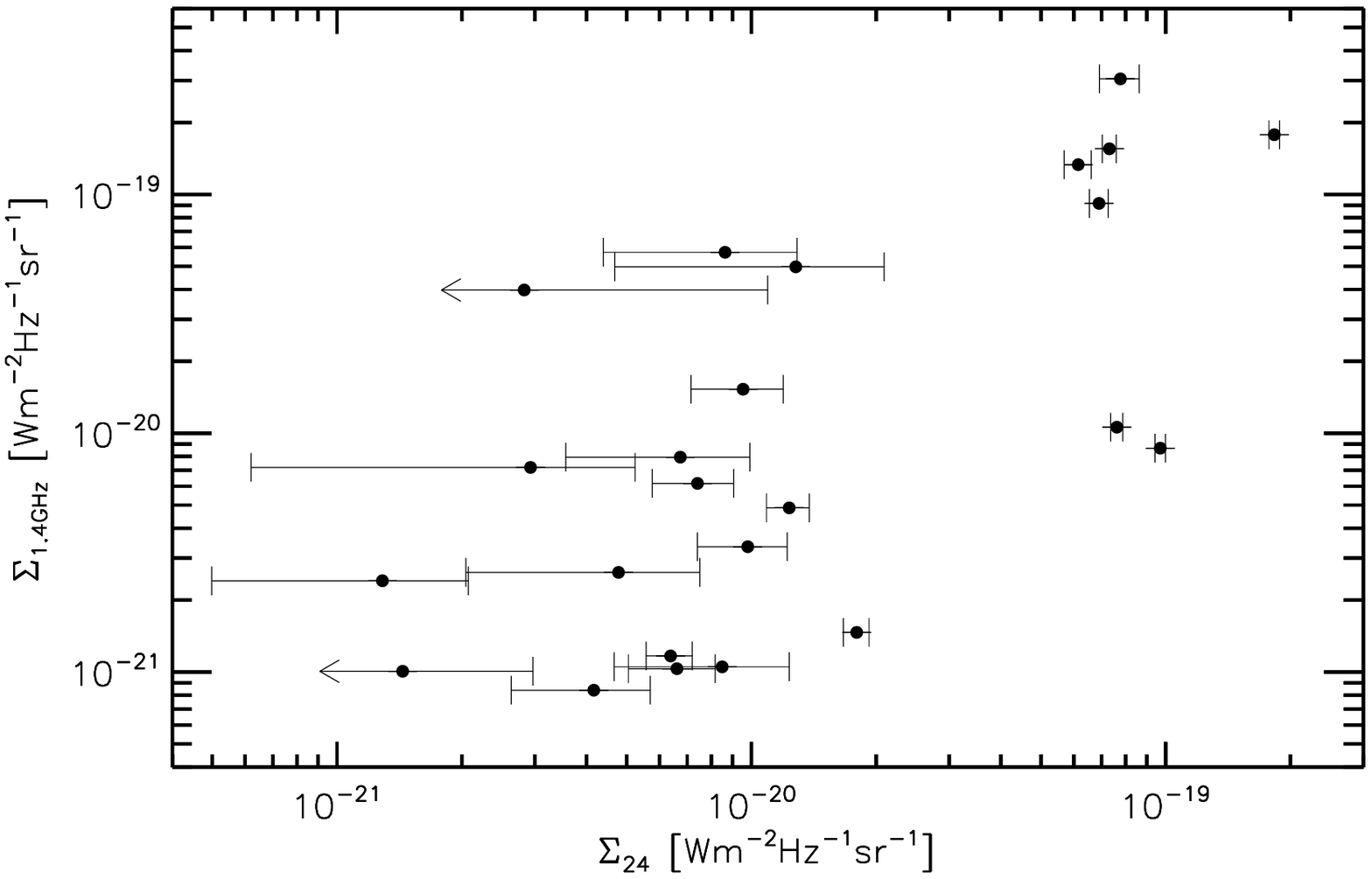,width=90mm}}  
}}
\caption[]{Top left panel: Relation between SNR $\Sigma_{1.4{\rm GHz}}$ and diameter of SNR, 
$\Sigma-D$ relation. Top right panel: Relation between SNR $\Sigma_{1.4{\rm GHz}}$ and the age of SNR. Bottom left: $\Sigma_{1.4{\rm GHz}}$ is weakly
correlated with $\Sigma_{160}$. Bottom right: $\Sigma_{1.4{\rm GHz}}$ has some correlation with $\Sigma_{24}$.\label{SeparationFluxes1}}
\end{figure*}

By performing linear fitting on radio data at 1.4 GHz, we have found the $\Sigma$--$D$ relation to be
\begin{equation} 
\Sigma_{1.4{\rm GHz}}\ =\ (1.2^{+3.7}_{-0.9})\times 10^{-17}\times D^{-2.1\pm0.4},
\end{equation}
where the diameter $D$ is in pc, with a correlation coefficient of $-0.7,$ using 26 SNRs (see Table~\ref{tbl2})
for which we have radio fluxes from \citet{Badenes10} catalogue and an estimation of the age in the literature. 
These data are presented in the top left panel of Figure~\ref{SeparationFluxes1}. This agrees well with that found 
in the literature \citep{Arbutina04, Urosevic03} and suggests a relatively constant radio
luminosity. Similarly, in the top right panel is the dependence of $\Sigma_{1.4{\rm GHz}}$ of the age of the remnant
for which we have found the relation
\begin{equation}  
 \Sigma_{1.4{\rm GHz}}\ =\ (0.27^{+2.29}_{-0.25})\times 10^{-16}\times{\rm age [yr]}^{-0.86\pm0.25},
\end{equation} with  correlation coefficient of $-$0.58. This is consistent with a time evolution of the diameter 
according to $D\propto{\rm age}^{0.41}$, or almost exactly the ${\rm age}^{2/5}$ as predicted for the Sedov phase (cf., e.g., \citealt{Badenes10}).

Similarly, we compared the surface brightness at FIR wavelengths ($\Sigma_{24}$, $\Sigma_{160}$, $\Sigma_{350}$
and $\Sigma_{500}$) and the diameters, but no correlation was found (see Table~\ref{tbl1}). It is most likely
that the FIR emission is dominated by the ISM rather than the SNR.

However, for the relation between FIR and radio surface brightness there is some tentative trend (see Figure~\ref{SeparationFluxes1}
bottom left for the $\Sigma_{160}$). This is a consequence of the dependence of
both the $\Sigma_{{\rm radio}}$ and $\Sigma_{{\rm FIR}}$ on the density of the ISM. We find
similar correlations between the $\Sigma_{1.4{\rm GHz}}$ and that at other FIR and submm wavelengths (Table~\ref{tbl1}). 
The correlation coefficients are similar for all these frequencies (0.55$-$0.7). Comparison between $\Sigma_{24}$ and $\Sigma_{{\rm 1.4GHz}}$ is given 
in Figure~\ref{SeparationFluxes1}, bottom right panel. Somewhat surprising is the much weaker correlation than the one obtained by \citet{Seok08} who found the correlation coefficient
between $\Sigma_{24}$ and $\Sigma_{{\rm radio}}$ to be 0.98, using only 8 clearly detected SNRs.

The 24-$\mu$m flux may have a contribution from line emission. Among the SNRs
believed to be more affected by this are N\,49 and N\,63A, where the line
contribution is estimated to be an exceptional $\sim80$\% \citep{Williams06}
and modest $\sim10$\% \citep{Caulet12}, respectively. We corrected $\Sigma_{24}$
for these two SNRs accordingly. In general, however, line emission
is thought to make a negligible contribution to the 24-$\mu$m flux
\citep{Williams10} and we found estimates in the literature only for these two SNRs.

\begin{figure}
\centerline{\vbox{
\psfig{figure=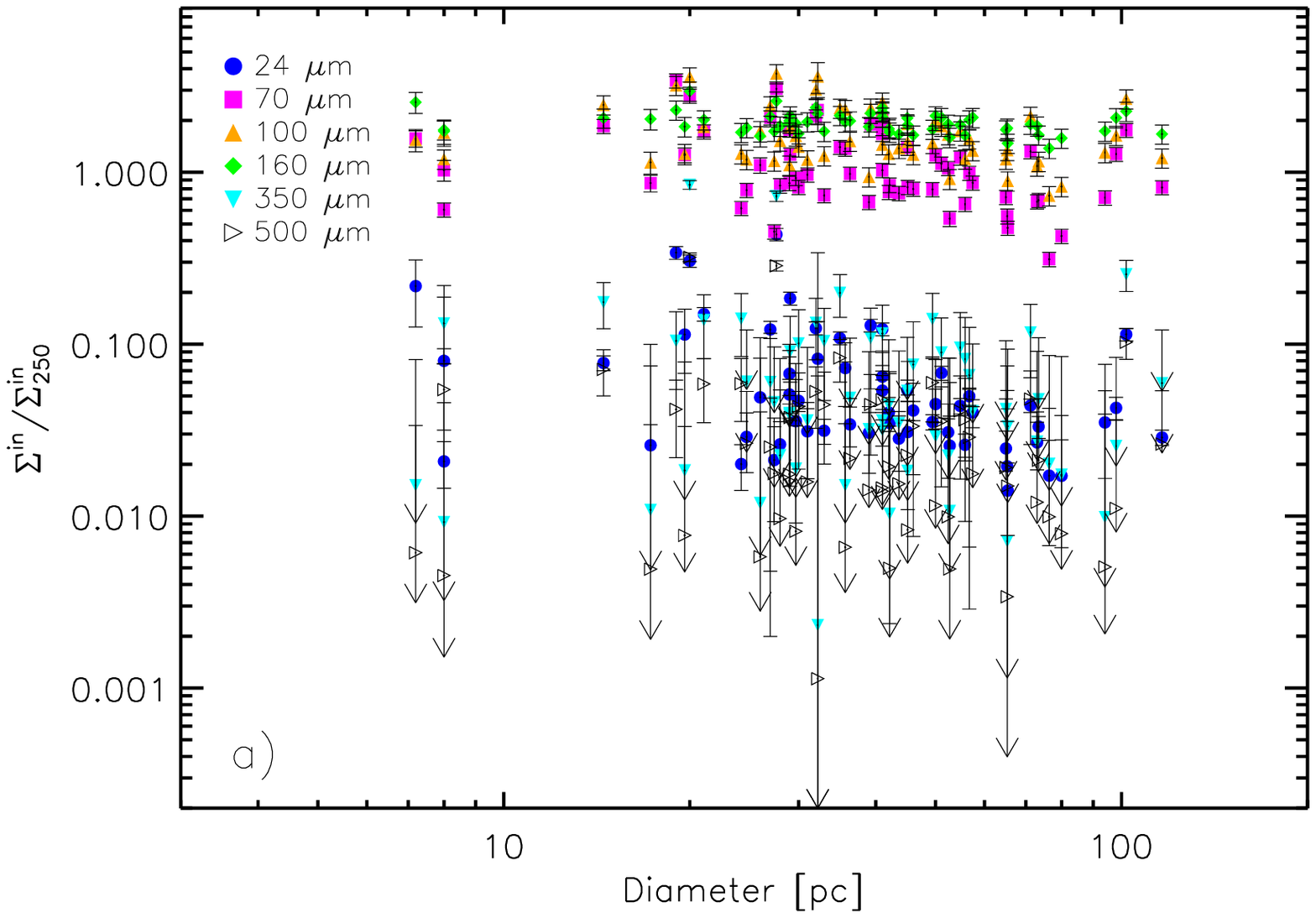,width=82mm}  
\psfig{figure=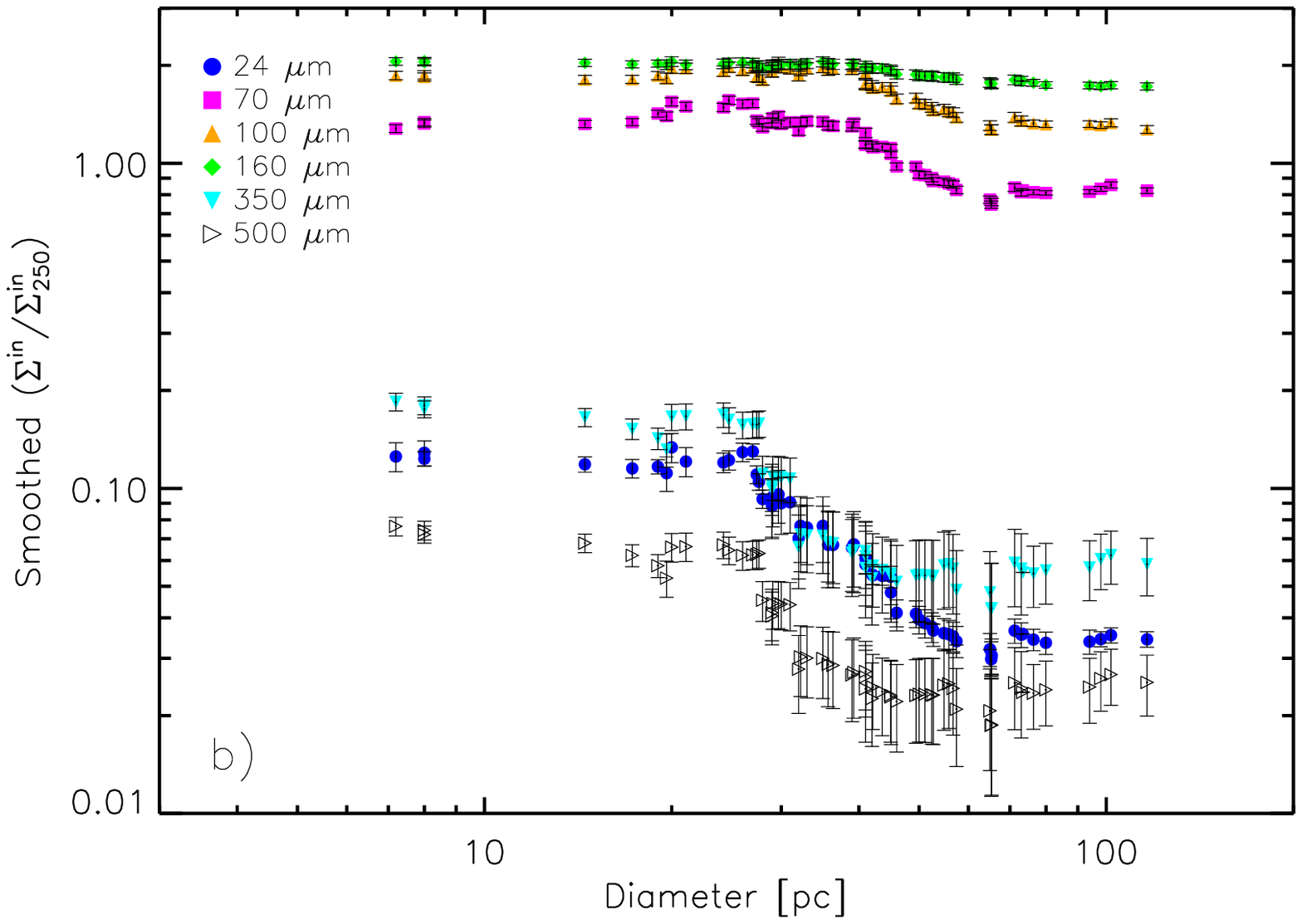,width=82mm} 
\psfig{figure=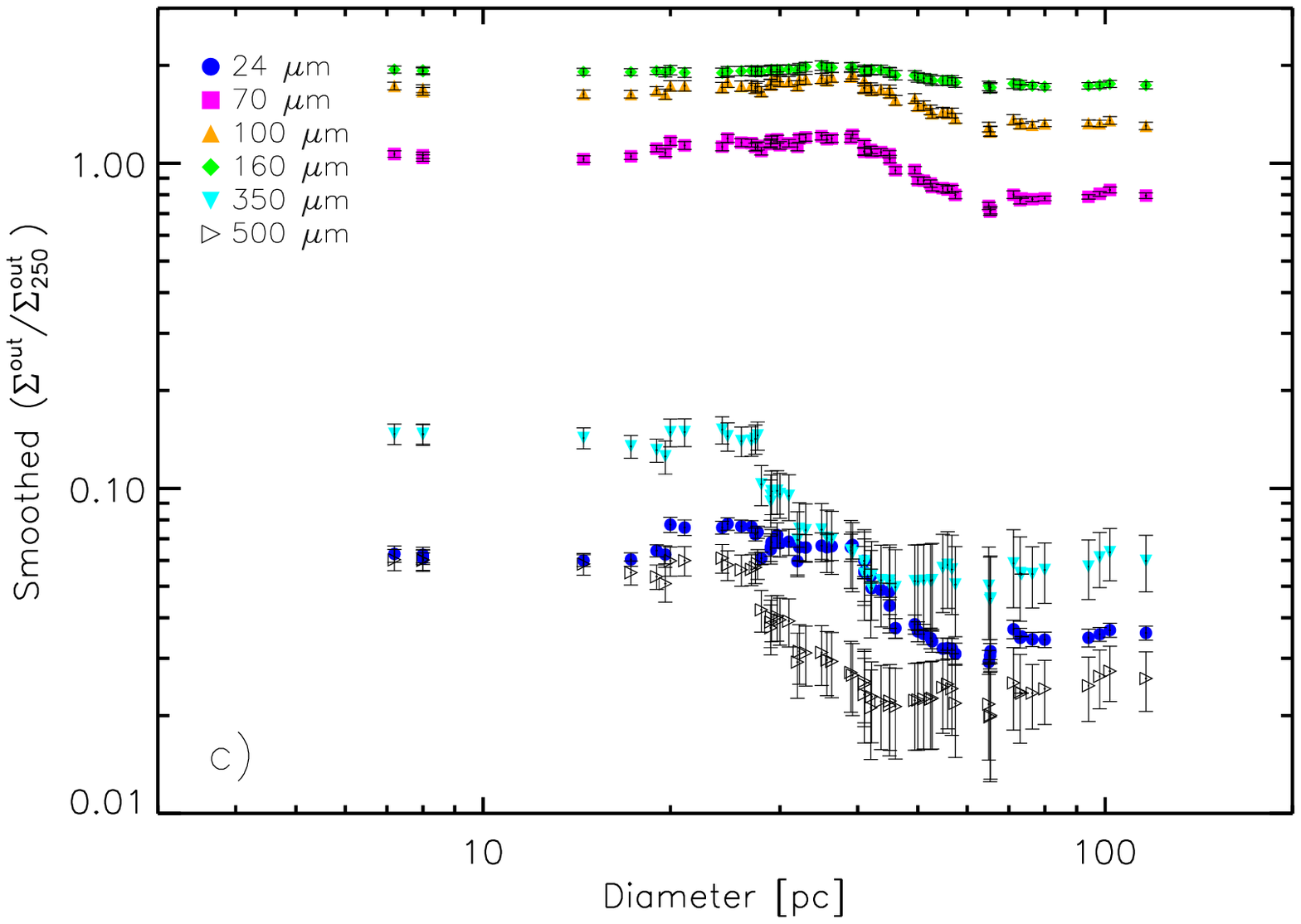,width=82mm}
\psfig{figure=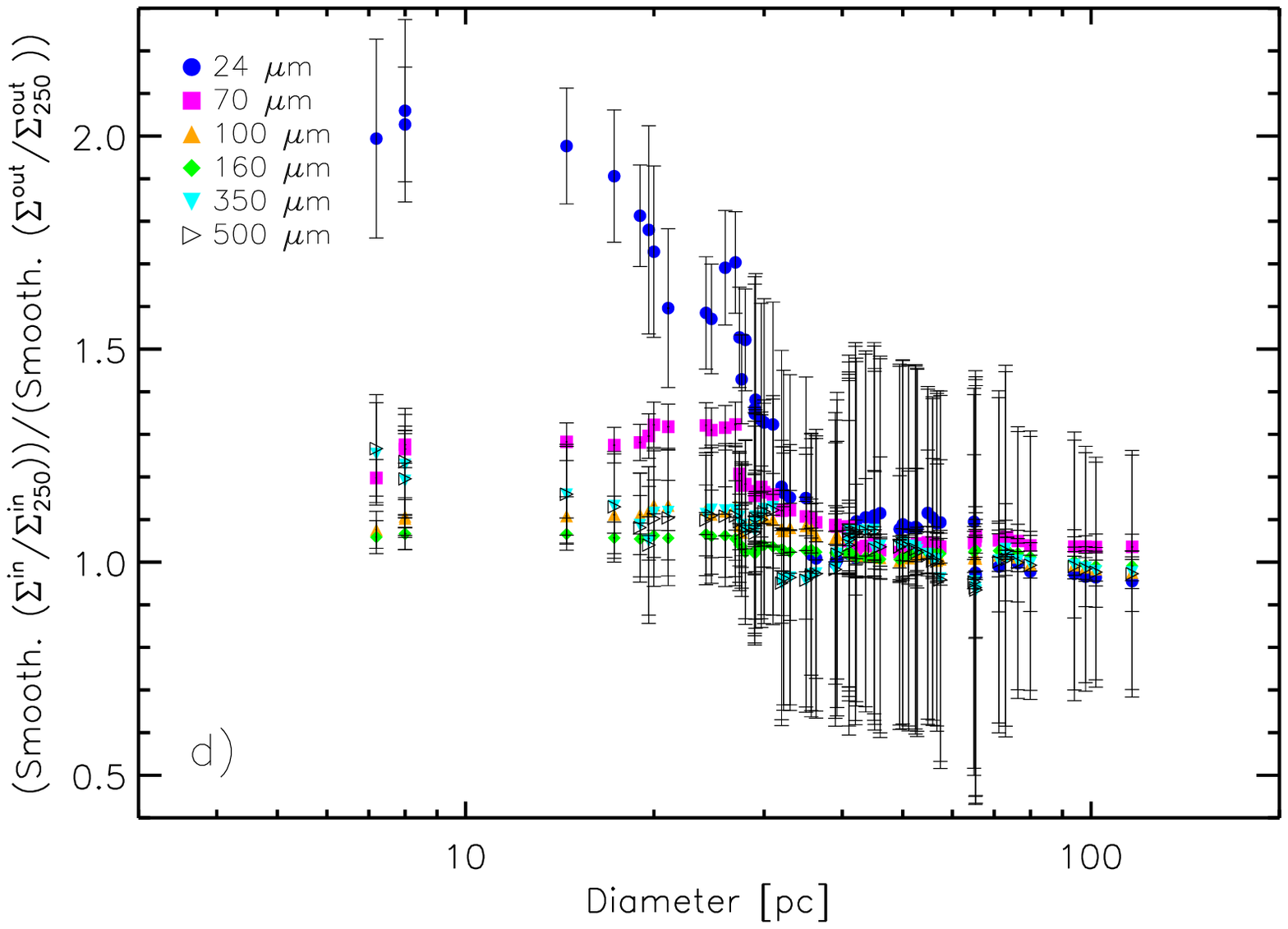,width=82mm}}
}
\caption[]{{\it a:} Evolution of $\Sigma$ (normalised to $\Sigma_{250}$) within SNRs at 24$-$500 $\mu$m 
depending on the diameter; {\it b:} The same as {\it a}, but smoothed; 
{\it c:} The same as previous, but in annuli around SNRs. {\it d:} Ratio of normalised and smoothed $\Sigma$ 
inside and outside of the SNR.  
\label{FluxesLMC_diam}}
\end{figure}

Since we have not found any fading of $\Sigma_{\rm FIR}$ with the diameter, we will try to find it in 
the following experiment. In Figure~\ref{FluxesLMC_diam}a we
first plot the surface brightness, normalised to $\Sigma_{250}$ to remove variations between SNRs for other
reasons than evolution, versus diameter. We then smooth the data over up to $\pm7$ consecutive measurements, 
revealing clear trends of fading with time (Figure~\ref{FluxesLMC_diam}b). It appears as though there is no further evolution for SNRs with $D>70$ pc. 

While the fading at 24 $\mu$m is the most
prominent evolution across the IR--submm range (a factor of three change over
our sample, compared to $\Sigma_{250}$), less expected is the fading at
submm wavelengths (350 and 500 $\mu$m), by a factor two compared to $\Sigma_{250}$ -- i.e., more pronounced than at 100 and 160 $\mu$m. Perhaps
free--free and/or synchrotron emission contributes at the longest wavelengths,
and diminishes during the evolution of the SNR. As the SED peaks shortward of 350 $\mu$m, estimates of the dust mass and temperature will 
not be greatly affected by such contributions at longer wavelengths.

Figure~\ref{FluxesLMC_diam}c shows the same as for the previous panel, but in the 20 pc thick
annuli just outside of SNRs. It suggests that the environment of SNRs also fades and that 
on average surface brightnesses within SNRs are not much different from the ones from annuli. This 
might be a sign of the cooling of the dust in the ISM. 

Finally, in Figure~\ref{FluxesLMC_diam}d, we derive the ratio of inner and outer normalised and 
smoothed $\Sigma$, finding that at most of the wavelengths SNRs show some decrease in $\Sigma$ compared to 
the surroundings. 

\subsection{Maps of flux ratio}\label{Ratiosx}

\begin{figure*}
\centerline{\hbox{
\psfig{figure=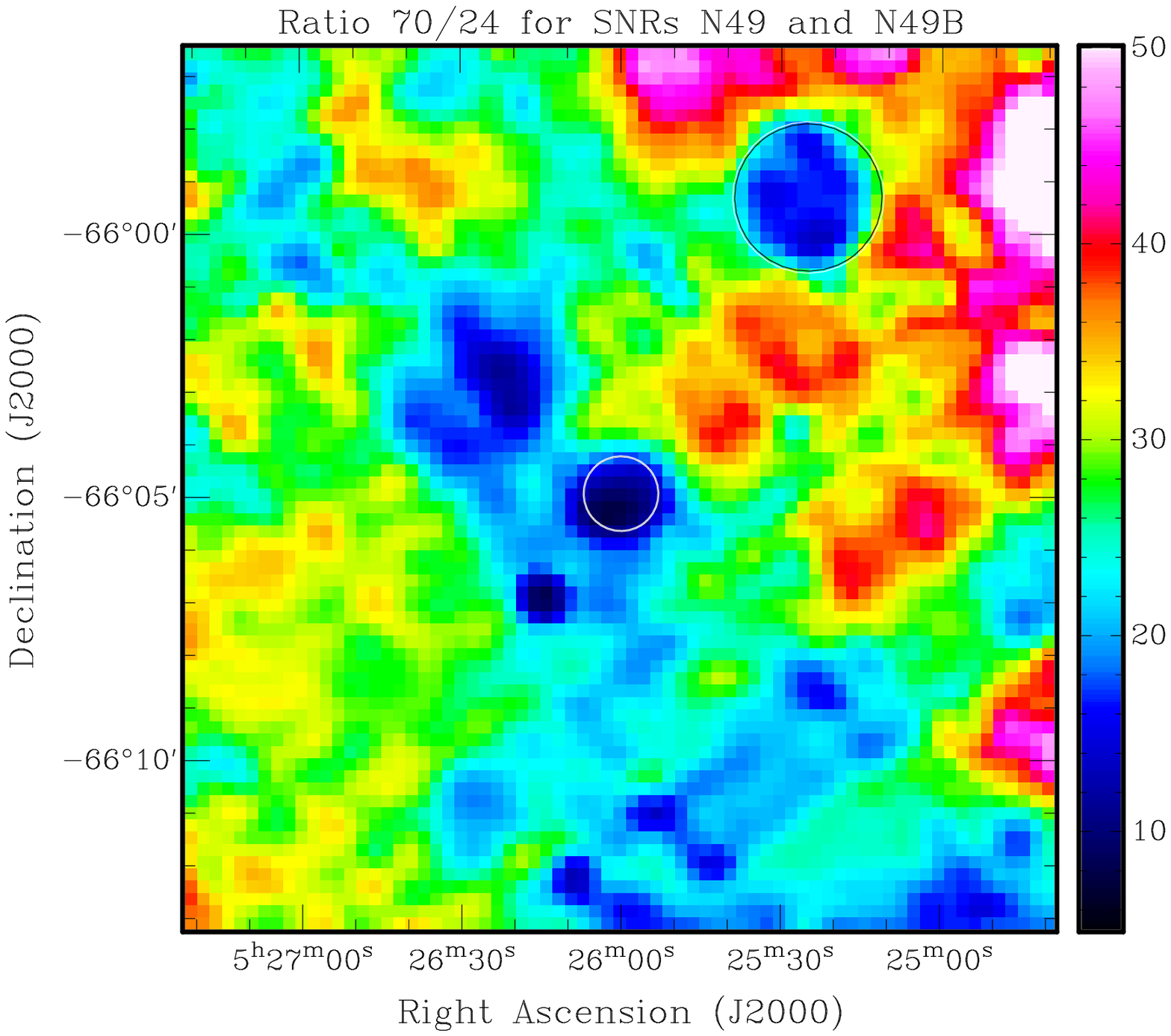,width=90mm}  
\psfig{figure=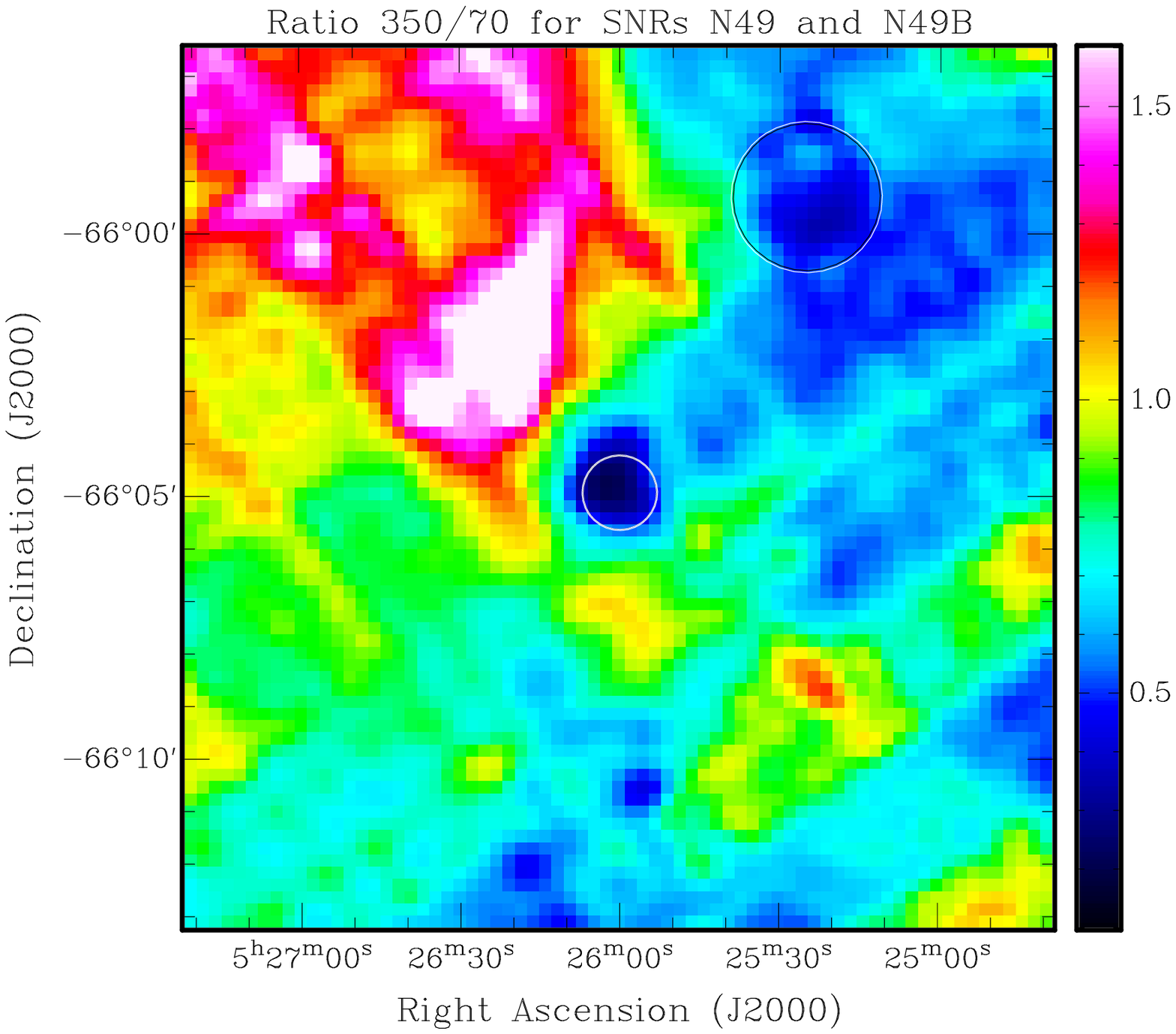,width=90mm}  
}}
\caption{Maps of flux ratios for N\,49 (central circle is the surface of the remnant) and N\,49B (circle on top right) {\it Left:} $R_{70/24}$; {\it Right:}
$R_{350/70}$. Colour correction was not applied. \label{ratimage}}
\end{figure*}

\begin{figure*}
\centerline{\vbox{
\hbox{
\psfig{figure=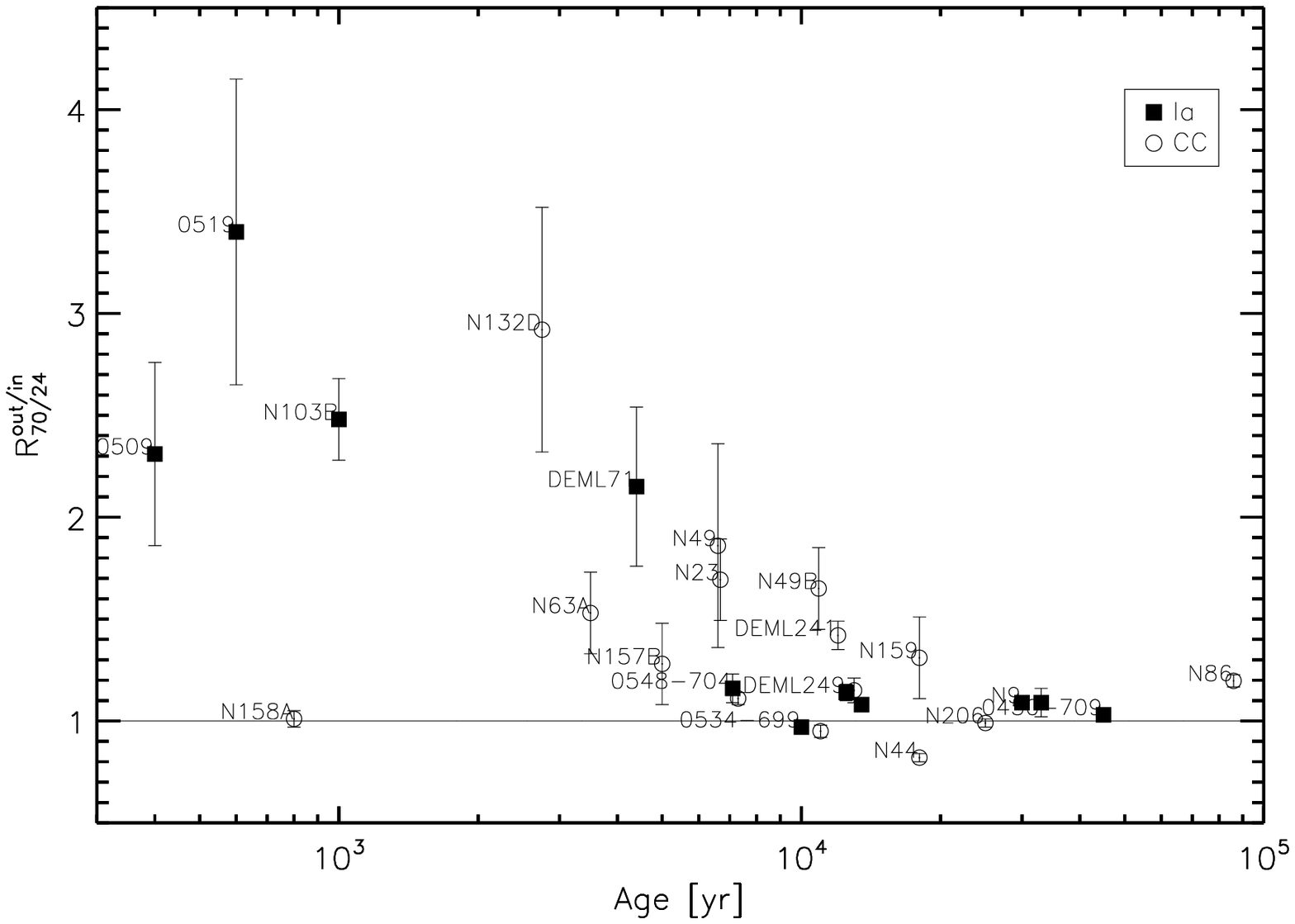,width=90mm}   
\psfig{figure=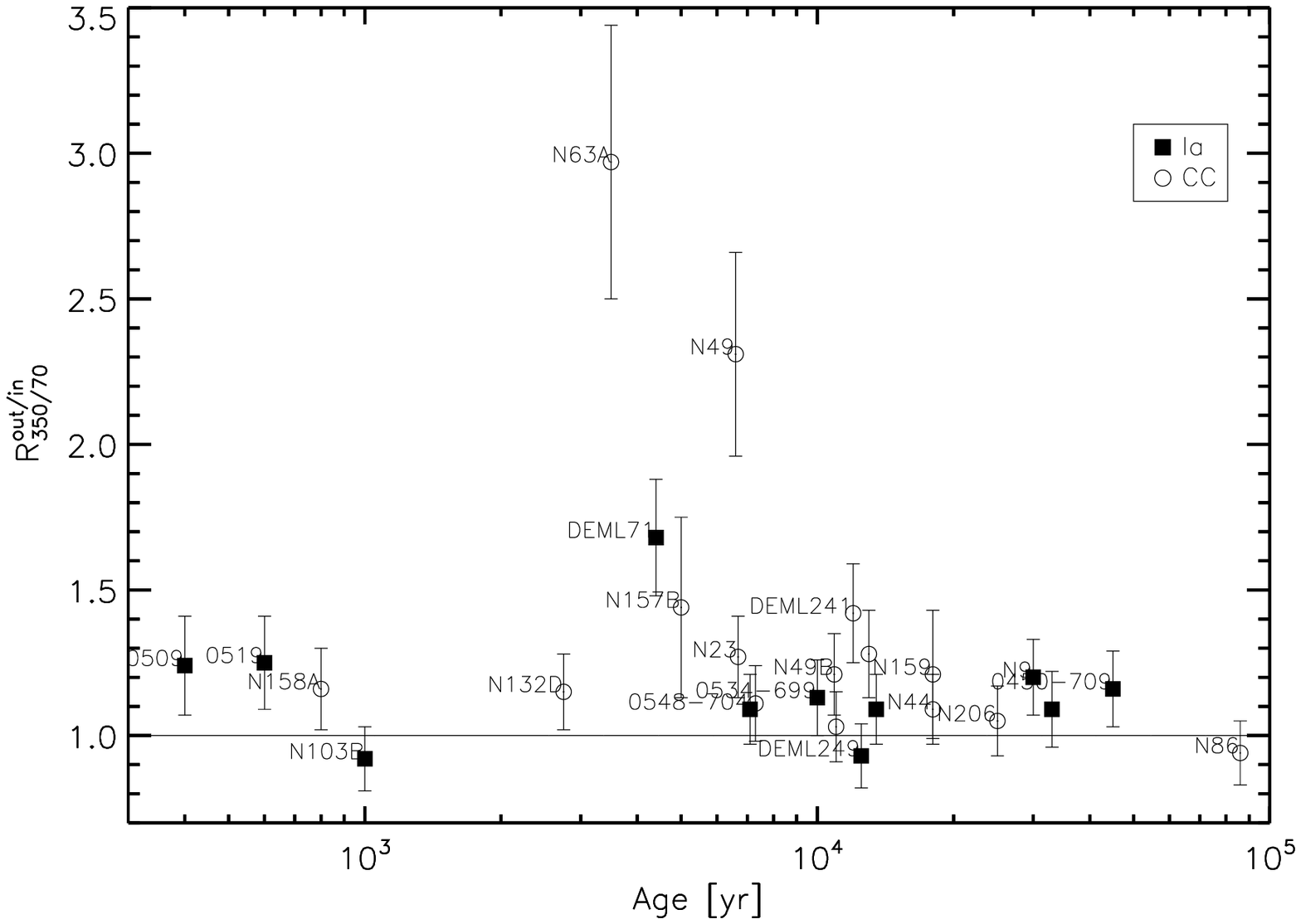,width=92mm}}  
\hbox{
\psfig{figure=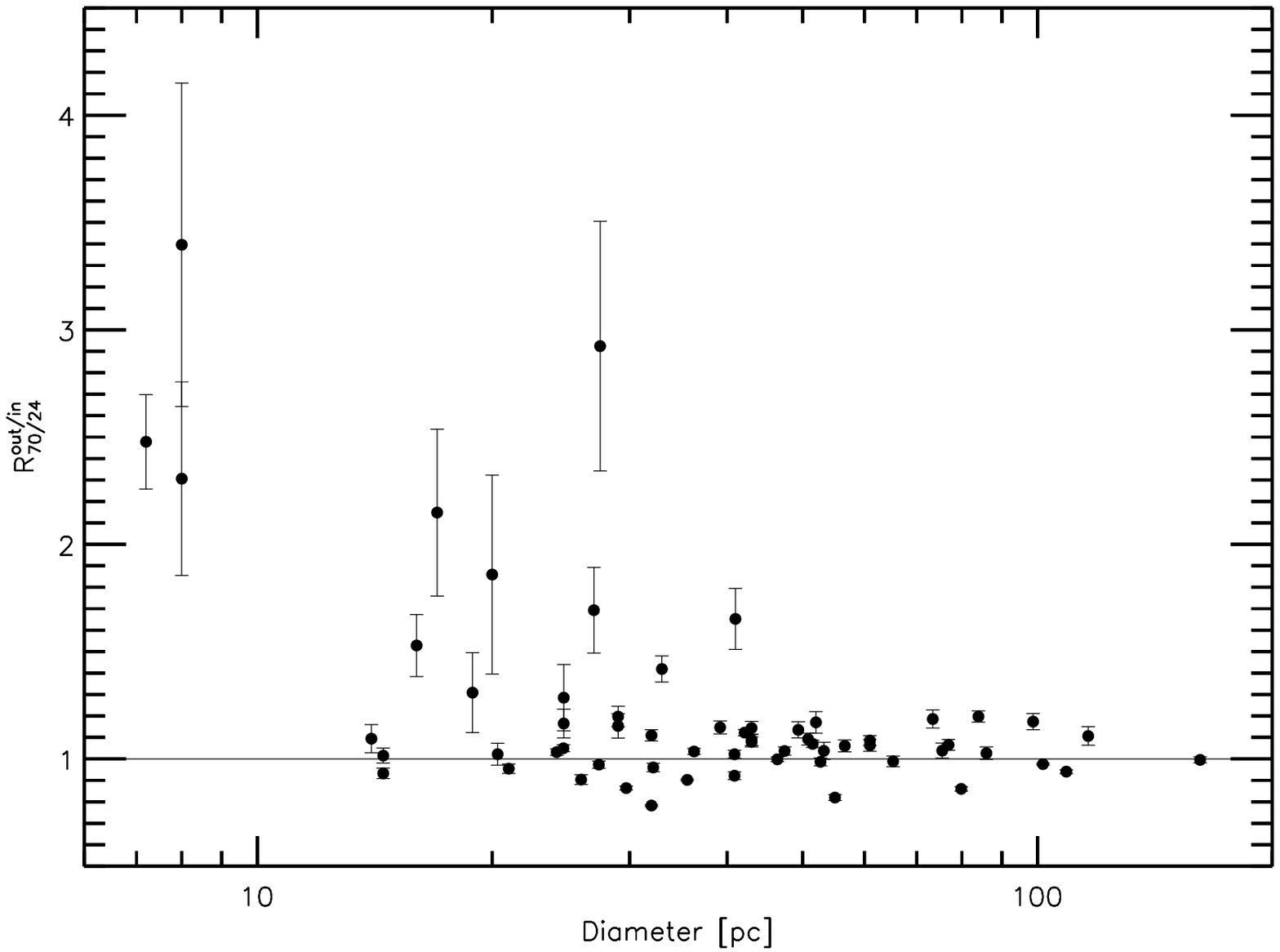,width=90mm}   
\psfig{figure=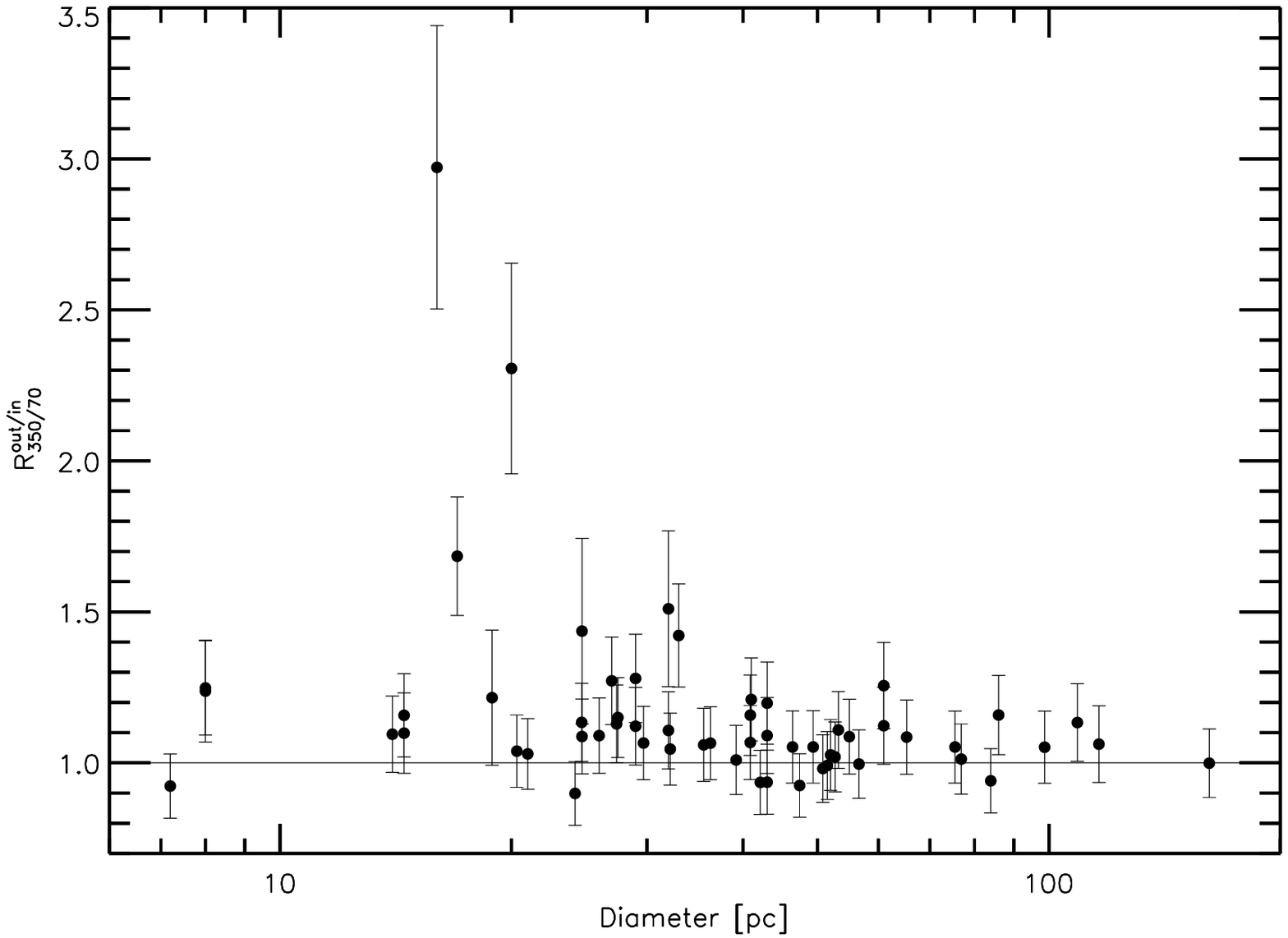,width=92mm}}  
}}
\caption[]{{\it Top:} Evolution of the average ratio of IR fluxes (70 and 24 $\mu$m)  
outside divided by the one within the SNR; {\it Left} and {\it Right} 
show ${\rm R^{out/in}_{70/24}}$ and ${\rm R^{out/in}_{350/70}}$ in time. {\it
Bottom:} Likewise for the dependence on diameter, for the entire sample of
SNRs.\label{evol7024}}
\end{figure*}

Following \citet{Sankrit10} and \citet{Williams10} who used IR flux ratios to
infer the sputtering and heating of dust in SNR shocks, we constructed maps of the flux ratios at 70 and 24 $\mu$m 
(hereafter $R_{70/24}$) and at 350 and 70 $\mu$m ($R_{350/70}$). Maps using other 
combinations of 24--500 $\mu$m fluxes where dividend flux is at longer wavelength than the one of the divisor
do not show any different behaviour -- ratios within SNRs are lower than in the surrounding medium. We show the images of these two ratios
for N\,49 and N\,49B in Figure~\ref{ratimage}. In making these maps we do not apply colour correction.

The differences between $R_{70/24}$ (and $R_{350/70}$) within and outside of
the SNRs result from differences in the warm dust contribution 
(and in some cases line emission around 24 $\mu$m), but also from the sputtering of the 
dust \citep{Sankrit10, Williams10}. This warm dust is
collisionally heated \citep{Williams10}, dominated by small grains, and
thus sensitive to the effect of thermal or non-thermal sputtering and shattering. 

We calculate the ratios between the average flux ratio outside (in 20 pc thick annuli) and within the SNR
-- which we shall call ${\rm R^{out/in}_{70/24}}$ and ${\rm R^{out/in}_{350/70}}$,
respectively -- and plot these against the ages of objects (if they are known) and diameters
(Figure~\ref{evol7024}). There is a clear trend of the $R_{70/24}$ ratio 
within the SNR to increase as the SNR evolves, until it reaches that of the surroundings. This suggests the
dust within/around the SNR cools in $\sim10^4$ yr. The older SNRs show little difference from surrounding 
ISM, yet the difference still lingers as the ratio-ratio stays
above unity for most SNRs, also in the $R_{350/70}$ ratio. The same observations are reflected 
in the dependence on diameter (Figure~\ref{evol7024}, bottom panels). ${\rm R^{out/in}_{70/24}}$ is more sensitive to the SNR 
temperature, while ${\rm R^{out/in}_{350/70}}$ is also sensitive to the dust mass that SNR interacts with. 

\subsection{Creating maps of dust mass and temperature} \label{making}

For producing mass and temperature maps, we use only the pixels with flux higher than 3$\sigma$ where the uncertainties are different for each 
wavelength and pixel. If the flux of a pixel is below that limit, then it is added to the fluxes of all other faint pixels in that image
and the value of average faint pixel is found, which is then fit using only 
the calibration uncertainties (as the background uncertainties become negligible). 

We assume a modified black body in the shape 
{\rm \begin{equation}
F_\nu\ \propto \nu^\beta\ B_\nu,
\end{equation}} using emissivity index $\beta=1.5$ \citep[see][]{Planck14}, where $B_{\nu}$ is 
Planck function and $\nu$ frequency. We use the emissivity $\kappa=0.1$ m$^2$ kg$^{-1}$ at $\lambda=1$ mm 
wavelength \citep{Mennella98}. The two free parameters are the temperature and mass, initially these were set to $T=20$ K \citep[see][who derived $T\approx18.7$
K]{Planck14} and $M=1$ M$_\odot$. We further constrained the temperature to be $\geq 3$ K (cosmic
microwave background) and $\leq 25,000$ K -- all grains will have sublimated
by $T\sim1500$ K (the solution for $T$ is always well below this physical limit); and the mass to be $\geq 10^{-5}$ M$_\odot$ and $\leq 180$ M$_\odot$. Each pixel 
was modelled separately, using a $\chi^2$-minimization procedure within IDL \citep[{\sc
mpfit},][]{Markwardt09}. 

Firstly we performed the fitting without colour correction, then we computed the 
colour corrections to that model, as an iterative search for a best dust model. 
Colour correction is computed using the {\sc IDL} wrapper distributed with the {\sc DustEM} 
code and described in \citet{Compiegne11}. Finally, we applied the fit to data that were divided by that 
colour correction. As a result we obtained temperature and mass maps for every SNR and surroundings.

\begin{figure}
\centerline{\vbox{
\psfig{figure=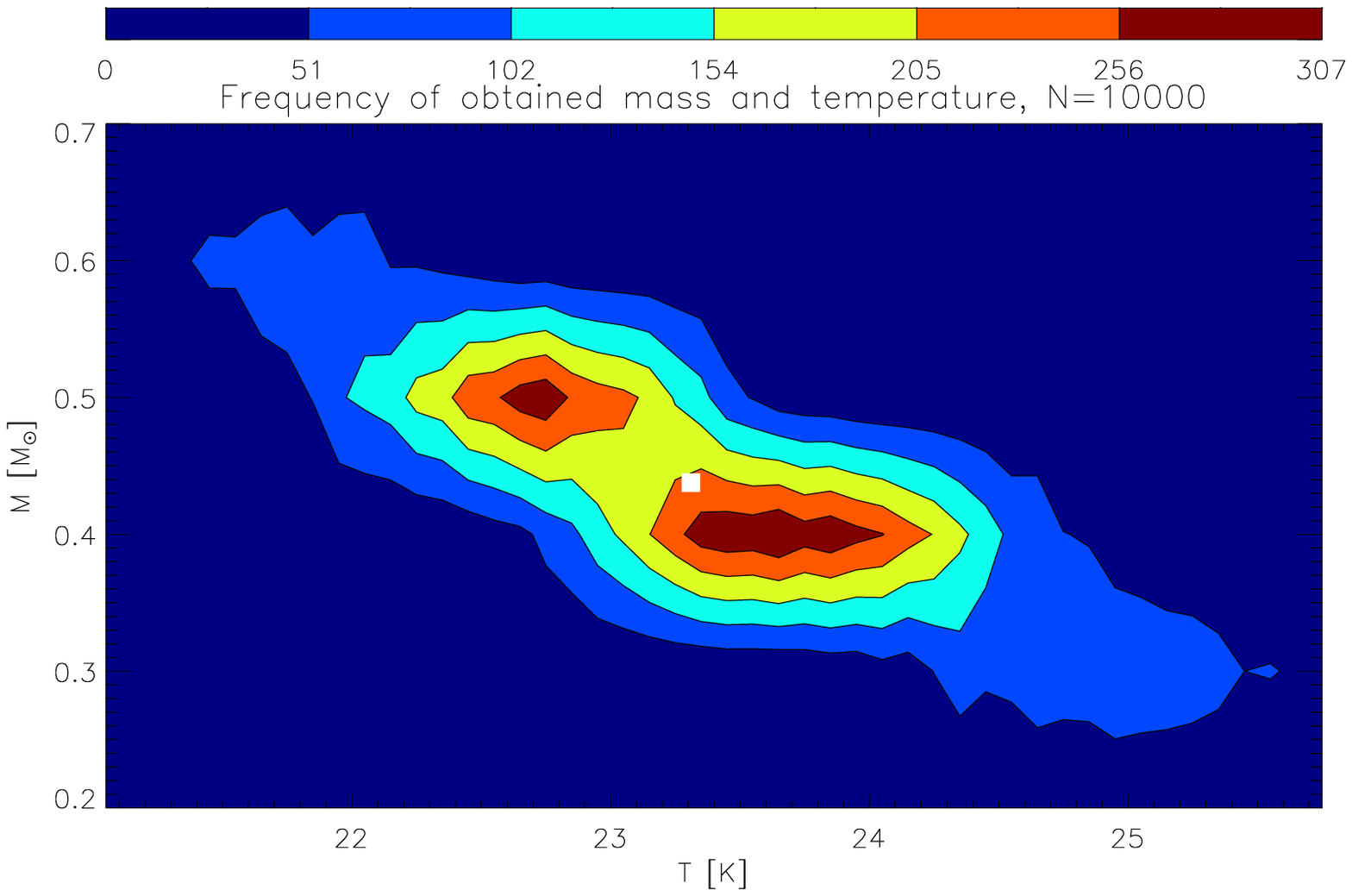,width=86mm} 
\psfig{figure=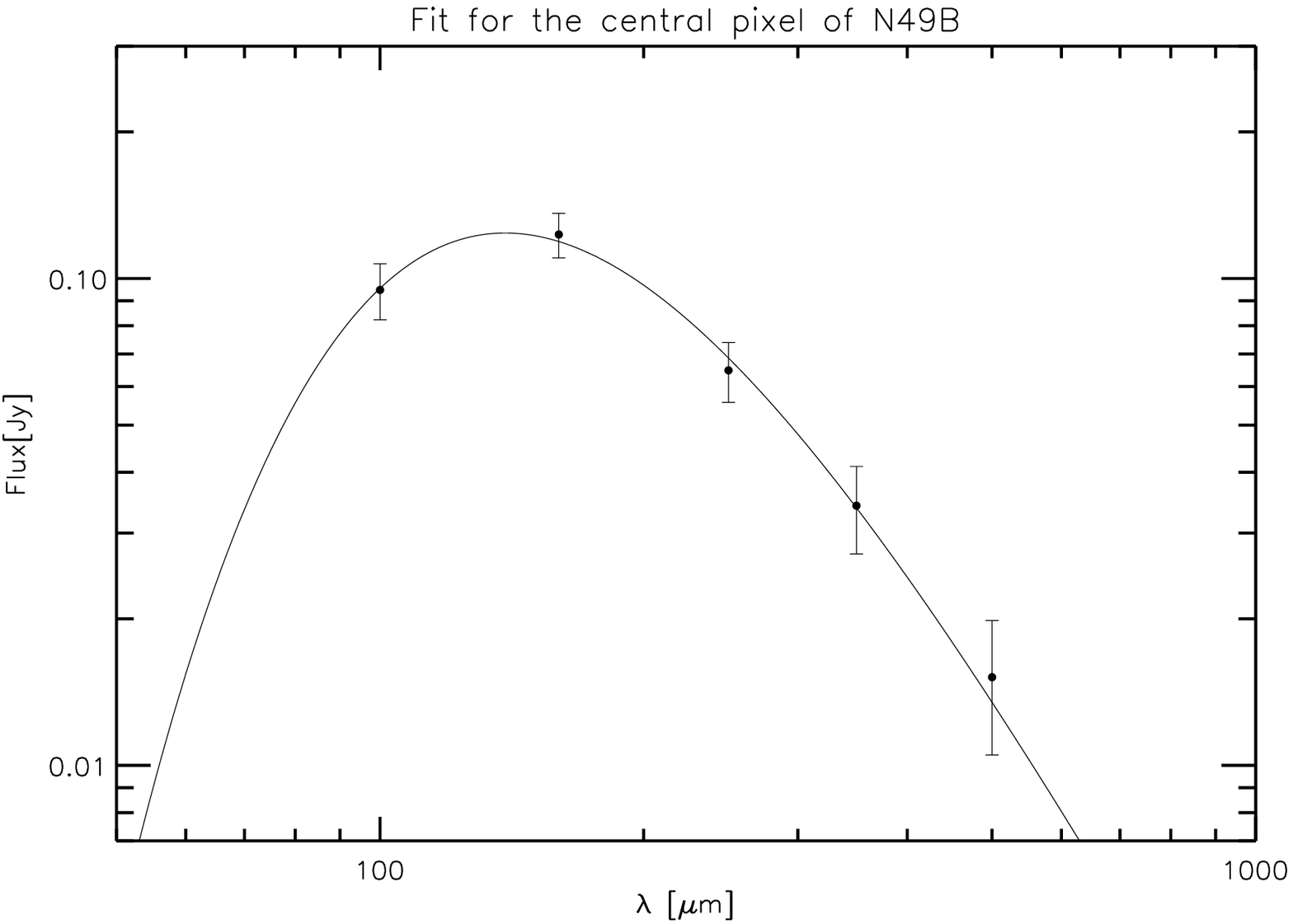,width=86mm}}} 
\caption[]{{\it Top:} The obtained values of mass and temperature for the central pixel 
in the image of N\,49B, based on fits to 10,000 simulated datasets. The white square is 
the actual result of the fit for that pixel. {\it Bottom:} Fitting of the SED of that particular pixel.
\label{N49cont1}}
\end{figure}

\begin{figure}
\centerline{\vbox{ 
\psfig{figure=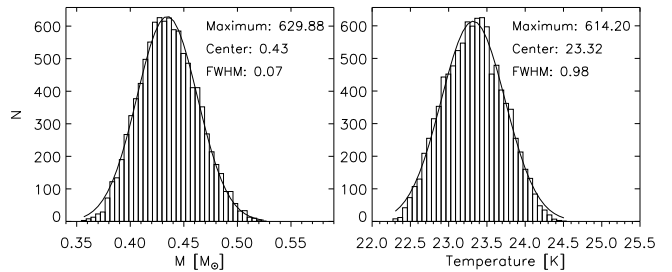,width=86mm}}}
\caption[]{Histograms of the ({\it left}) mass and ({\it right}) temperature obtained for the 10,000 
simulated datasets around the central pixel of SNR N\,49B. From FWHM we can find the values of $\sigma$.
\label{bb}}
\end{figure}

\begin{figure}
\centerline{\vbox{
\psfig{figure=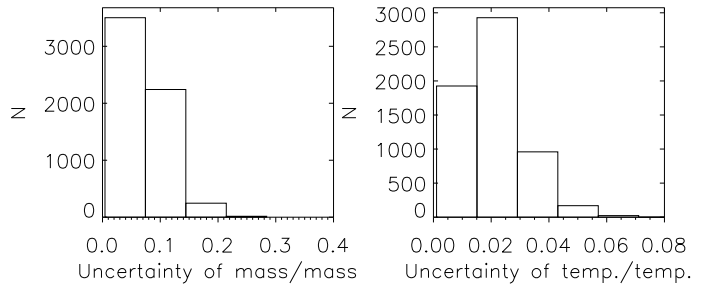,width=86mm}}} 
\caption[]{Histograms presenting the ratio of uncertainties of parameters and the parameters themselves for pixels in the N\,49B image, for ({\it Left:}) mass 
and ({\it Right:}) temperature.
\label{rr1}}
\end{figure}


\begin{figure}
\centerline{\vbox{
\psfig{figure=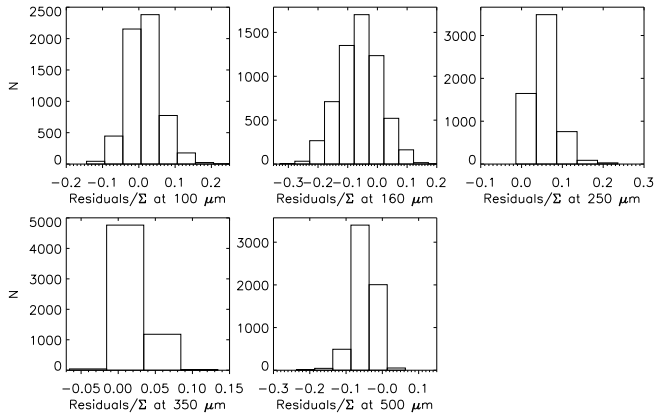,width=86mm}}} 
\caption[]{Histograms of the residual/$\Sigma$ (for all pixels) on all wavelengths for SNR N\,49B and
its surroundings. These residuals are of order of the surface brightness uncertainties ($\sim$10\%)
and therefore our adopted model describes the data well.
\label{rr}}
\end{figure}

The errors of these maps can be estimated using Monte Carlo simulations or using {\sc mpfit}. Monte Carlo simulation is done by adding random Gaussian noise 
with standard deviation equal to the error on the fluxes, and fitting the resulting SED. In the contour plot in Figure~\ref{N49cont1}, top, is the distribution of values of mass 
and temperature using 10,000 simulated datasets. The white square marks the actual fit for the given pixel. In Figure~\ref{N49cont1}, 
bottom, is the actual fit for the given pixel. 

In Figure~\ref{bb} we show the actual histograms of obtained values of mass and temperature for the same 10,000 data sets.
From the width of the Gaussians we find the errors of the mass and the temperature for that pixel of 6.9\% and 1.8\% respectively. 
Average $\chi^{2}$ was 3.75. With five data points and two fit parameters, the number of degrees of freedom is three. These errors are different from pixel to pixel and from remnant to remnant. For low density SNRs, like 
0506$-$675 and DEM\,L71 we can not use this method for finding errors since the flux is too low (hardly above 
3$\sigma$). 

Because Monte Carlo simulation for every pixel and remnant would be time consuming, we find the uncertainties of mass and temperature from the {\sc mpfit} procedure itself. In Figure~\ref{rr1} we give 
the distributions of the ratios of the uncertainties and corresponding parameters, for mass and temperature, respectively, for SNR N\,49B and surroundings, using 9120 pixels, or 320$\times$320 pc$^{2}$. 
The maps of N\,49B and surroundings are representative for the LMC because there are enough faint and bright pixels to estimate the error distribution. These uncertainties are $<$40\% for the mass and 
$<$13\% for the temperature. The pixel averaged $\chi^{2}$ was 1.57. 

In Figure~\ref{rr} we show the histograms of the residuals (model minus data) divided by corresponding surface brightnesses at all wavelengths for 
SNR N\,49B and the surroundings. The residuals 
are of order of the surface brightness uncertainties ($\sim$10\%)
and thus, our adopted model describes the data well. More 
sophisticated model and error analysis can be found in \citet{Gordon14}, who used the same data but different $\kappa$ and $\beta$ as free parameter.

\subsection{Maps of dust mass and temperature}\label{MTmaps}

Here we show and describe the actual maps for a selection of SNRs, with the remainder presented
in Appendix A. In Section~\ref{discussion} we will give our conclusions from these data 
interpreting the general lack of dust in the mass maps at the 
position of most of the SNRs by sputtering of the dust by shocks, and notice that the temperature is warmer in the direction of SNRs.

\subsubsection{N\,49}

\begin{figure*}
\centerline{
\vbox{
\hbox{
\psfig{figure=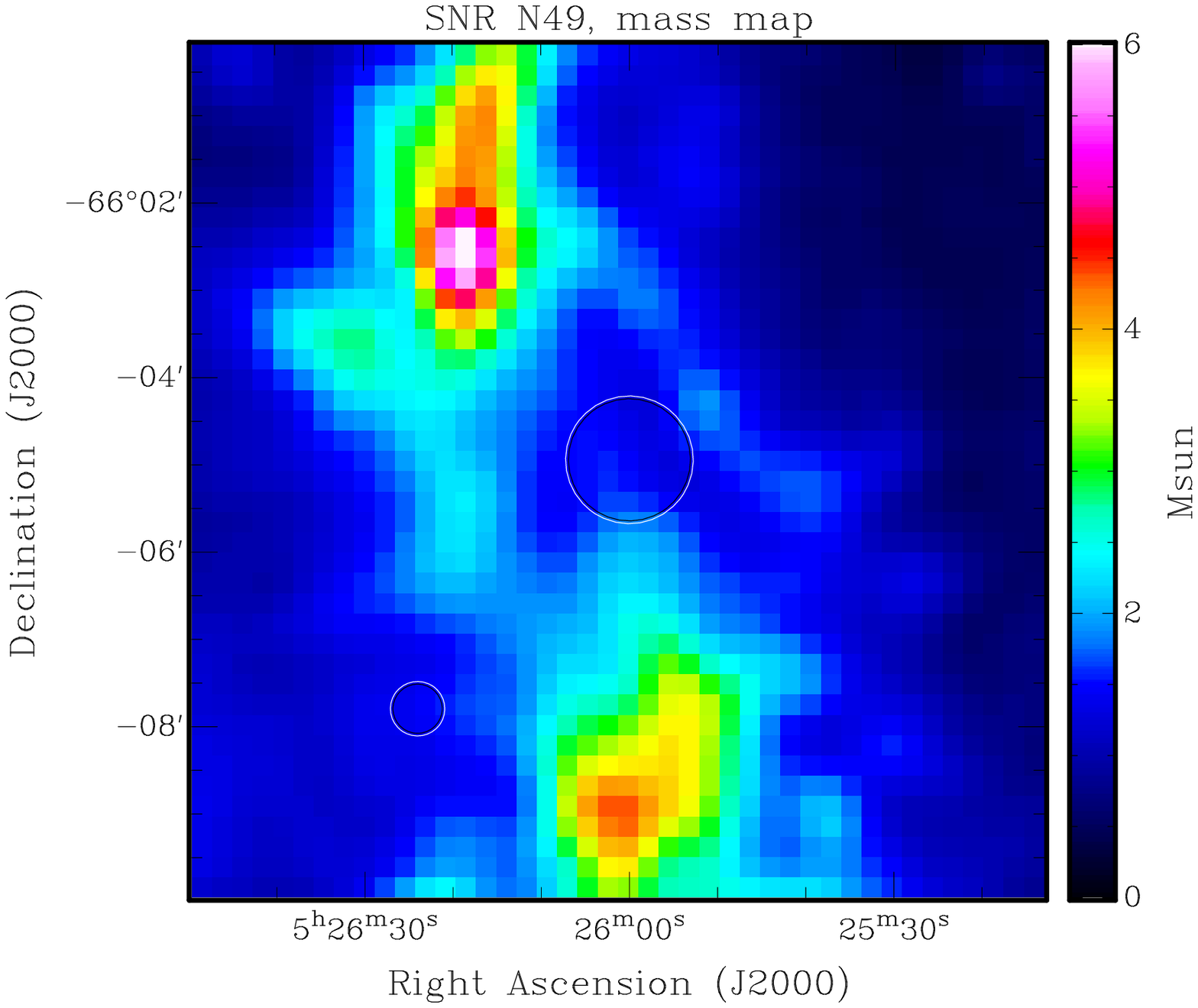,width=66mm}  
\psfig{figure=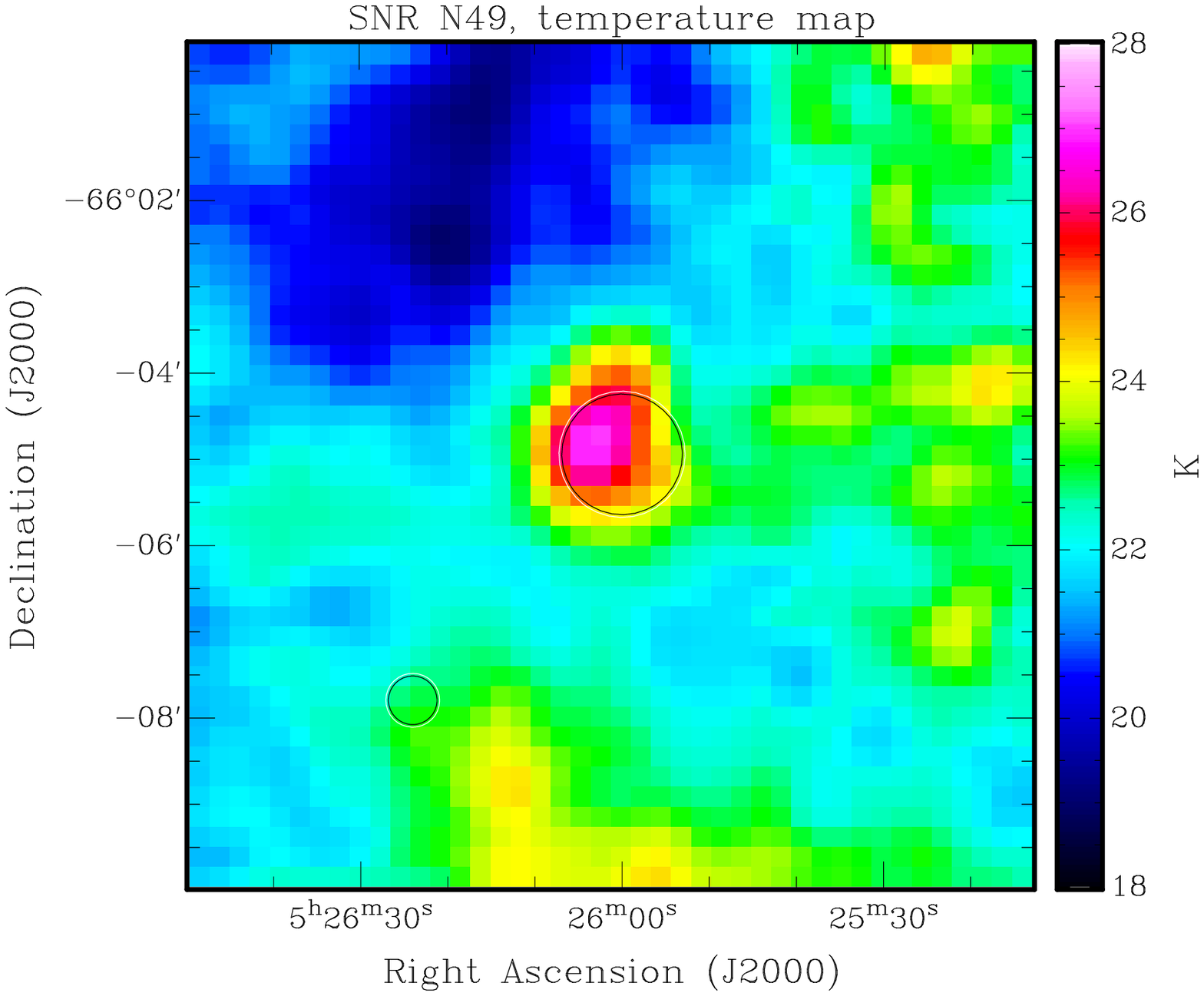,width=66mm}}
\hbox{
\psfig{figure=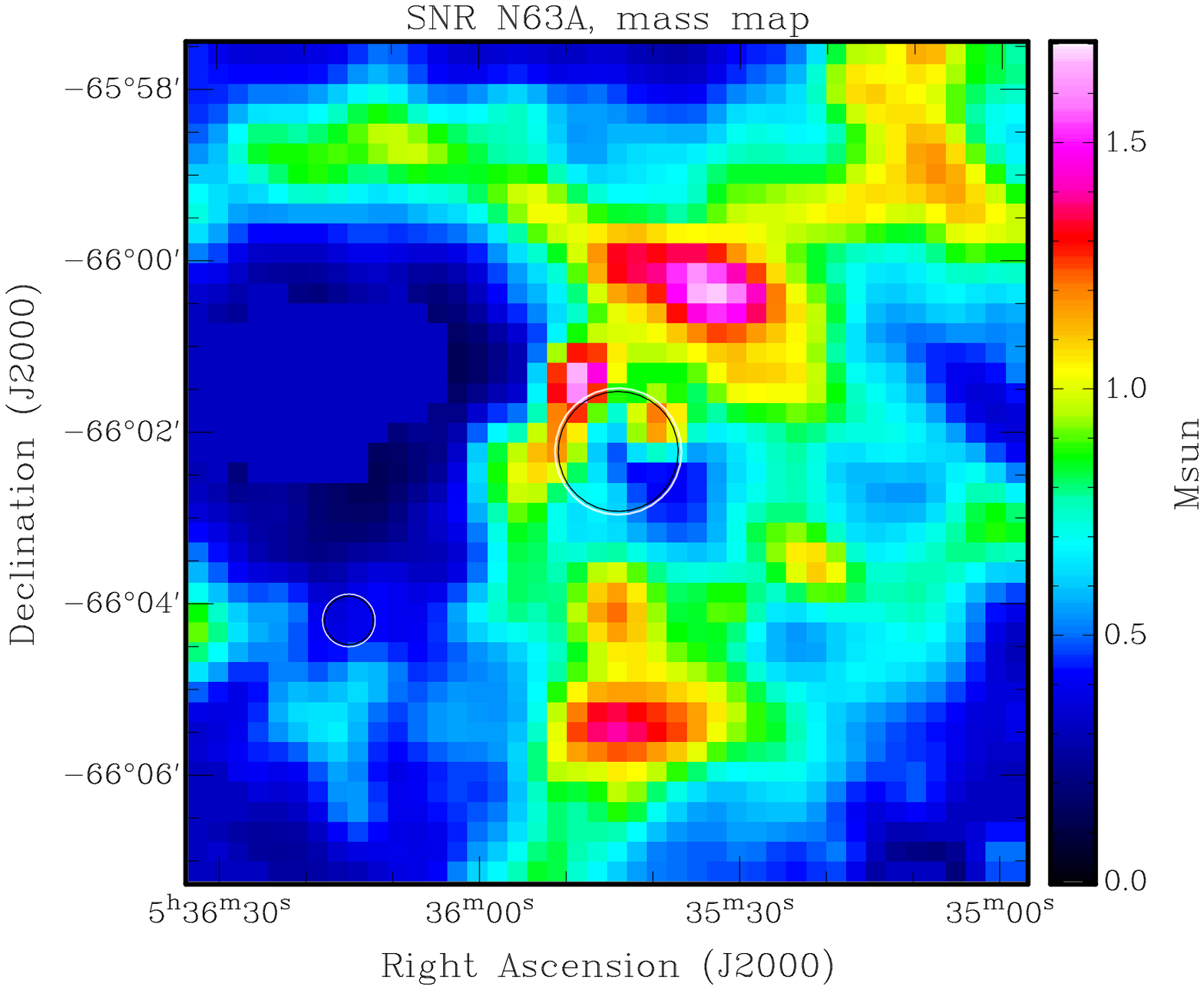,width=66mm}  
\psfig{figure=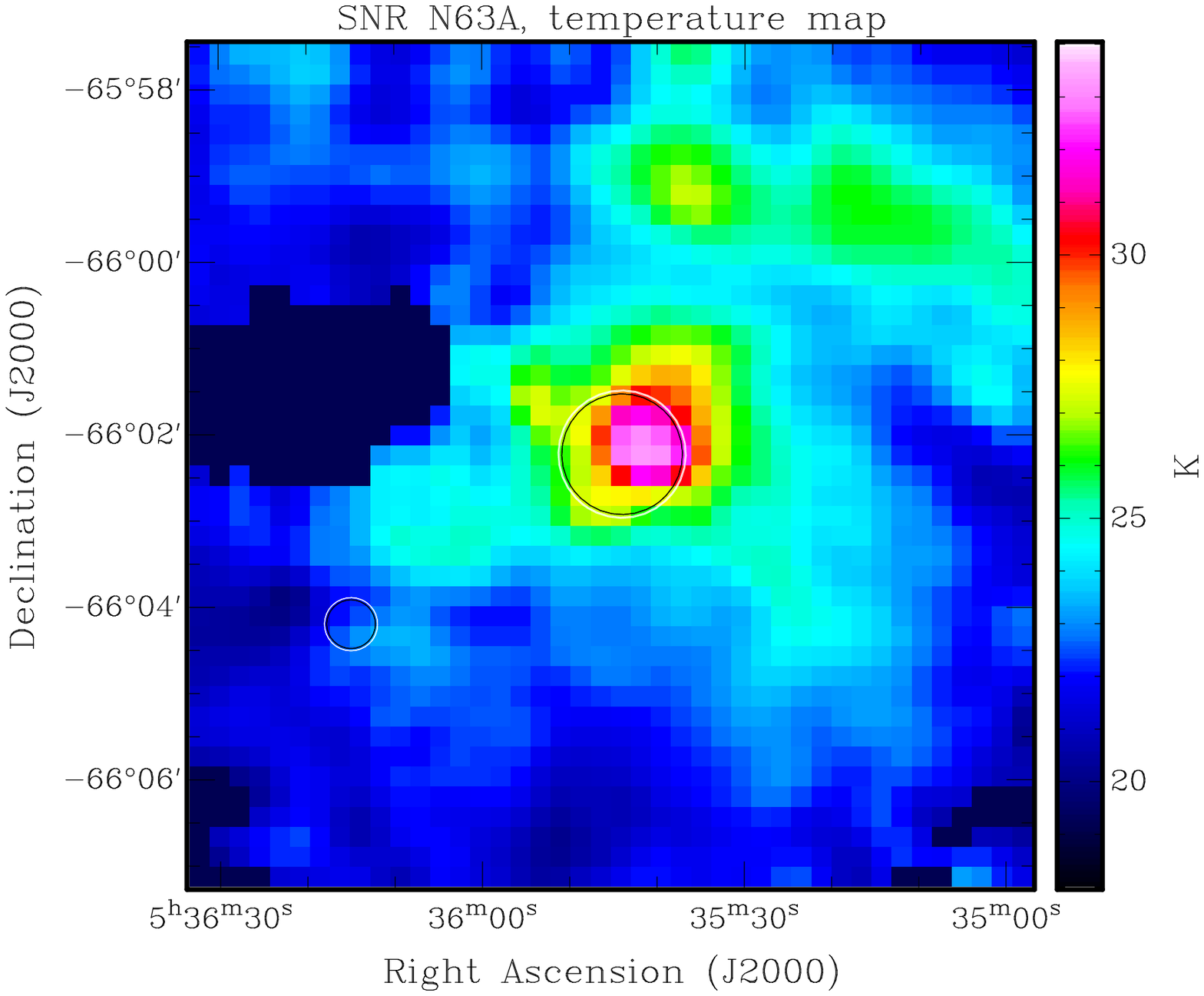,width=66mm}} 
}}
\caption[]{{\it Left:} dust mass maps; {\it Right:} dust temperature maps. The circle in
the centre of the maps represents the SNR, while the smaller circle to the
lower left represents the beam size. {\it Top:} SNR N\,49, {\it Bottom:} SNR N\,63A \label{N49}}
\end{figure*}

\begin{figure*}
\centerline{\vbox{
\hbox{
\psfig{figure=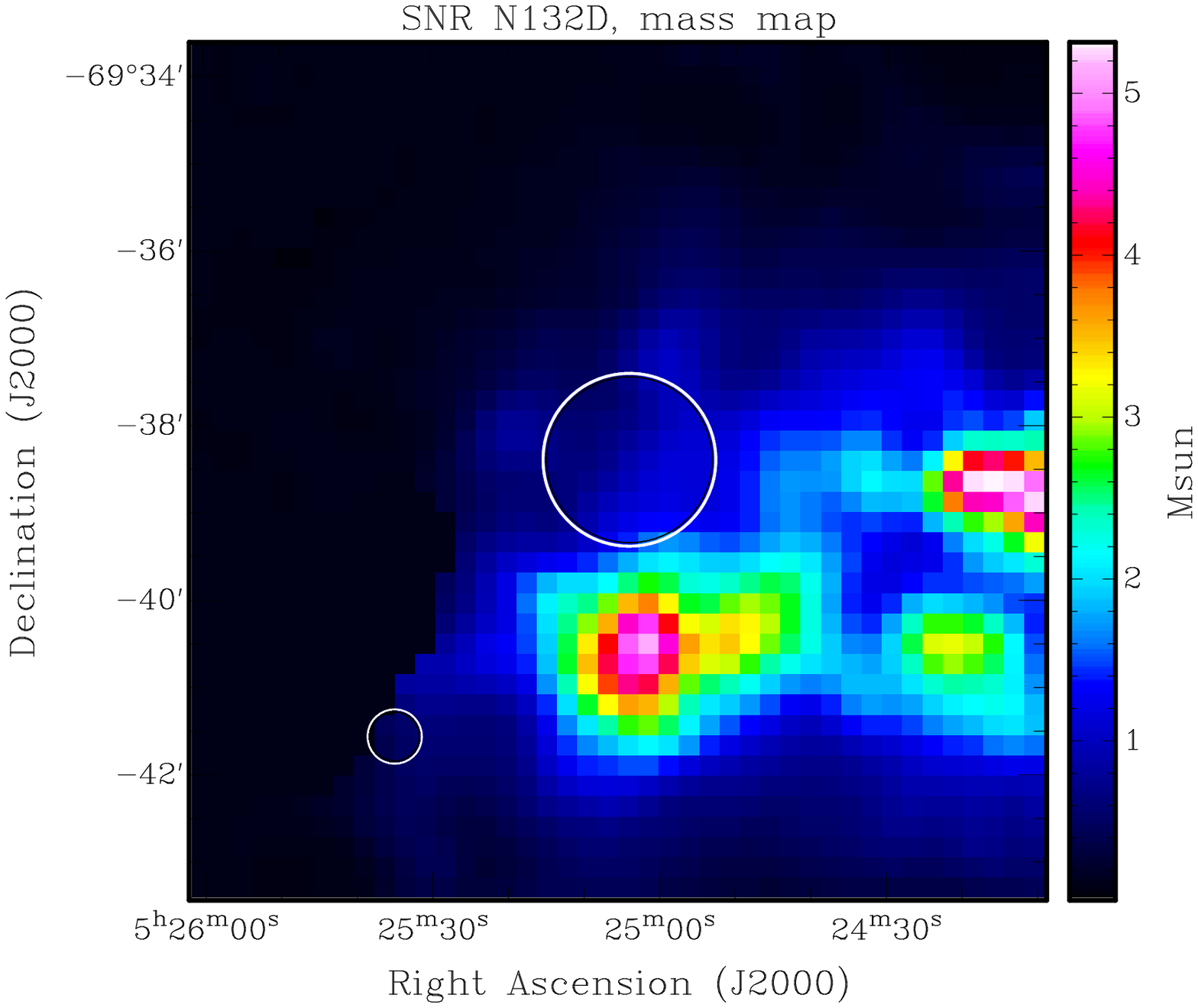,width=66mm}  
\psfig{figure=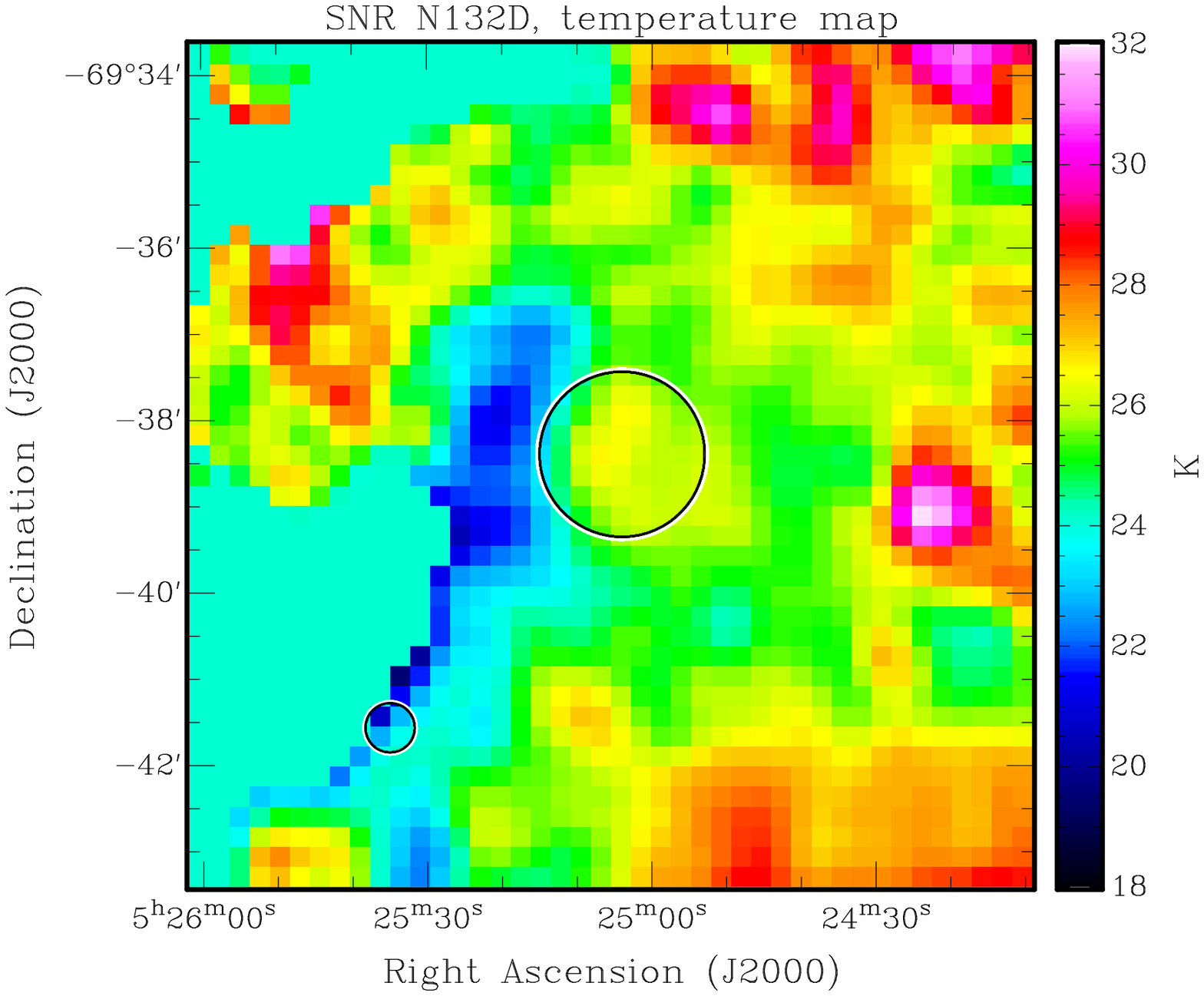,width=66mm}} 
\hbox{
\psfig{figure=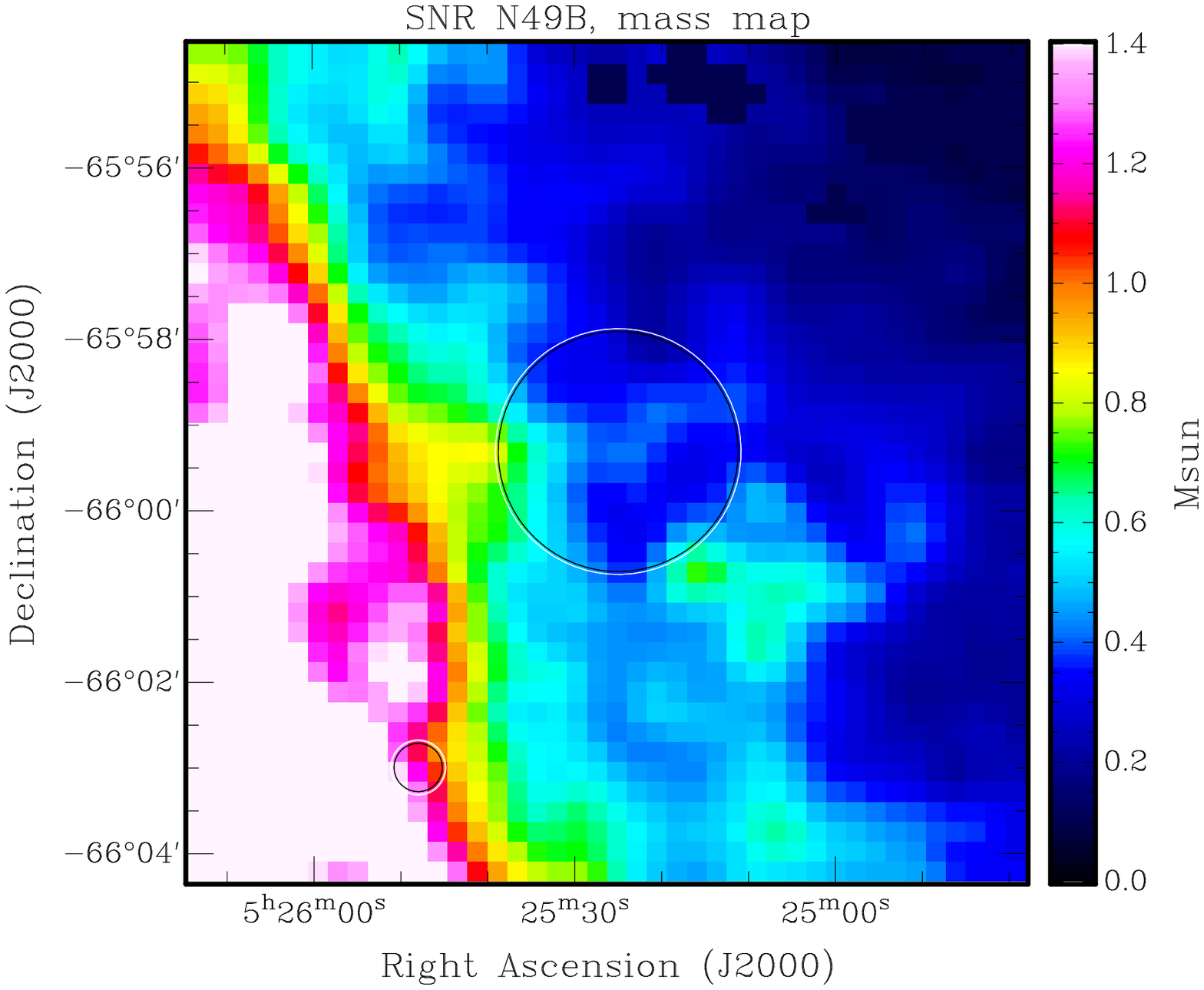,width=66mm}  
\psfig{figure=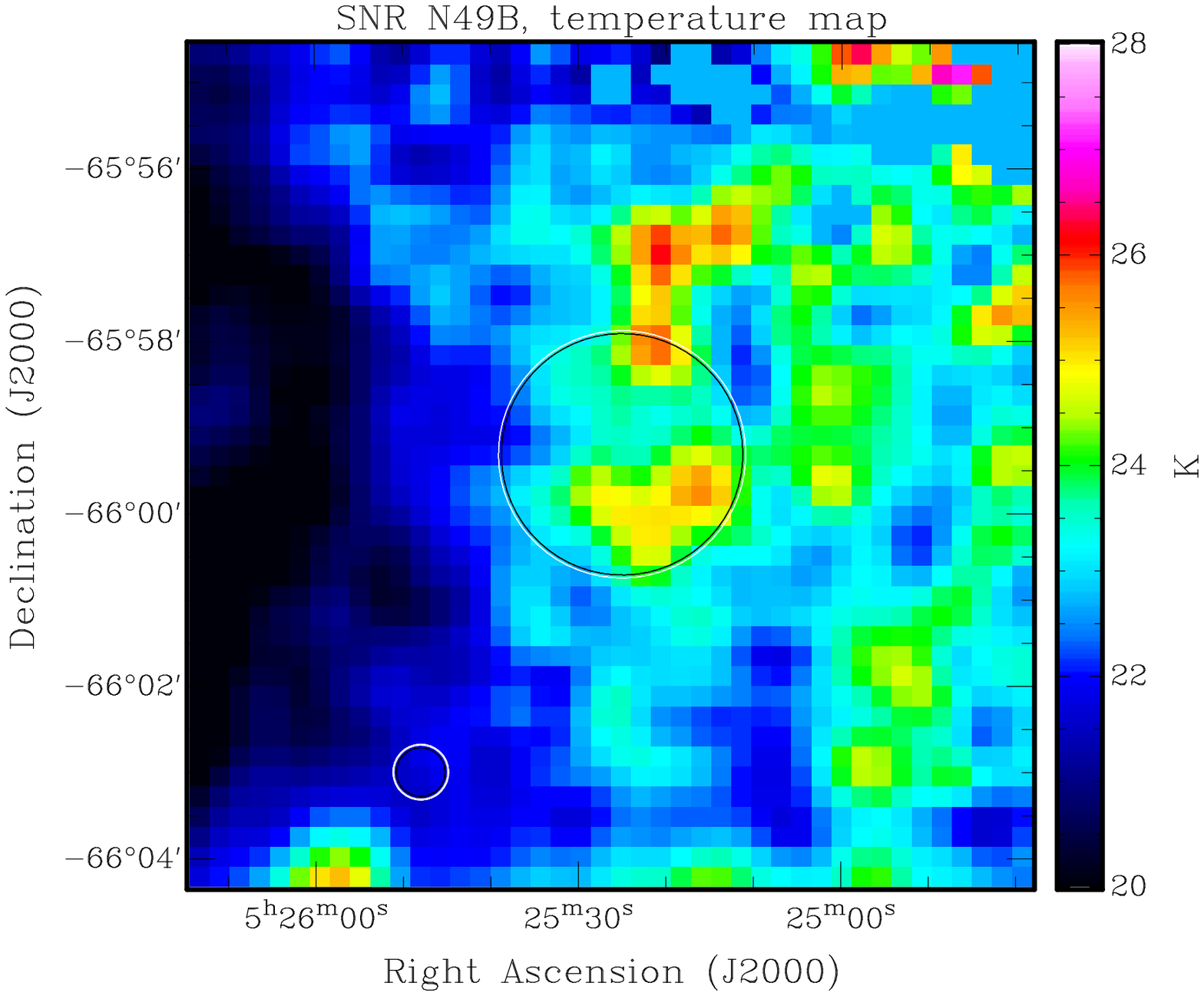,width=66mm}} 
}}
\caption[]{Like Figure~\ref{N49}, but for ({\it Top:}) SNR N\,132D, and ({\it Bottom:}) SNR N\,49B. \label{N491}}
\end{figure*}

The maps of mass and temperature of the dust in and
around N\,49 are displayed in Figures~\ref{N49}, top. The progenitor has an estimated
mass of 20 M$_{\odot}$ \citep{Hill95} -- this is a remnant resulting
from core collapse. Elevated temperatures are seen at the location of the
'blob', an interstellar dust cloud which is heated by the SNR shock; this bright
cloud is visible on the original {\it Spitzer} and {\it Herschel} images of the object 
\citep{Williams06,Otsuka10}. We conclude that we do
not see this cloud so prominently because it is massive, but because it is
heated. 

\subsubsection{N\,63A}

The maps of mass and temperature in and around N\,63A are displayed in Figures~\ref{N49}, bottom. 
The progenitor is likely to have been massive \citep{Hughes98}, and there is a large H\,{\sc ii} region to its
North--Western side. The warmest, North--Western part of the SNR corresponds
to the shocked lobes, which have a high contribution from line emission at the 24-$\mu$m flux \citep{Caulet12}. 
We also notice the lack of dust in this SNR. The 
remnant is detected with {\it Spitzer} \citep{Williams06}.

\subsubsection{N\,132D}

The maps of mass and temperature for N\,132D are displayed in Figures~\ref{N491}, top. 
This is a young, O-rich SNR, thought to have a Ib progenitor \citep{Vogt11}. \citet{Rho09} mapped IR
spectral lines arising from the ejecta of this SNR. The remnant was detected
with {\it Spitzer} \citep{WilliamsB06}. \citet{Tappe06} reported the destruction of PAHs/grains in 
the supernova blast wave via thermal sputtering. N\,132D has warmed up the little dust at the location of the SNR itself.  

\subsubsection{N\,49B}

This SNR was detected with {\it Spitzer} by \citet{WilliamsB06}. The maps for
N\,49B are displayed in Figures~\ref{N491}, bottom. Despite its mature age of $\approx10,000$
yr, its influence through the heating and lack of dust is clear. 

\subsubsection{DEM\,L71}

\begin{figure*}
\centerline{\vbox{
\hbox{
\psfig{figure=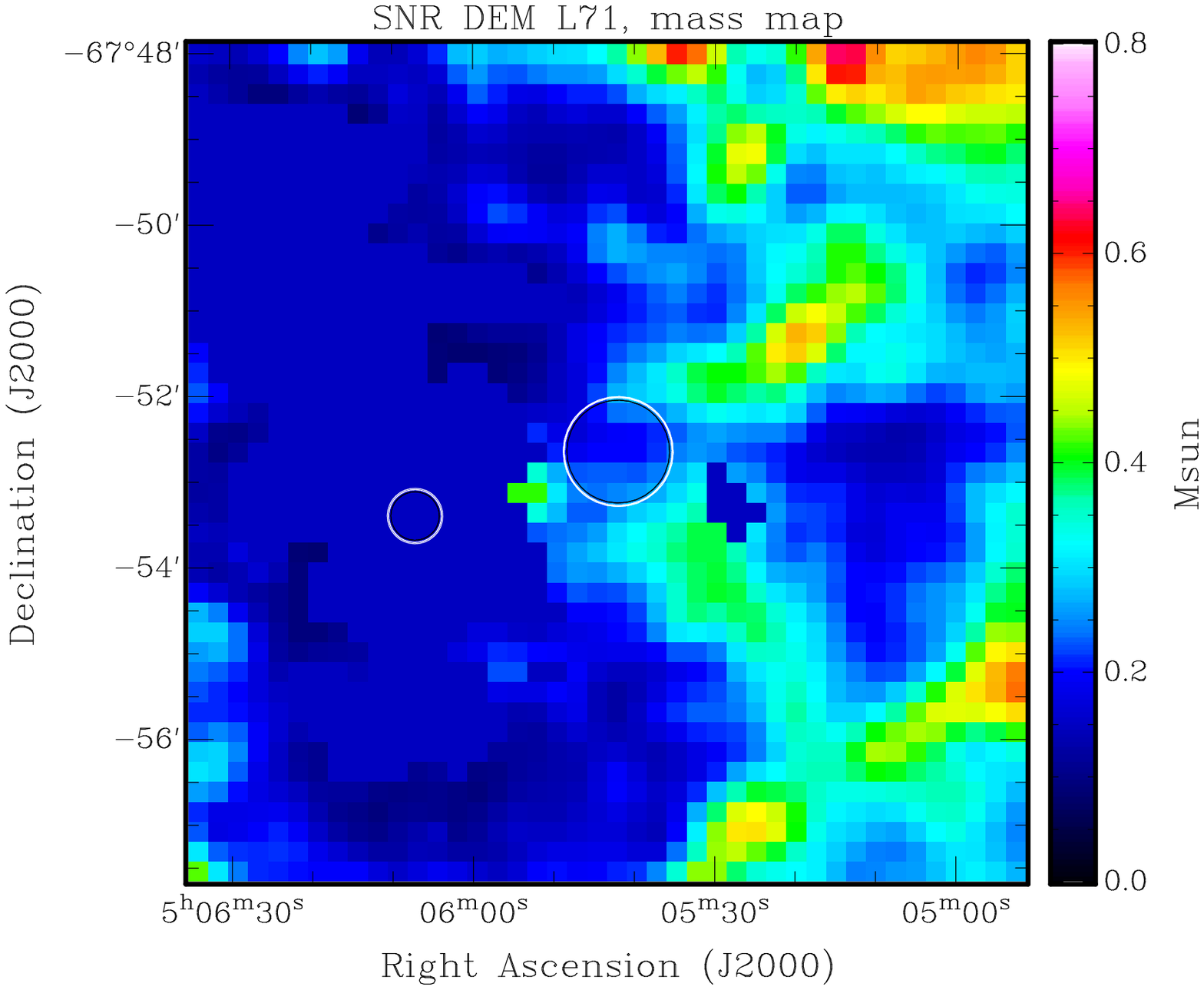,width=68mm}   
\psfig{figure=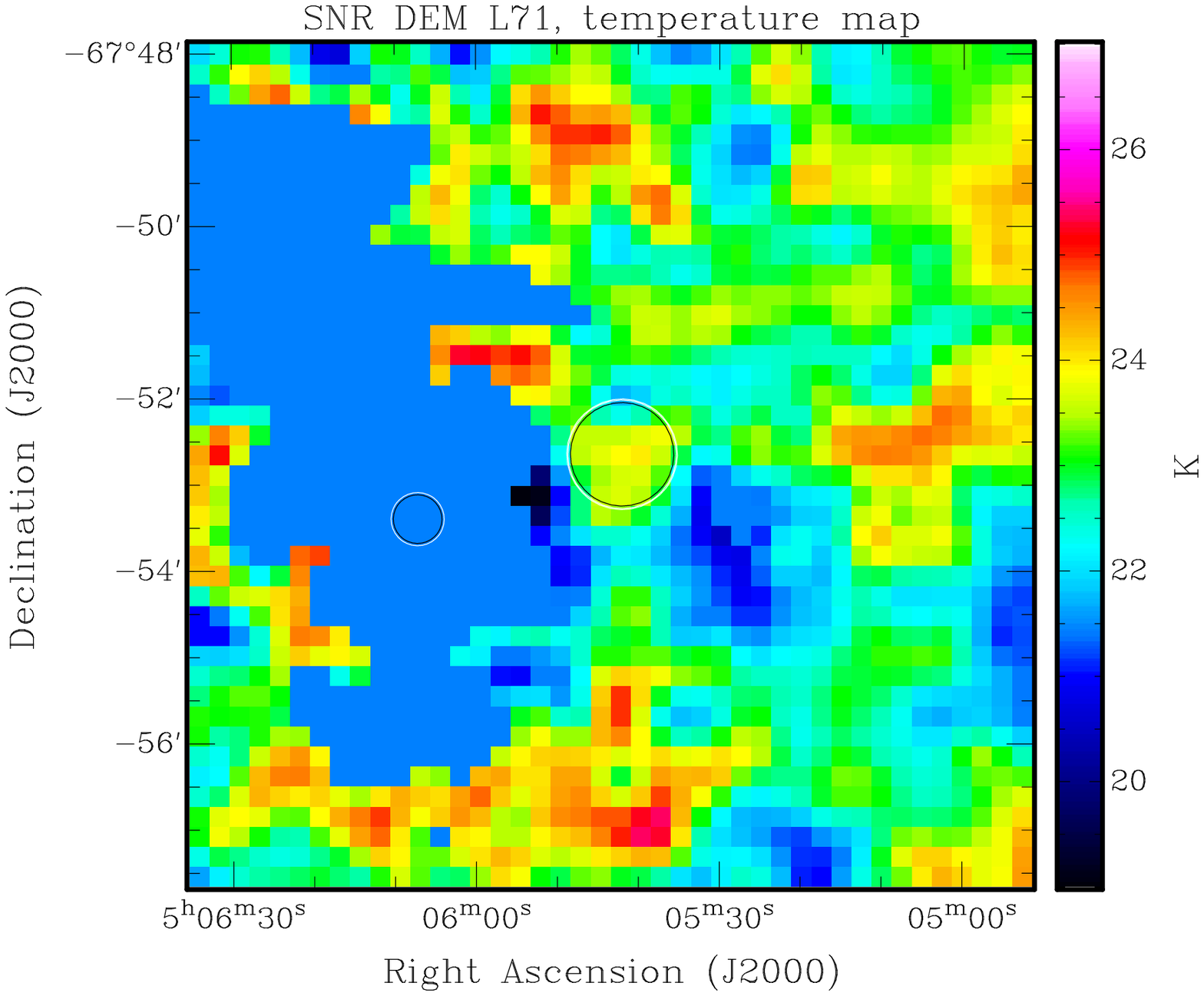,width=68mm}}  
\hbox{
\psfig{figure=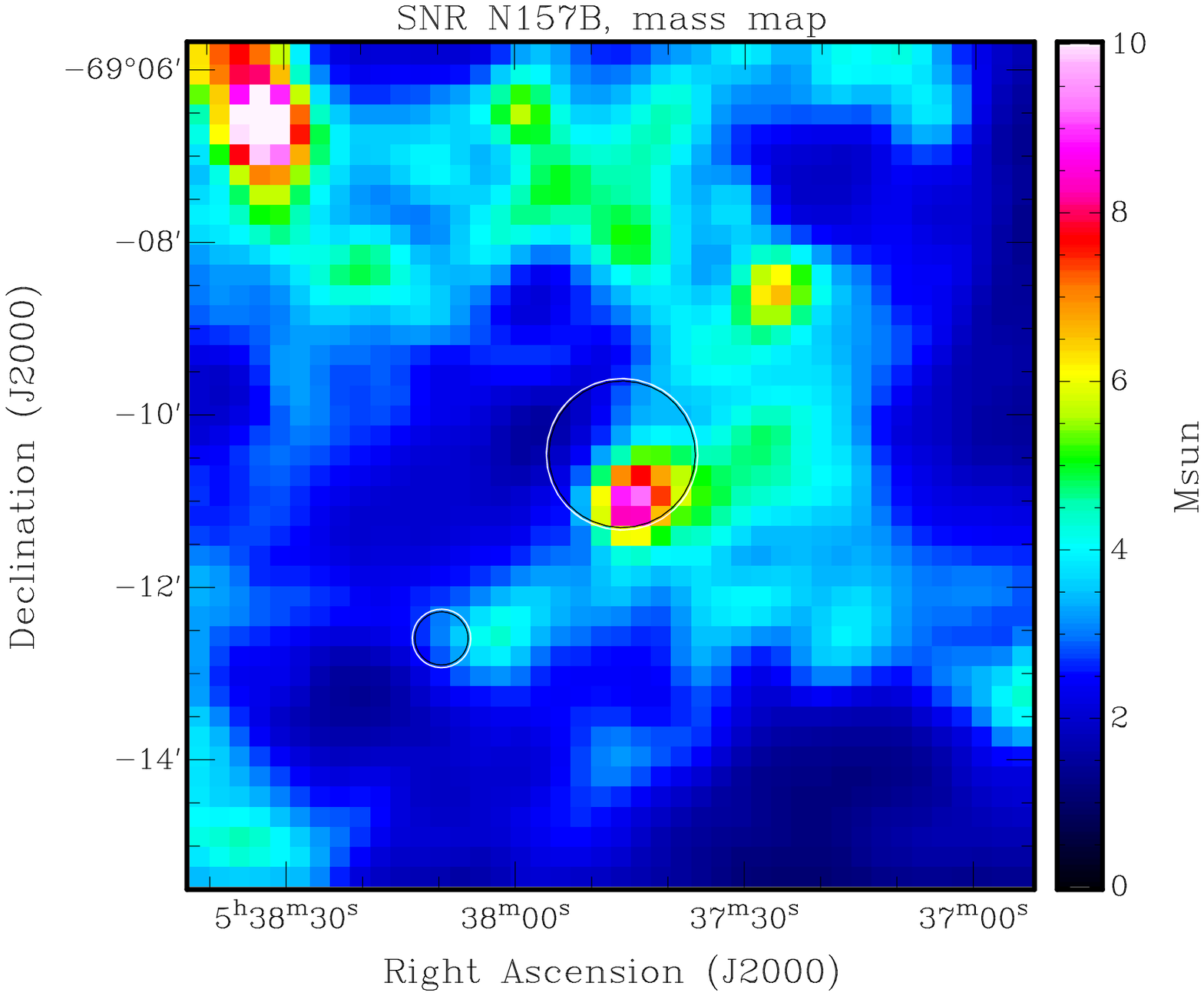,width=68mm}   
\psfig{figure=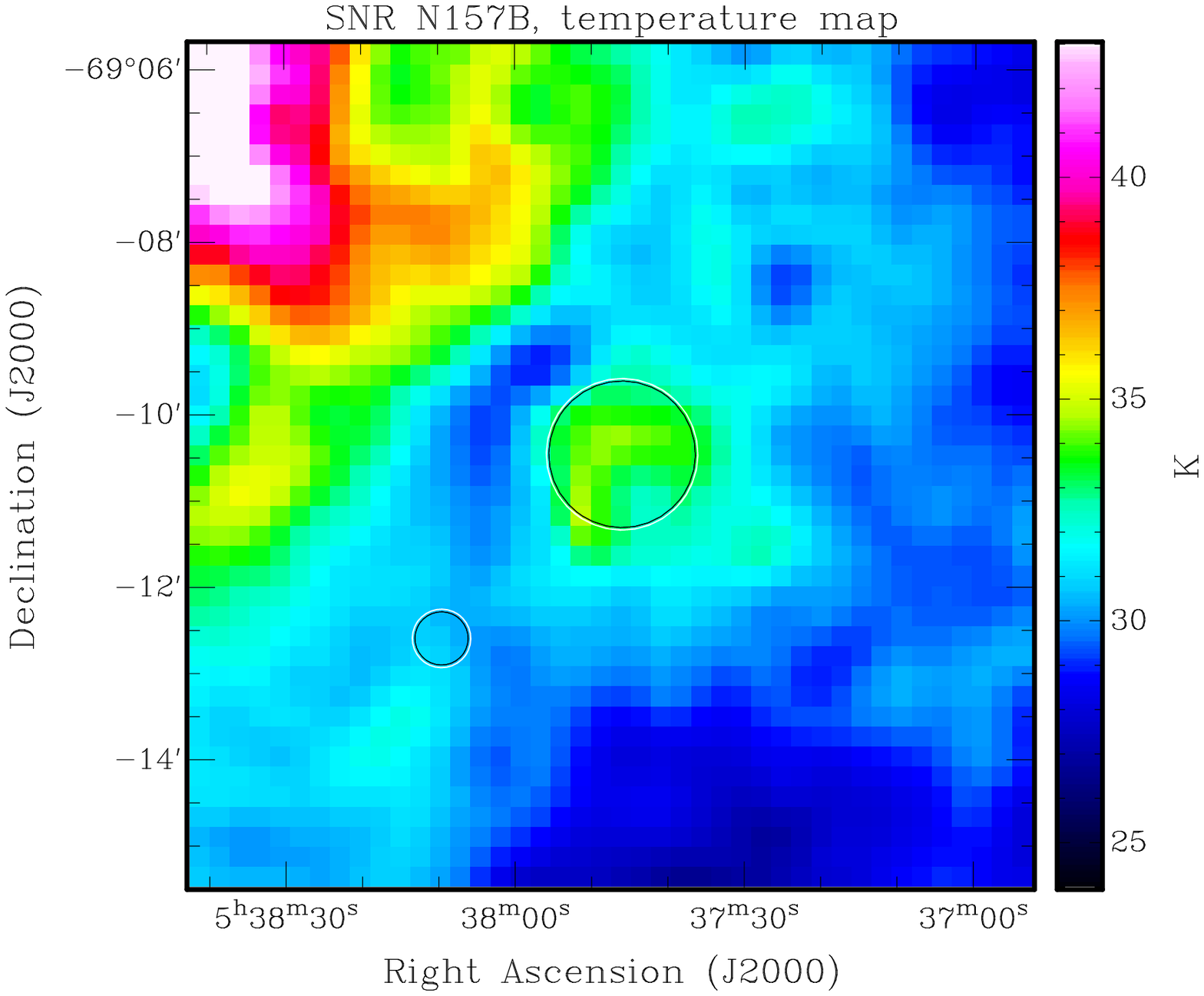,width=68mm}} 
\hbox{
\psfig{figure=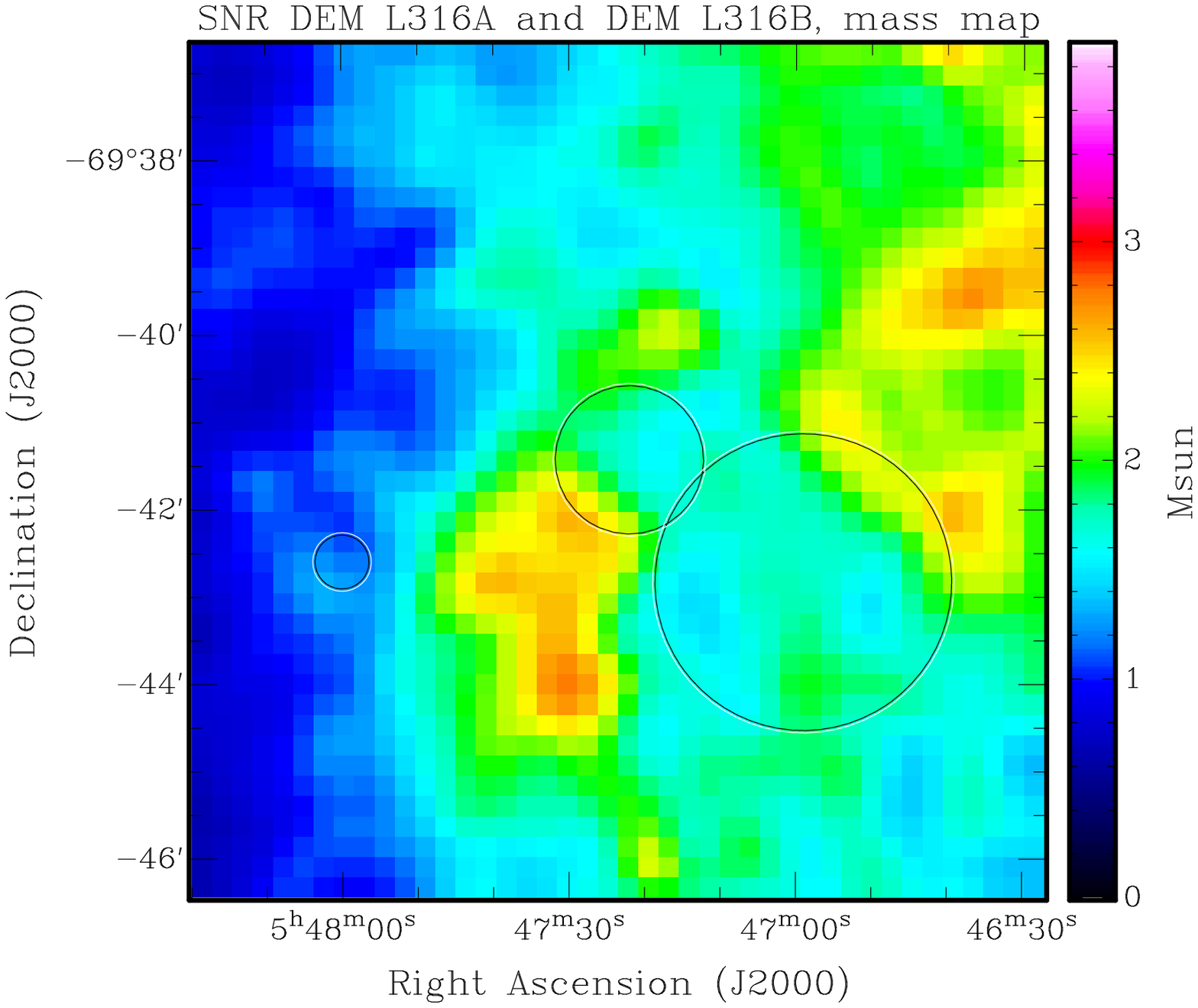,width=68mm}   
\psfig{figure=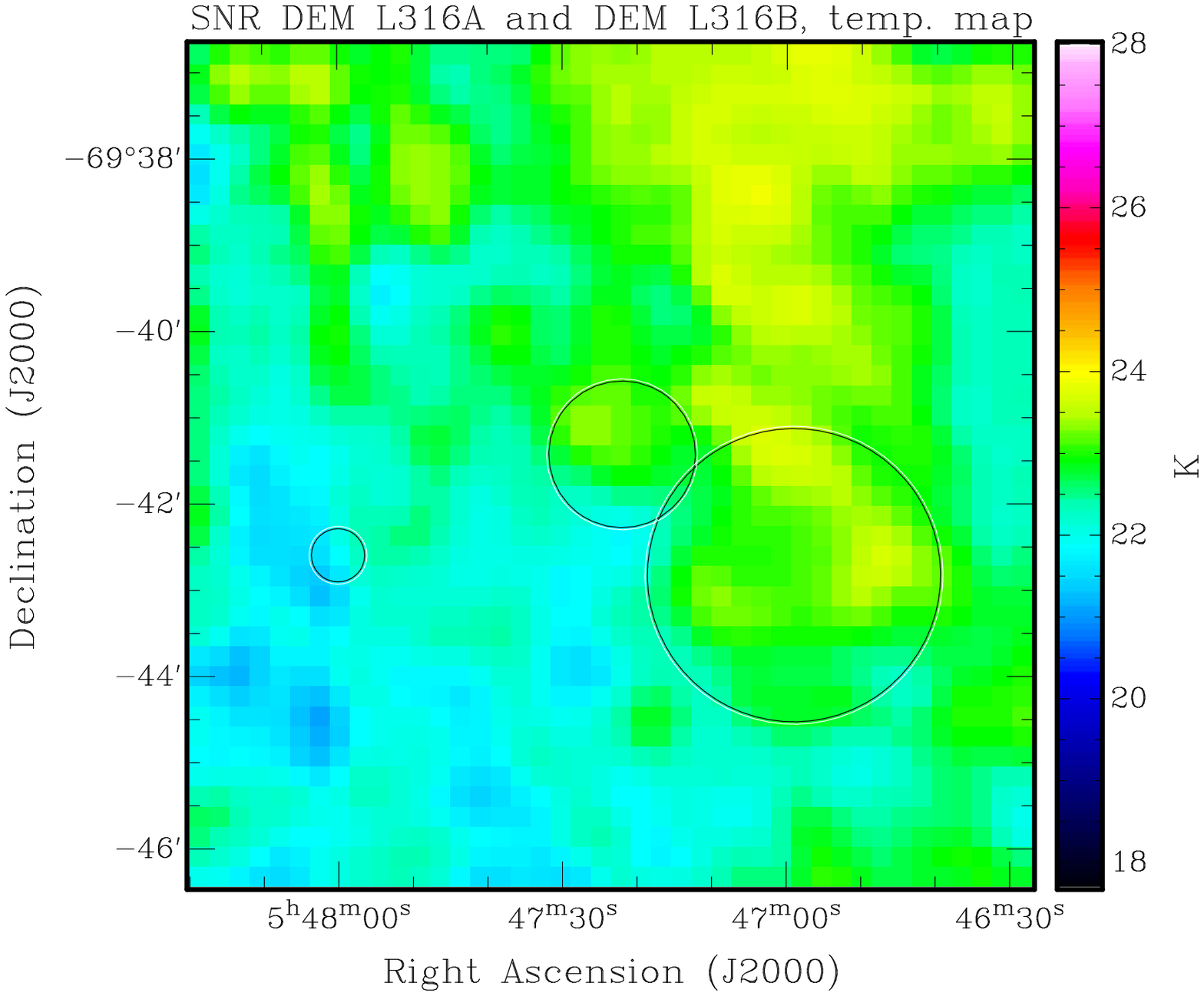,width=68mm}}  
}}
\caption[]{Like Figure~\ref{N49}, but for ({\it Top:}) SNR DEM\,L71, ({\it Middle:}) SNR N\,157B, and
({\it Bottom:}) SNRs DEM\,L316A and DEM\,L316B. \label{maps}}
\end{figure*}

The maps for DEM\,L71 are displayed in Figures~\ref{maps}, top. This is a young type Ia
remnant; 0.034 M$_\odot$ of warm dust in this SNR was measured by \citet{Williams10}. 

\subsubsection{N\,157B}

For N\,157B the maps are displayed in Figures~\ref{maps}, middle. This SNR, near to the
Tarantula Nebula mini-starburst, has resulted from the core collapse of a
massive progenitor, 20--25 M$_\odot$ \citep[][who detected it with {\it
Spitzer}]{Micelotta09}. It would be interesting to find out if the massive dust structure 
in the direction of this object is connected to the remnant.

\subsubsection{DEM\,L316A and DEM\,L316B}

These two objects are shown in Figures~\ref{maps}, bottom. There was an unresolved question 
whether these two objects are interacting and how close they are \citep{Williams2005}. These data suggest that they do belong to the 
similar FIR environment, although the first one is of Ia and the second one of CC origin. 

\subsubsection{Other SNRs of note}

Many SNRs show signs of dust removal and/or heating. One of the `hidden'
remnants, the Honeycomb is devoid of dust compared to its
surroundings whilst lacking any signs of heating. Other SNRs showing signs of
dust removal and/or heating within the SNR include 0453$-$68.5 \citep[detected with {\it Spitzer};][]{WilliamsB06}, N\,4,
the type Ia SNR\,0519$-$69.0 \citep[detected with {\it Spitzer};][]{Borkowski06a}, DEM\,L241, N\,23, B\,0548$-$70.4 \citep[heated dust; 
detected with {\it Spitzer};][]{Borkowski06a}, SNR\,0520$-$69.4 (probably removal and heating
on the edges), DEM\,L204 devoid of dust, DEM\,L109 and a
nearby compact SNR candidate, J\,051327$-$6911 \citep{Bojicic07} appearing
to heat a nearby dust cloud.

Several SNRs appear to interact with interstellar clouds at their
periphery. Young N\,158A (0540$-$69.3) (progenitor of 20--25 M$_\odot$, \citealt{Williams08}) is possibly interacting
with a dense cloud to the North. \citet{Williams08} detected its PWN with {\it Spitzer} at
wavelengths $\le$ 24 $\mu$m, but did not find any IR detection of the shell. They found dust synthesized
in the SNR, heated to 50--60 K by the shock wave generated by the PWN. The data we have only show possible heating of 
surrounding medium, but not an obvious influence by this SNR. The young SNR N\,103B
is not resolved; only R$_{70/24}$ shows some impact of this SNR. Other
SNRs interacting (or about to interact) with surrounding clouds include N\,9,
B\,0507$-$7029, SNR\,0450$-$709 and DEM\,L299. 

Some SNRs are confused. DEM\,L218 (MCELS\,J\, 0530$-$7008), 
of type Ia \citep{deHorta12} is covered with the object that \citet{Blair06} 
called SNR\,0530$-$70.1. N\,159 (0540$-$697) is in the complicated region, close to the black-hole
high-mass X-ray binary LMC\,X-1 and a bright H\,{\sc ii} region to the South--West \citep{Seward10}. 
SN\,1987A is utterly unresolved and barely visible on our mass map (we remind the reader that our maps are
constructed at the angular resolution of the 500-$\mu$m data).

\subsubsection{Pulsar wind nebul{\ae}}  

Some SNRs contain a PWN, which could make a synchrotron
contribution to the FIR spectrum or provide an additional mechanism to heat
dust. The SNRs which have been connected with pulsars in the LMC are N\,49 \citep{Park12}, 
N\,206, 0453$-$68.5, B\,0540$-$693 in N\,158A, N\,157B, DEM\,L241 \citep{Hayato06}, and DEM\,L214
\citep[J\,0529$-$6653;][]{Bozzetto12}. On the other hand, while all PWNe are
powered by pulsars, PWNe are known without a detected pulsar
\citep{Gaensler03a}, e.g., N\,23 \citep{Hayato06}. We did not find signatures of PWNe on any 
of our maps.

\section{Discussion}\label{discussion}

In Section~\ref{popmass} and Section~\ref{poptemp} we discuss the influence of the population of SNRs
on the interstellar dust mass and temperature. In Section~\ref{Produced} we argue for a lack of evidence for large
amounts of dust having formed in SNRs and survived. Given the extant literature which consistently indicates 
significant destruction of dust within SNRs by the reverse and forward 
shocks and hot gas \citep{Barlow78,Jones94,Borkowski06a,Bianchi07,Nozawa07,Nath08,Silvia10,Sankrit10} we interpret the 
removal of dust as due to sputtering, although we can not prove conclusively on the basis 
of our data that some dust may not be simply pushed out of the way. We attempt to quantify the amount of sputtered and/or pushed dust 
in Section~\ref{Sputtered}, based on the difference of column density between SNR surroundings and SNRs. 
We end with a brief discussion about the thickness of the interstellar dust layer
(Section~\ref{thickness}). 

\begin{figure}
\centerline{\vbox{
\psfig{figure=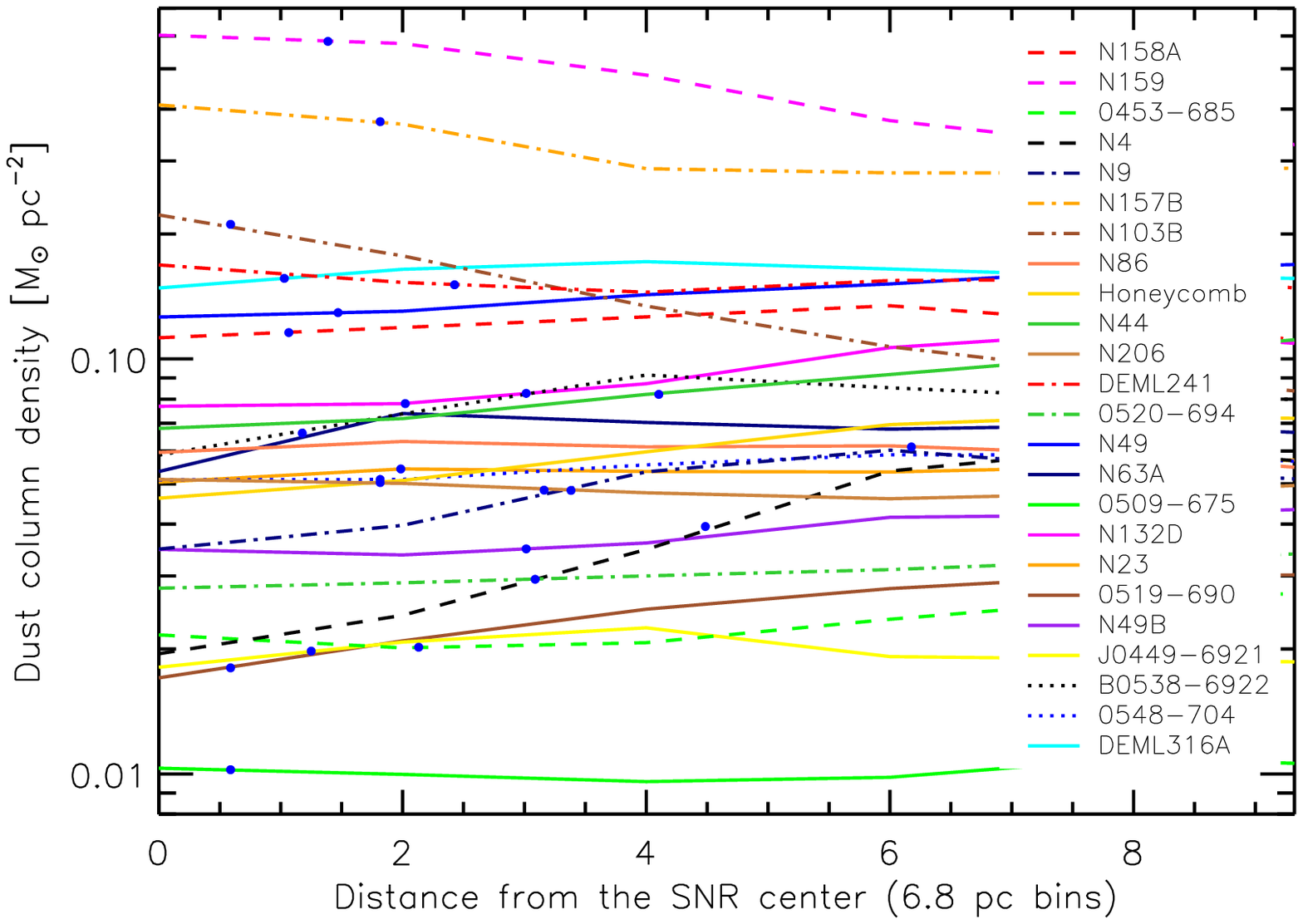,width=86mm}  
\psfig{figure=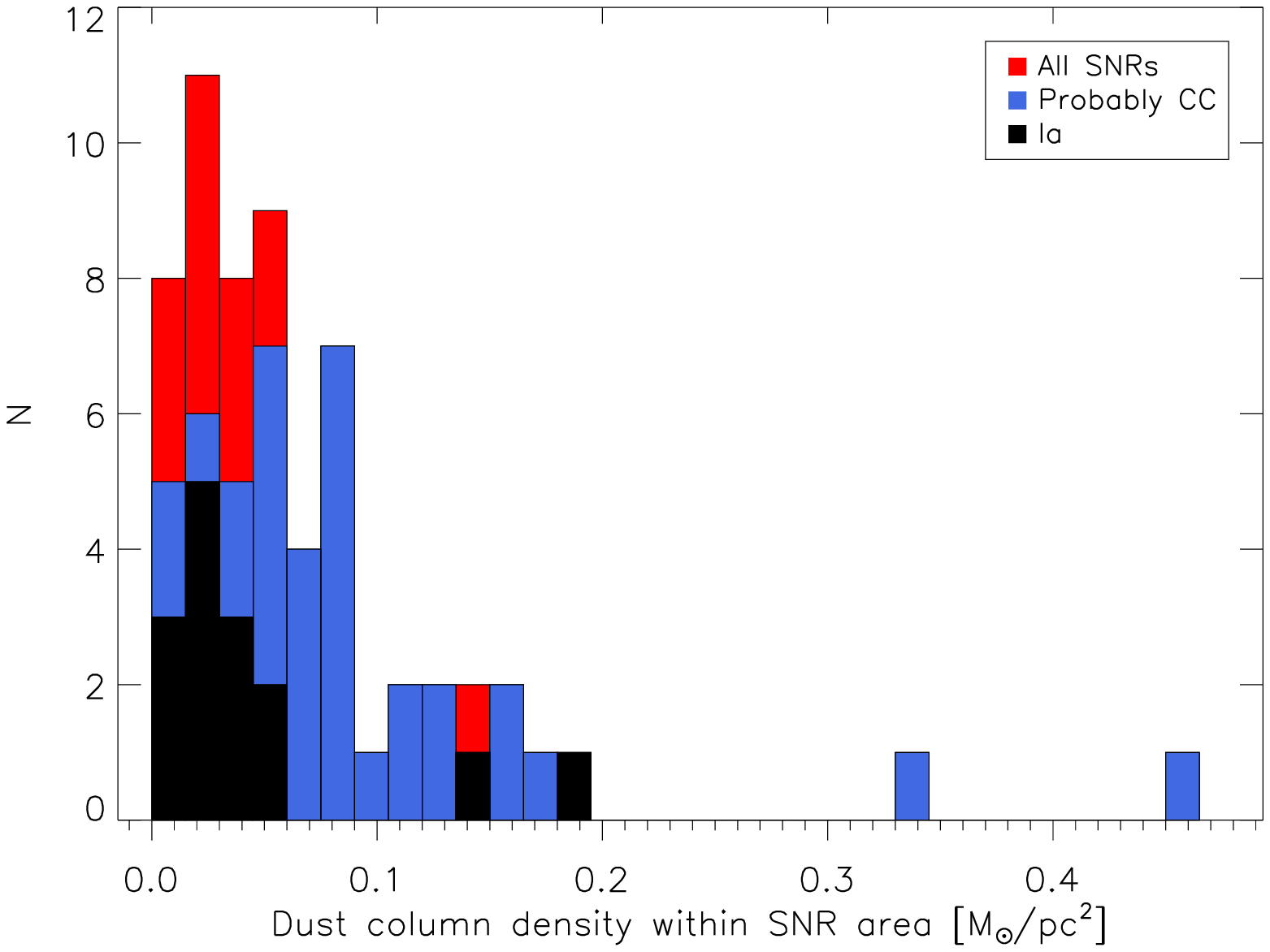,width=86mm}}  
}
\caption[]{{\it Top:} Radial profiles of dust mass distribution across SNRs
and their environment. The blue circles mark the SNR radius. {\it Bottom:} 
Histogram of the distribution of the average dust mass [${\rm M_{\odot}}$ pc$^{-2}$] within one 
diameter of Ia, core-collapse and all SNRs together.\label{MassesSNRs}}
\end{figure}

\subsection{Influence of SNRs on interstellar dust mass} \label{popmass}

In Figure~\ref{MassesSNRs} (top panel) we present the radial profiles, the average dust column density in and around SNRs as a function of radial distance 
to the centre of the SNR, for a sub-sample of 22 SNRs, where each successive bin comprises 
a 6.8 pc wide annulus. The blue circles mark where the SNR radius end. While often less dust is seen towards the SNR -- although sometimes 
the opposite is seen -- the radial profiles are generally fairly flat; this indicates that the 
amount of (cold) dust that is removed is relatively small compared to the amount of dust in 
the LMC in that direction. These profiles also reflect a general
relation between SN progenitor type and ISM density, as core-collapse SNe are usually seen in regions of 
recent star formation whereas SNe Ia often bear no memory of their natal environment. On this basis,
N\,157B and N\,159 stand out in terms of ISM density; they are bright 
at FIR and radio wavelengths and their progenitors are probably massive. 

To examine the difference in ISM surrounding CC and Ia SNRs, in the bottom panel 
of Figure~\ref{MassesSNRs} we show the distributions of average dust mass 
within the SNR diameter, separately for CC, Ia and 
all SNRs together (for typing see Section~\ref{sample}). Ias are usually in dust-poor environments, but there are exceptions such as 
the probable prompt Ia SNRs DEM\,L316A and N\,103B, that are seen in the direction of
dense ISM.

Some of the SNRs that we claim here to sputter and/or push away dust are already claimed to be
dust destroyers based on {\it Spitzer} observations. With {\it Herschel} data the lack of dust can actually 
be observed for SNRs that have enough dust to interact with, since they enable us to see the colder dust that surrounds SNRs. 

If dust were pushed out of the way, we should see dust piling up around the rims of 
SNRs -- which we do not see, but we might not be able to resolve it. For most of the remnants (D$\gtrsim 10$ pc) the dust formed in the progenitor 
envelope and CSM must have been significantly sputtered and/or pushed away by shocks, in early stages. Only if the progenitor
star was an early-type star, it could have formed an interstellar bubble of a size $\sim$30 pc \citep{Castor75}\footnote{The winds of red supergiant SN progenitors
are slow, $v\ll$100 km s$^{-1}$, and such CSM is quickly over-run by the SNR (see, e.g. \citealt{vl10}).}, otherwise the dust around remnants is not 
directly connected with the CSM of the progenitor but with the pre-existing ISM dust, whose density may be related indirectly to the progenitor type. 

As \citet{Forest88} have shown there seems to be a significant association between SNRs and H\,{\sc ii}
regions, suggesting the preponderance of core-collapse SNRs in the LMC. The effect that they observed might also be
attributed to SNRs being rendered visible by virtue of their interaction with
the ISM (or CSM). At our dust mass maps many SNRs are close to, or at the 
rim of a dusty structures (clouds). That could hide the influence of SNRs to be visible for the observer in two ways:
if the cloud is irregular and inhomogeneous, we would not see a lack of dust in SNRs; if the SNR is next to the cloud 
from our perspective, then it is more difficult to claim that it has removed any dust and not possible
to estimate how much. 

Also, large SNRs ($D>100$ pc) subtend a larger area on the sky than the annulus, which tends to
reduce the difference between inner and outer mass of the dust. On the other hand, the annuli of the smallest SNRs
are more likely to be more massive than areas within remnants from the same reason.

\subsection{Influence of SNRs on interstellar dust temperature} \label{poptemp}

\begin{figure}
\centerline{\vbox{
\psfig{figure=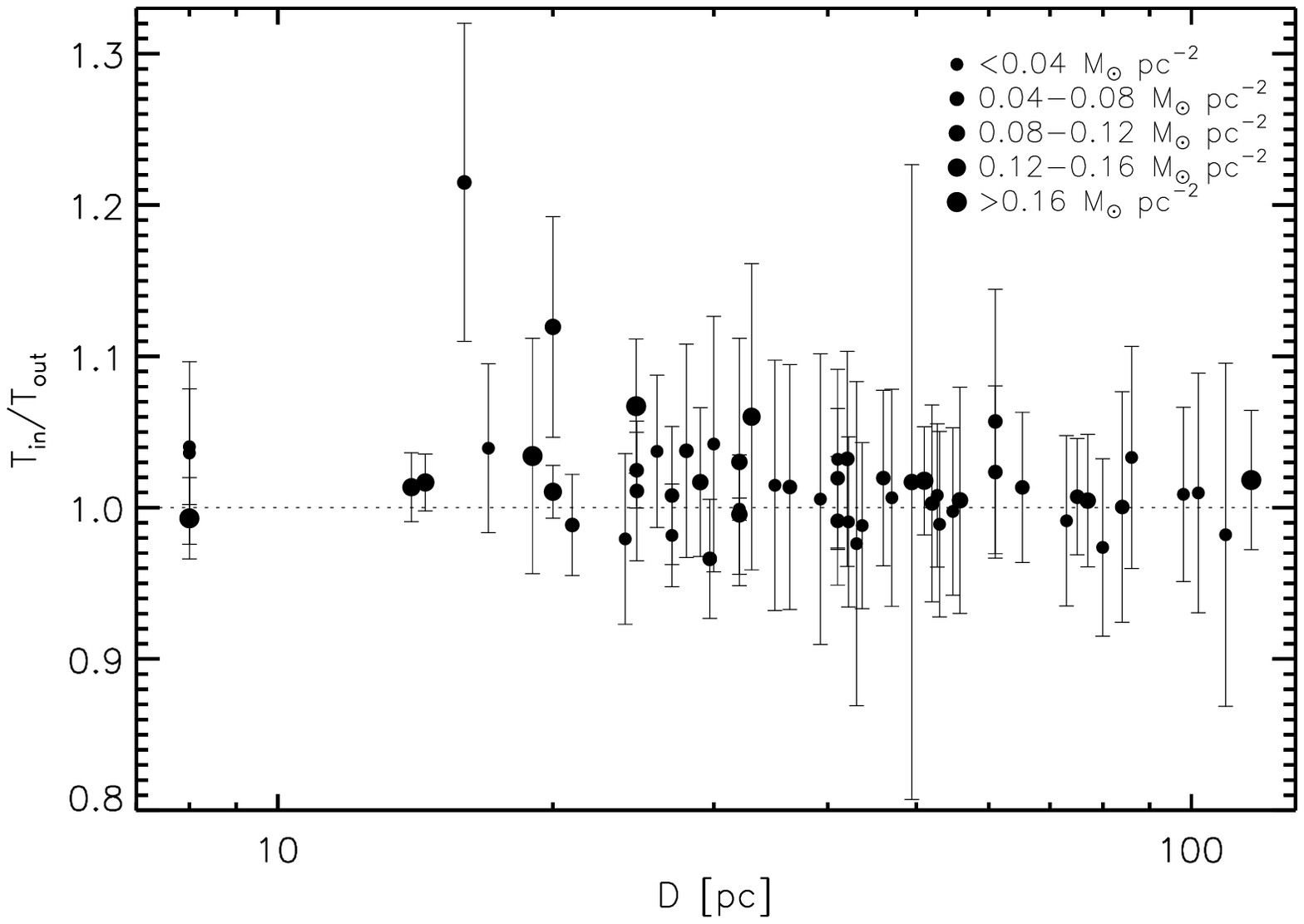,width=86mm}  
\psfig{figure=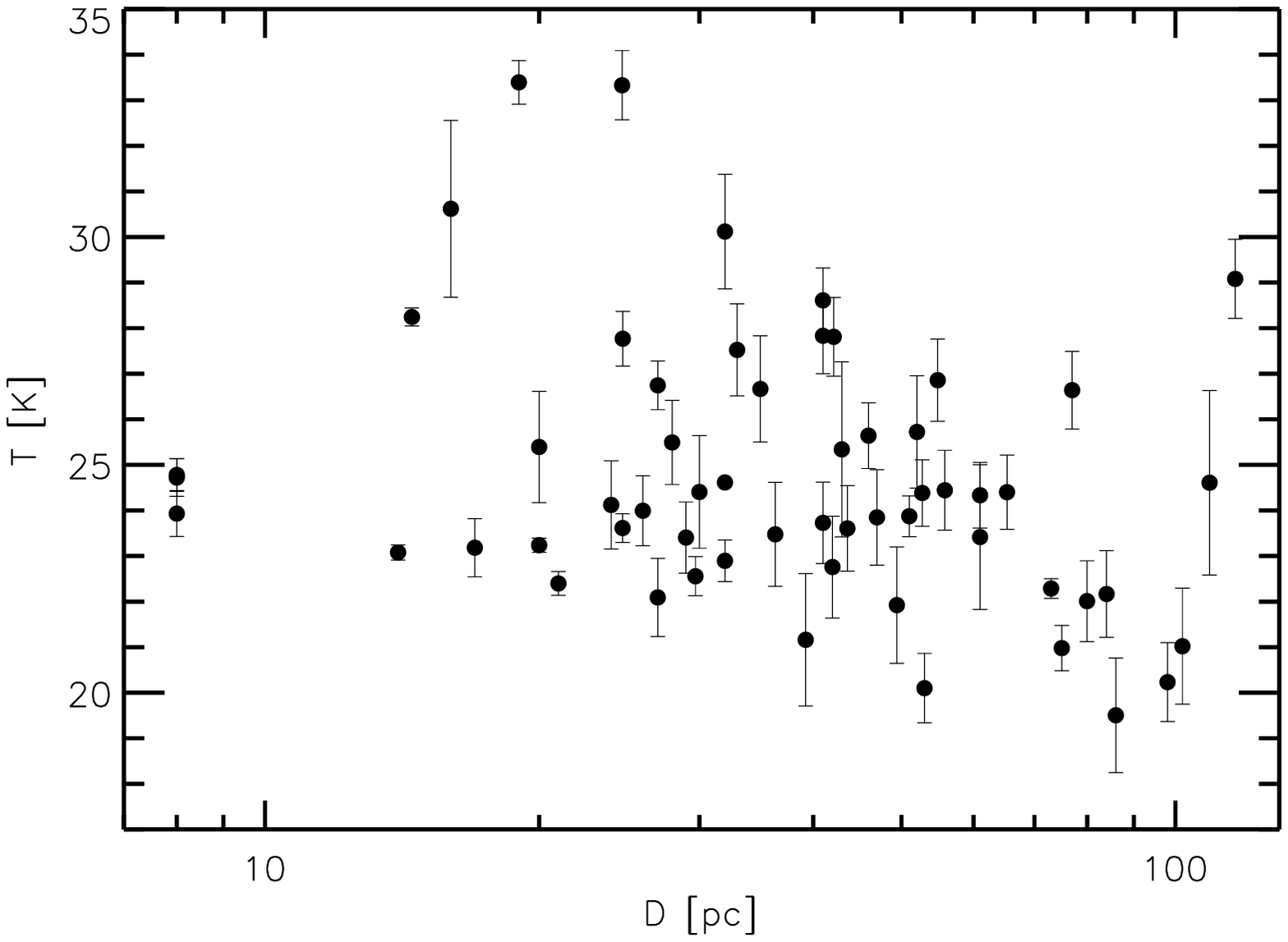,width=86mm}  
}}
\caption{{\it Top:} ratio of average dust temperature within and outside the SNR vs.\ SNR diameter. The size of
the symbol is proportional to the average dust column density in the direction of the SNR. {\it Bottom:} 
Temperature within SNRs vs. diameters. \label{T_mass}}
\end{figure}

Within SNRs there is usually a higher temperature -- even for the older remnants. In older remnants the dust
temperature often peaks near the edges or outside of the remnants which can probably
also be a sign that the dust has been removed. In Figure~\ref{T_mass}a we show the ratio of inner and outer (from annuli 20 pc thick) temperature and in Figure~\ref{T_mass}b 
we show the temperatures for the individual SNRs (both derived from temperature maps while 
the errors are standard deviations) which indicates that most of the SNRs do heat up the surroundings and that the temperature of this dust 
is somewhat higher than the temperature of the interstellar radiation field. The heating of the dust is most prevalent in the
more compact, presumably younger SNRs as well as the ones with a higher dust
content.

The temperature maps generally show heating of dust on SNR locations, usually in SNRs
that are detected with {\it Spitzer}, although here we use only {\it Herschel} data. Since
the dust is being eroded and cooled, this heating is only seen in young and sufficiently dense SNRs. 

We do not exclude the possibility that the heated dust belongs to the closer ISM surroundings of SNRs. 

\subsection{Production of dust in SNRs}\label{Produced}

It is possible that all SNRs create some dust in their ejecta, but it is not possible to recognise with the maps we 
have. The youngest SNR in our sample is SN\,1987A. 
It has produced $\sim0.6$ M$_\odot$ of cold dust \citep[cf.\ ][]{matsuura,lakicevic,indebetouw}. It is
unresolved, stands out on the {\it Spitzer} and short-$\lambda$ {\it Herschel} images but not at $500$ $\mu$m. On our mass map 
the mass is slightly higher at the place of that object, but this is barely visible -- only because the surrounding ISM is not dense. 

Another object that is probably in free expansion phase (other objects are in later stages -- most of
the material associated with them is swept up ISM) is N\,103B. N\,103B might not be 
resolved on our maps; it has reduced ${\rm R_{70/24}}$ compared to the surrounding ISM, but other ratios do not 
show that characteristic. It may indeed be in a rare environment, behind (or in front of) the massive cloud \citep{Dickel95}. We see no dust production here 
distinguishable, but we do not expect it, since it is a Ia remnant.  

Almost all other SNRs are larger than $50^{\prime\prime}$, and cover $>50$ pc$^{2}$.
Even if 1 M$_\odot$ of dust were present in the ejecta, even if it is not spread on more than $\sim$3 arcsec$^{2}$, in an evolved SNR this would correspond to a
surface density $<0.009$ M$_\odot$ pc$^{-2}$, i.e.\ often below that of the surrounding ISM ($\sim 0.01-0.6$ M$_{\odot}$ pc$^{-2}$; Fig.~\ref{MassesSNRs}).

\subsection{How much dust have SNRs removed?}\label{Sputtered}

\begin{table}  
\caption{Name of SNR, dust column densities within ($N_{{\rm in}}$) and outside ($N_{{\rm out}}$) the SNR in M$_\odot$ pc$^{-2}$ 
and the total amount of dust which could be removed by the SNR ($M$[M$_{\odot}$]).\label{tbl3}}
\begin{tabular}{lllll}
\hline\hline
Name             &  $N_{{\rm in}} \pm\Delta N_{{\rm in}}$   & $N_{{\rm out}} \pm \Delta N_{{\rm out}}$   & $M \pm \Delta M$   \\
\hline
B\,0519$-$690&$0.015\pm0.004$&$0.024\pm0.007$&$0.9\pm0.3$\\      
DEM\,L71&$0.019\pm0.005$&$0.021\pm0.006$&$0.20\pm0.07$\\              
B\,0509$-$675&$0.009\pm0.003$&$0.01\pm0.003$&$0.04\pm0.02$\\          
N\,103B&$0.19\pm0.05$&$0.13\pm0.04$&-\\                              
0548$-$704&$0.05\pm0.02$&$0.06\pm0.02$&$4\pm2$\\                      
DEM\,L316A&$0.15\pm0.04$&$0.16\pm0.05$&-\\                                  
N\,9&$0.05\pm0.02$&$0.05\pm0.02$&-\\                                   
0534$-$699&$0.034\pm0.009$&$0.033\pm0.009$&-\\                        
DEM\,L238&$0.009\pm0.003$&$0.01\pm0.003$&-\\                          
DEM\,L249&$0.04\pm0.01$&$0.031\pm0.008$&-\\                                  
0520$-$694&$0.027\pm0.007$&$0.033\pm0.009$&$11\pm3$\\                   
DEM\,L204&$0.008\pm0.003$&$0.011\pm0.003$&-\\                          
0450$-$709&$0.031\pm0.008$&$0.035\pm0.009$&-\\                           
HP99498&$0.019\pm0.005$&$0.017\pm0.005$&-\\                              
DEM\,L218&$0.023\pm0.006$&$0.018\pm0.005$&-\\                           
N\,23&$0.05\pm0.02$&$0.06\pm0.02$&$4\pm2$\\                              
N\,132D&$0.08\pm0.02$&$0.10\pm0.03$&-\\                                  
N\,157B&$0.34\pm0.09$&$0.29\pm0.08$&-\\                            
N\,44&$0.09\pm0.03$&$0.11\pm0.03$&-\\                                    
N\,158A&$0.13\pm0.04$&$0.13\pm0.04$&-\\                                   
N\,206&$0.05\pm0.02$&$0.05\pm0.02$&-\\                                   
N\,120&$0.08\pm0.03$&$0.10\pm0.03$&-\\                                  
N\,49B&$0.035\pm0.009$&$0.04\pm0.02$&$12\pm4$\\                         
N\,49&$0.12\pm0.04$&$0.14\pm0.04$&$6\pm2$\\                              
N\,11L&$0.07\pm0.02$&$0.08\pm0.03$&$4\pm2$\\                             
N\,86&$0.06\pm0.02$&$0.04\pm0.02$&-\\                               
0453$-$685&$0.019\pm0.005$&$0.024\pm0.007$&-\\                           
N\,63A&$0.06\pm0.02$&$0.07\pm0.02$&-\\                                 
DEM\,L203&$0.07\pm0.02$&$0.08\pm0.02$&-\\                                  
DEM\,L241&$0.15\pm0.04$&$0.15\pm0.04$&-\\                                
DEM\,L299&$0.09\pm0.03$&$0.10\pm0.03$&-\\                                  
DEM\,L109&$0.08\pm0.03$&$0.07\pm0.02$&-\\                                 
MCELS\,J0506-6541&$0.025\pm0.007$&$0.026\pm0.007$&-\\                        
0507$-$7029&$0.031\pm0.008$&$0.04\pm0.01$&-\\                             
0528$-$692&$0.012\pm0.004$&$0.013\pm0.004$&$2.0\pm0.8$\\                       
DEM\,L214&$0.003\pm0.001$&$0.003\pm0.001$&-\\                          
0532$-$675&$0.07\pm0.02$&$0.07\pm0.02$&-\\                                
Honeycomb&$0.05\pm0.02$&$0.06\pm0.02$&$10\pm3$\\                            
0536$-$6914&$0.17\pm0.05$&$0.15\pm0.04$&-\\                                 
DEM\,L256&$0.09\pm0.03$&$0.08\pm0.02$&-\\                                   
N\,159&$0.5\pm0.2$&$0.4\pm0.2$&-\\                                        
DEM\,L316B&$0.15\pm0.04$&$0.16\pm0.05$&-\\                                 
J0550$-$6823&$0.05\pm0.02$&$0.05\pm0.02$&-\\                                
B\,0450$-$6927&$0.14\pm0.04$&$0.11\pm0.03$&-\\                              
0454$-$7005&$0.012\pm0.004$&$0.014\pm0.004$&-\\                            
DEM\,L214&$0.009\pm0.003$&$0.011\pm0.003$&-\\                              
MCELS\,J04496921&$0.12\pm0.03$&$0.10\pm0.03$&-\\                           
N\,186D&$0.05\pm0.02$&$0.05\pm0.02$&-\\                                    
0521$-$6542&$0.04\pm0.01$&$0.04\pm0.02$&-\\                                
MCELS\,J0448$-$6658&$0.018\pm0.005$&$0.020\pm0.006$&$3.0\pm0.8$\\           
N\,4&$0.05\pm0.02$&$0.06\pm0.02$&-\\                                       
RXJ0507$-$68&$0.025\pm0.007$&$0.028\pm0.008$&-\\                            
B\,0528$-$7038&$0.010\pm0.003$&$0.017\pm0.005$&-\\                        
0538$-$693&$0.08\pm0.03$&$0.08\pm0.02$&-\\                                 
0538$-$6922&$0.08\pm0.03$&$0.08\pm0.03$&-\\                               
B\,0449$-$693&$0.11\pm0.03$&$0.12\pm0.03$&$10\pm3$\\                       
J0508$-$6830&$0.04\pm0.02$&$0.08\pm0.03$&$27\pm7$\\                        
J0511$-$6759&$0.023\pm0.006$&$0.028\pm0.008$&$3.0\pm0.8$\\               
J0514$-$6840&$0.022\pm0.006$&$0.021\pm0.006$&-\\                            
J0517$-$6759&$0.05\pm0.02$&$0.06\pm0.02$&-\\                                
\hline
\end{tabular}
\end{table}

We estimate the amount of dust that is removed by the SNRs by comparing the dust column 
density towards the SNR with that of within a 20 pc thick annulus. Assuming that
both reflect the ISM dust in those directions, the difference will
correspond to the amount of dust that was removed (or added) by the SNR. In
Table~\ref{tbl3} we list the determinations for nearly all (60) SNRs. The average dust column density within the SNR 
is $N_{{\rm in}}$ and within an annulus surrounding the SNR is $N_{{\rm out}}$. We
only quote the values for the removed dust mass $M$ for those cases we are
reasonably confident about (for these SNRs we believe that they did remove dust
and that the difference in dust mass is not caused by the accidental position of the SNR
next to the cloud). We find that the latter (15 SNRs) show a wide range in dust removal, but with
a mean of 6.5 M$_\odot$ (and a median of $\sim4$ M$_\odot$) well above the
typically inferred amounts of dust that are produced in the ejecta ($<1$
M$_\odot$). If this is representative of the SNR population as a whole we
would obtain $M_1\sim390$ M$_\odot$. If instead of the mean, we use the median value
of these 15 SNRs, we will have $M_1^{\prime}=240$ M$_\odot$. In Table~\ref{tbl3}, the errors
for $N_{{\rm in}}$ and $N_{{\rm out}}$ include the uncertainties of the fitting as well as the uncertainty of $\kappa$
which is $\sim$25\% \citep{Gordon14}. The uncertainties of $M$ are found by adding in quadrature the combined 
uncertainties of the fitting of $N_{{\rm in}}$ and $N_{{\rm out}}$ together with the uncertainty of $\kappa$ and multiplying
with the areas of the objects.

\begin{figure}
\centerline{\psfig{figure=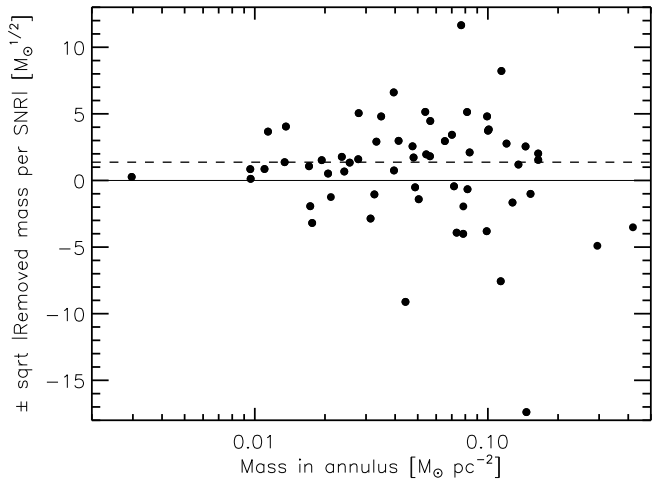,width=86mm}}  
\caption[]{Removed dust mass vs.\ dust mass in an annulus surrounding the
SNR. Square-root values are plotted to limit the dynamical range, with
negative masses ($M<0$) represented by $-\sqrt{-M}$. The median value is
indicated by the dashed line. \label{sputt_rr}}
\end{figure}

Because this may be biased, we attempt to estimate the result for the entire
sample of SNRs, including negative values, as follows:
\begin{equation}
M_2\ =\ \sum\limits_{i=1}^N\
        \left[\pi (D/2)^{2}\times(N_{{\rm out}}-N_{{\rm in}})\right],
\end{equation}
where $D$ is the SNR diameter (in pc), and the sum is over $N$ objects. 
Using this approach we obtain $M_2=-13.6$ M$_\odot$ of dust removed by $N=60$ SNRs in the
LMC. This result is driven by a few severe outliers for which the individual
estimates are particularly uncertain (Figure~\ref{sputt_rr}), yet 40 out of 60 SNRs have less dust within
their diameter than in the annulus. 

Our third estimate is based on the median rather than the sum (or average):
\begin{equation}
M_2^\prime\ =\ 60\times{\rm Median}_{i=1}^N\
                \left[\pi (D/2)^{2} \times(N_{{\rm out}}-N_{{\rm in}})\right].
\end{equation}
Now, we obtain $M_2^\prime\sim113$ M$_\odot$. This, on the other hand, may
exclude rare, but real, more prominent contributors to dust removal.

\begin{figure}
\centerline{\psfig{figure=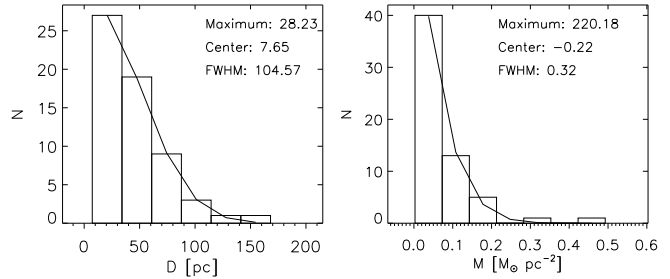,width=86mm}}  
\caption[]{{\it Left:} Distribution of the diameters of the SNRs in our sample. {\it Right:}
Distribution of the mass within the SNRs in our sample. \label{Dpc}}
\end{figure}

An alternative way of estimating the combined effect of the SNR population
within the LMC is based on an empirical Monte Carlo simulation. We generate
10,000 sets of values for mass ($m$) and for diameter ($D$). These are
drawn from the positive domain of a Gaussian, where the
width is set by $\sigma= 0.13$ M$_\odot$ pc$^{-2}$ for $m$ and $\sigma=44.4$ pc for $D$ (Figure~\ref{Dpc}). For each of these pairs of values
($m$, $D$) we derive the removed mass, from which we obtain the average
removed mass per SNR and thence
\begin{equation}
M_3\ =\ 60\times\left\langle\pi (D/2) ^{2}
          \times m\times0.09\times\frac{40}{60}\right\rangle,
\end{equation}
where $40/60$ is the probability that the SNR is removing the dust that we
see in its direction -- 40 is the number of SNRs with $(N_{{\rm out}}/N_{{\rm in}})>1$, in our sample
of 60 SNRs; 0.09 is the average fraction of removed dust mass as compared to
the dust mass that is seen in the direction of the SNR ($\langle
N_{{\rm out}}/N_{{\rm in}}\rangle=1.09$). We thus obtain $M_{3}\sim 389$ M$_\odot$. The results of these five
estimates are given in Table~\ref{tbl6}. 

\begin{table}
\begin{center}
\caption{Removed mass by SNRs in the LMC according to various methods (see text).\label{tbl6}}
\begin{tabular}{llr}
\hline\hline
ID                &  Method         &  Removed mass M$_\odot$\\
\hline
M$_{1}$           &  mean, good     &  390                     \\
M$_{1}^{\prime}$  &  median, good   &  240                     \\
M$_{2}$           &  mean, all      &  $-$13                    \\
M$_{2}^{\prime}$  &  median, all    &  113                     \\
M$_{3}$           &  Monte Carlo    &  389                     \\
\hline
\end{tabular}
\end{center}
\end{table}

While extrapolation of the 'cleanest sample', M$_{1}$ and Monte Carlo
give high values, M$_{2}$ is too influenced by the fore/background of 3$-$4 very massive SNRs 
that have $N_{{\rm out}}<N_{{\rm in}}$. We will adopt the average of the values in Table~\ref{tbl6}, i.e.\ 224 M$_\odot$ removed by the whole sample.

The different approaches that we followed in the estimation of the removed dust give a range of a factor of 
$\pm$2, but the uncertainty of $\kappa$ causes also a similar factor (we compared $\kappa$ from this work with the one
from \citealt{Gordon14} and found that our dust masses are $\sim$60\% of the masses for their SMBB model due to $\kappa$ uncertainty). Therefore, our 
estimate of the total error of removed mass is a factor of three.  

For a SN rate in the LMC of $dN/dt=10^{-2}$ yr$^{-1}$ \citep{Filipovic98} and a mean lifetime
$\tau_{\rm SNR}=10^4$ yr for SNRs to be visible \citep{vandenbergh04}, the
number of SNRs existing in the LMC should be $N=\tau_{\rm SNR}\times dN/dt=100$, i.e.\ double the sample considered here.
Correcting for this, we deduce a total mass of removed dust of $M=373$
M$_\odot$, within the range 124$-$1119 M$_{\odot}$. Thus, under the assumption that the dust is sputtered, we derive 
a dust destruction rate by SNRs in the LMC of 
$dM/dt=M/\tau_{\rm SNR}=0.037$ M$_\odot$ yr$^{-1}$ within a range of 0.012--0.11 M$_\odot$ yr$^{-1}$. For a total interstellar dust
mass in the LMC of $M_{\rm dust}=7.3\times10^5$ M$_\odot$ \citep{Gordon14} this
would imply an interstellar dust lifetime of $\tau_{\rm dust}=M_{\rm
dust}/(dM/dt) \sim 2 \times10^7$ yr within a range 0.7$-$6 $\times10^7$ yr. Of course, not all of the interstellar dust
is affected by SNRs to the same degree, and some dust may survive a lot
longer.

\subsection{Thickness of the dust layer in the LMC}\label{thickness}

We can use the $N_{{\rm out}}/N_{{\rm in}}$ values from Table~\ref{tbl3} to estimate the thickness,
$d$, of the dust layer within the LMC -- and hence the average
density -- if we assume that all dust has been removed from within the SNRs.
In that case, and assuming SNRs are spherical with diameter $D$ and fully
embedded within the dust layer, the volume in a column with area
$A=\pi(D/2)^2$ that contains dust is $Ad$ outside the SNR and
$Ad-\frac{4}{3}\pi(D/2)^3$ in the direction of the SNR. Dividing by $A$, we
obtain column lengths of $d$ and $d-\frac{2}{3}D$, respectively. Assuming a
constant density, then the surface mass densities would compare as
$N_{{\rm in}}/N_{{\rm out}}=1-\frac{2D}{3d}$ and hence we could estimate a thickness (in pc)
\begin{equation}
d\ =\ \frac{2}{3}D\left(1-\frac{N_{{\rm in}}}{N_{{\rm out}}}\right)^{-1},
\end{equation}
and average dust volume density $\rho=N_{{\rm out}}/d$ in ${\rm M_{\odot}}$ pc$^{-3}$.

We thus obtain a median value for the thickness of the dust layer within the LMC to be $d\sim107$ pc.

\section{Conclusions}\label{conclusions}

We present the first FIR and submm analysis of the population of
61 SNRs in the LMC, based on {\it Herschel} images from the HERITAGE survey
at 100, 160, 250, 350 and 500 $\mu$m in combination with {\it Spitzer}
24- and 70-$\mu$m images. These {\it Herschel} data allow us to estimate the 
mass of the cold interstellar dust. To that aim, we produce maps of 
dust mass and temperature. We reach the following conclusions:

\begin{itemize}
\item{Although the FIR surface brightness of SNRs is very similar to the one of the ISM, it is slowly decreasing with
time, meaning that SNRs cool down and/or dust is removed from
them, or the dust is sputtered. The radio surface brightness is weakly correlated with that of the FIR.}

\item{There is no evidence for large amounts of dust having formed and
survived in SNRs. In fact, most of the dust seen in our maps is pre-existing.}

\item{If SNRs are ``empty'' in terms of dust, then we estimate a typical
thickness of the ISM dust layer within the LMC of $\sim107$ pc.}

\item{The ISM is generally denser around core-collapse SNRs than type Ia, but
significant variations are seen between individual SNRs of either type.}

\item{We argue that SNRs sputter and heat cold interstellar dust by their hot plasma
and shocks, which is evident from the dust and temperature maps. The data presented here do not however 
exclude the possibility that dust is pushed out of the sightline.}

\item{The amount of removed dust for all SNRs in LMC is estimated to
be $\sim373^{+746}_{-249}$ M$_\odot$ (3.7$^{+7.5}_{-2.5}$ M$_{\odot}$ per SNR). Under the
assumption that all of that dust is sputtered, we derive a dust destruction rate of $0.037^{+0.075}_{-0.025}$ M$_\odot$ yr$^{-1}$ and 
thus a lifetime of interstellar dust in the regions close to SNRs of $2^{+4.0}_{-1.3}\times10^7$ yr.}
\end{itemize}

\acknowledgments{We would like to thank the referee for her/his constructive report and Dr. Eli Dwek for 
helpful advice. ML acknowledges an ESO/Keele studentship. DU acknowledges support from the Ministry of Education, Science and 
Technological Development of the Republic of Serbia through project No. 176005.}


\appendix

\section{FIR atlas of SNRs in the Large Magellanic Cloud: maps of dust mass and temperature}

We present the maps of dust mass and temperature for the remaining SNRs 
from this work in the online supported material, here: \\
http://www.napravisajt.net/MashaLakicevic \\
or here: \\
http://www.astro.keele.ac.uk/$\sim$jacco/papers/Appendix\_Lakicevic2015.pdf 

\end{document}